\documentclass[12pt,psamsfonts,reqno]{amsart} 

\theoremstyle{definition}

\theoremstyle{remark}

\newtheorem*{remark*}{Remark}

\newcommand {\U} {\mathrm{U}}

\newcommand{\ket}[1]{\left |  #1 \right \rangle}

\newcommand {\CalE} {\mathcal E}

\newcommand {\CalO} {\mathcal O}
\newcommand {\CalZ} {\mathcal Z}
\newcommand {\CalN} {\mathcal N}

\newcommand {\CalL} {\mathcal L}

\newcommand {\CalM} {\mathcal M}

\newcommand {\CalS} {\mathcal S}
\newcommand {\CalT} {\mathcal T}

\newcommand {\CalX} {\mathcal X}

\newcommand {\BI}   {\mathbb I}

\newcommand {\BR}   {\mathbb R}
\newcommand {\BZ}   {\mathbb Z}
\newcommand {\BC}   {\mathbb C}

\newcommand{\bV}{\mathbf{V}}

\newcommand{\bH}{\mathbf{H}}
\newcommand{\bM}{\mathbf{M}}
\newcommand{\bN}{\mathbf{N}}
\newcommand{\bK}{\mathbf{K}}
\newcommand{\bQ}{\mathbf{Q}}

\newcommand{\bX}{\mathbf{X}}

\newcommand{\bY}{\mathbf{Y}}

\newcommand{\msS}{\mathscr{S}}
\newcommand{\msW}{\mathscr{W}}

\newcommand{\sk}{\mathsf{k}}

\newcommand{\sP}{\mathsf{P}}
\newcommand{\sQ}{\mathsf{Q}}

\newcommand{\sT}{\mathsf{T}}

\newcommand{\sY}{\mathsf{Y}}

\newcommand{\fq}{\mathfrak{q}}
\newcommand{\fQ}{\mathfrak{Q}}

\DeclareMathOperator{\ch}{ch}
\DeclareMathOperator{\sch}{sch}

\DeclareMathOperator{\Ker}{Ker}

\DeclareMathOperator{\Hom}{Hom}

\DeclareMathOperator{\tr} {tr}
\DeclareMathOperator{\Tr} {Tr}
\DeclareMathOperator{\str} {str}
\DeclareMathOperator{\sdet} {sdet}
\DeclareMathOperator{\rk} {rk}

\DeclareMathOperator{\Id} {Id}

\numberwithin{equation}{section}

\usepackage[margin=1in]{geometry}

\usepackage{amsmath}
\usepackage{amssymb}
\usepackage{amsxtra}
\usepackage{amscd}
\usepackage[mathscr]{euscript}

\usepackage[pagebackref=false]{hyperref}
\hypersetup{colorlinks = false, ,extension = notused,linkcolor = blue, anchorcolor = red,citecolor = blue,filecolor = red,pagecolor = red,urlcolor = blue}
\usepackage[numbers,sort&compress]{natbib}
\usepackage{hypernat}

\usepackage{iftex}
\ifxetex
        \usepackage{fontspec}
\else
\fi

\usepackage{float}
\usepackage{tikz-cd}
\usetikzlibrary{decorations.pathreplacing}
\usepackage[mathscr]{euscript}

\begin{document}

\title{Super instanton counting and localization}

\author{Taro Kimura}
\author{Vasily Pestun}

\address{Taro Kimura, Keio University, Japan}
\address{Vasily Pestun, IHES, France} 

 \begin{abstract}
  We study the super instanton solution in the gauge theory with U$(n_{+}| n_{-})$ gauge group.
  Based on the ADHM construction generalized to the supergroup theory, we derive the instanton partition function from the super instanton moduli space through the equivariant localization.
  We derive the Seiberg--Witten geometry and its quantization for the supergroup gauge theory from the instanton partition function, and study the connection with classical and quantum integrable systems.
  We also argue the brane realization of the supergroup quiver gauge theory, and possible connection to the non-supergroup quiver gauge theories.
 \end{abstract}

\maketitle 

\tableofcontents

\parskip=4pt

\section{Introduction}\label{sec:Introduction}

Supergroup and superalgebra provide a natural framework to describe supersymmetric quantum field theories, involving both bosonic and fermionic symmetries~\cite{Kac:1977em,Frappat:1996pb,Quella:2013oda}.
They are typically used for the global symmetries of quantum field theories, however one may utilize them for the local gauge symmetry.
It has been known that the supergroup gauge theory becomes inevitably non-unitary due to violation of spin-statistics theorem, therefore it has not been seriously thought of as a physically important system.
Even with such a difficulty, studying the supergroup gauge theory seems quite important to understand the profound structure of quantum field theories.
Actually recent studies on the supergroup gauge theory elucidate its interesting properties, e.g., D-brane realization~\cite{Vafa:2001qf,Okuda:2006fb,Nekrasov:2017cih,Nekrasov:2018xsb}, non-unitary holography~\cite{Vafa:2014iua}, supergroup Chern--Simons theory~\cite{Mikhaylov:2014aoa}, dynamical space-time signature change~\cite{Dijkgraaf:2016lym}.

In this paper, we study the anti-selfdual Yang--Mills connection in the supergroup gauge theory, that we call the super instanton, and non-perturbative aspects of supersymmetric gauge theory originating from the super instanton.
The instanton has a systematic construction, a.k.a., the Atiyah--Drinfeld--Hitchin--Manin (ADHM) construction~\cite{Atiyah:1978ri}.
We first generalize the ADHM construction to the supergroup gauge theory, and define the corresponding moduli space of the super instanton defined with the ADHM variables.
We show that two instanton numbers, $k_+$ and $k_-$, are necessary to characterize the super instanton because the vector space $K = \BC^k$ is replaced with the graded one $K = \BC^{k_+|k_-}$ in the ADHM construction for the supergroup gauge theory.
The physical instanton charge is given by $\sk = \operatorname{sdim}_\BC K =  k_+ - k_-$.
Here the minus sign for negative instanton comes from using $\operatorname{str} F \wedge F$ to define the topological charge.
Hence these instanton numbers are actually interpreted as the topological charges for {\em positive} and {\em negative} instantons.
We remark that the negative instanton is an anti-selfdual configuration with a negative topological charge, while {\em the anti-instanton} is a selfdual configuration with a negative charge.

We then derive the instanton partition function from the super instanton moduli space through the equivariant localization.
We obtain the super-analog of the Losev--Moore-Nekrasov--Shatashvili (LMNS) integral formula~\cite{Losev:1997tp,Moore:1997dj}, and also the Nekrasov-type combinatorial formula~\cite{Nekrasov:2002qd,Nekrasov:2003rj} for the instanton partition function.
We use this partition function to derive the Seiberg--Witten geometry describing the Coulomb branch of the moduli space of the vacua for 4d $\CalN = 2$ supergroup gauge theory~\cite{Seiberg:1994rs,Seiberg:1994aj}.
It has been pointed out that 4d $\CalN = 2$ supergroup gauge theory is equivalent to the non-supergroup gauge theory with a specific matter content~\cite{Dijkgraaf:2016lym}.
However, we have to be careful about this point, in particular, if turning on the equivariant parameters for the space-time rotation denoted by $(q_1,q_2) = (e^{\epsilon_1}, e^{\epsilon_2}) \in \mathrm{U}(1)^2 \subset \mathrm{SO}(4)$.
We show that a proper assignment of the equivariant parameters is necessary for the supergroup gauge theory, and the Seiberg--Witten geometry with generic equivariant parameters is described by the $qq$-character, which is the double quantum deformation of the character of representations associated with quiver, in a similar manner to the non-supergroup gauge theory~\cite{Nekrasov:2015wsu,Nekrasov:2016qym,Nekrasov:2016ydq,Nekrasov:2017rqy,Nekrasov:2017gzb,Kimura:2015rgi,Kimura:2016dys,Kimura:2017hez,Bourgine:2015szm,Bourgine:2016vsq}.

For 4d $\CalN = 2$ gauge theory, there is a geometric correspondence between its Coulomb branch and the phase space of the classical integrable system~\cite{Gorsky:1995zq,Martinec:1995by,Donagi:1995cf,Seiberg:1996nz}, which is further promoted to the quantum integrable system through the equivariant deformation of the gauge theory, a.k.a., the Bethe/Gauge correspondence~\cite{Nekrasov:2009zz,Nekrasov:2009rc,Nekrasov:2013xda}.
We study such a geometric correspondence to the algebraic integrable systems for the supergroup gauge theory, and discuss possible implications for those.
We also discuss the brane construction of 4d $\CalN=2$ supergroup quiver gauge theory, and provide another argument to describe the supergroup gauge theory in terms of the non-supergroup gauge theory.

The remaining part of this paper is organized as follows:
In Sec.~\ref{sec:ADHM0}, we start with the ADHM construction for U$(n)$ gauge theory, and then generalize to the U$(n_+|n_-)$ supergroup gauge theory by replacing the vector spaces appearing in the construction with the graded ones.
In Sec.~\ref{sec:counting}, we derive the instanton partition function based on the ADHM construction for the supergroup gauge theory through the equivariant localization.
We derive both the LMNS-type integral formula and the Nekrasov-type combinatorial formula of the partition function from the super instanton counting.
In Sec.~\ref{sec:qq-ch}, we discuss the Seiberg--Witten geometry and its quantization from the super instanton partition function.
We show that the argument becomes essentially parallel with the non-supergroup gauge theory by replacing polynomial functions with rational functions for the supergroup gauge theory.
In Sec.~\ref{sec:Bethe/Gauge} we discuss the Bethe/Gauge correspondence for the supergroup gauge theory.
We study the asymptotics of the instanton partition function in the Nekrasov--Shatashvili (NS) limit $q_2 \to 1$ $(\epsilon_2 \to 0)$.
The resultant saddle point equation turns out to be equivalent to the Bethe equation of the corresponding quantum integrable system.
We provide a possible interpretation of the Bethe equation obtained from the supergroup gauge theory.
In Sec.~\ref{sec:brane}, we discuss the Hanany--Witten-type brane construction of 4d $\CalN = 2$ supergroup quiver gauge theory together with the negative brane.
We show the connection to the non-supergroup gauge theory through the gauging/Higgsing procedure.
In Sec.~\ref{sec:decoupling}, we provide another argument to describe the supergroup gauge theory in terms of the non-supergroup theory through the decoupling trick.
In Sec.~\ref{sec:outlook}, we conclude with possible future directions.

\subsection*{Acknowledgements}

We would like to thank Misha Bershtein, Heng-Yu Chen, Norton Lee, and Nikita Nekrasov for correspondence and comments.
TK is grateful to Insitute des Hautes \'Etudes Scientifiques for kind hospitality where a part of this work was carried out.
The work of TK was supported in part by 
JSPS Grant-in-Aid for Scientific Research (No.~JP17K18090), the MEXT-Supported Program for the Strategic Research Foundation at Private Universities ``Topological Science'' (No.~S1511006), JSPS Grant-in-Aid for Scientific Research on Innovative Areas ``Topological Materials Science'' (No.~JP15H05855), and ``Discrete Geometric Analysis for Materials Design'' (No.~JP17H06462).
The research  on this project has received funding from the European Research Council (ERC) under the European Union's Horizon 2020 research and innovation program (QUASIFT grant agreement 677368).

\section{ADHM construction}\label{sec:ADHM0}

The ADHM construction is a systematic method to construct the instanton configuration 
on spacetime $\BR^4$ compactified to $S^4$~\cite{Atiyah:1978ri}.
Let $\CalS$ denote the spacetime $\BR^{4} \simeq \BC^2$.
In this Section, we first review the construction with the ordinary non-supergroup theory.
Then we discuss a generalization to the supergroup gauge theory.

\subsection{U$(n)$ theory}

Let $K$ and $N$ be the vector spaces, $K = \BC^k$ and $N = \BC^n$.
The $k$-instanton configuration for U$(n)$ gauge theory is constructed with the ADHM matrices, $B_{1,2} \in \Hom(K,K)$, $I \in \Hom(N,K)$, $J \in \Hom(K,N)$, obeying the  ADHM equations
\begin{align}
 \mu_\BR = 0 \, , \qquad \mu_\BC = 0
 \, ,
 \label{eq:ADHMeq}
\end{align}
where the moment maps $(\mu_\BR, \mu_\BC)$ are 
\begin{subequations} \label{eq:mom_maps}
\begin{align}
 \mu_\BR & := [ B_1, B_1^\dag ] + [ B_2, B_2^\dag ] + I I^\dag - J^\dag J
 \, ,
 \\
 \mu_\BC & := \left[ B_1, B_2 \right] + IJ
 \, ,
\end{align}
\end{subequations}
modulo $\U(k)$ action of the form
\begin{align}
 (v) \cdot (B_1, B_2, I, J)
 = (v B_1 v^{-1}, v B_2 v^{-1}, v I , J v^{-1})
\end{align}
for $v \in \mathrm{U}(k)$.
The resulting ADHM moduli space also receives the induced action of U$(n)$ from
\begin{align}
 (\nu) \cdot (B_1, B_2, I, J)
 = (B_1 ,  B_2 ,  I \nu, \nu J )
\end{align}
for  $\nu \in \mathrm{U}(n)$.

Let $(z_1, z_2) \in \CalS = \BC^2$ be the spacetime coordinate.
We define the Dirac operator $D^\dag : K \otimes \CalS \oplus N \to K \otimes \CalS$ as
\begin{align}
 D^\dag =
 \begin{pmatrix}
  B_1 - z_1 & B_2 - z_2 & I\\
  - B_2^\dag + \bar{z}_2 & B_1^\dag - \bar{z}_1 & - J^\dag
 \end{pmatrix}
 \label{eq:Dirac_op}
\end{align}
Due to the ADHM equation \eqref{eq:ADHMeq}, we have $D^\dag D : K \otimes \CalS \to K \otimes \CalS$
\begin{align}
 D^\dag D = \Delta \otimes \Id_\CalS
\end{align}
where $\Delta : K \to K$ behaves $\Delta \sim |z|^2$ as $z \to \infty$.
We consider the normalized zero modes $\Psi$ in the kernel of the Dirac operator $\Psi \in \Ker D^\dag$,
so that $D^\dag \Psi = 0$  and  the normalization of the zero mode $\Psi : N \to K \otimes \CalS \oplus N$ is fixed $\Psi^\dag \Psi = \Id_N$.

We remark that
\begin{align}
 \Psi \Psi^\dag = \Id_{K \otimes \CalS \oplus N} - D \frac{1}{\Delta} D^\dag =: P
\end{align}
is a projector from $K \otimes \CalS \oplus N$ onto $N$ with $P^2 = P$.
Then the connection constructed with the Dirac zero mode
\begin{align}
 A = \Psi^\dag d \Psi
 \label{eq:ADHM_connection}
\end{align}
is anti-selfdual, and the Yang--Mills action is evaluated using Osborn's formula~\cite{Osborn:1981yf} as follows:
\begin{align}
 - \frac{1}{16 \pi^2} \int d^4 x \, \tr_N F_{\mu\nu} F^{\mu\nu}
 = \frac{1}{16 \pi^2} \int d^4 x \, \partial^2 \partial^2 \tr_K \log \Delta^{-1}
 = k > 0
 \, ,
\end{align}
which proves the ADHM construction gives the $k$-instanton anti-selfdual configuration for U($n$) gauge theory. We remark that a selfdual configuration, not defined by the presented ADHM construction, gives rise to a negative instanton number $k < 0$, while the anti-selfdual solution gives $k>0$.

  \subsection{U$(n_+|n_-)$ theory}

  The ADHM construction is a consequence of the Nahm transformation, which is a correspondence between $k$-instanton solution in U($n$) gauge theory on a four-torus $T^4$ and $n$-instanton solution in U($k$) gauge theory on the dual torus $\check{T}^4$.
  Taking the large radii limit of the torus, we obtain the four-dimensional U($n$) instanton on $\BC^2$ and the zero-dimensional ADHM equation, respectively.
 From this point of view, to preserve Nahm duality in the supergroup case, it is
 natural to allow graded $(k_+|k_-)$-instanton configuration for $\U(n_{+}| n_{-})$ gauge theory.

  We define the graded vector spaces
 \begin{align}
  K = \BC^{n_+|n_-} = K^+ \oplus K^-
  \, , \qquad
  N = \BC^{n_+|n_-} = N^+ \oplus N^-  
  \, ,
  \label{eq:supervect_sp}
 \end{align}
 where
 \begin{align}
  K^\sigma = \BC^{k_\sigma}
  \, , \qquad
  N^\sigma = \BC^{n_\sigma}
  \qquad \text{for} \qquad
  \sigma = \pm
  \, .
 \end{align}
 Then the remaining process is totally parallel with the previous case:
 Define the ADHM variables
 \begin{align}
  B_{1,2} \in \Hom(K,K)
  \, , \qquad
  I \in \Hom(N,K)
  \, , \qquad
  J \in \Hom(K,N)
  \, .
 \end{align}
 Here, since $K$ and $N$ are defined as \eqref{eq:supervect_sp}, these variables are represented as supermatrices, whose diagonal blocks consist of commuting variables, and off-diagonal ones consist of anti-commuting (Gra{\ss}mann) variables.
 
 We impose the ADHM equations \eqref{eq:ADHMeq} with the same moment maps \eqref{eq:mom_maps}.
 In this case, we have supergroup transformations, U($k_+|k_-$) and U($n_+|n_-$), acting on the ADHM variables,
 \begin{align}
 (v,\nu) \cdot (B_1, B_2, I, J)
 = (v B_1 v^{-1}, v B_2 v^{-1}, v I \nu, \nu^{-1} J v^{-1})
\end{align}
for $v \in \mathrm{U}(k_+|k_-)$, $\nu \in \mathrm{U}(n_+|n_-)$.
 We define formally the same Dirac operator \eqref{eq:Dirac_op}, and consider the connection constructed from the zero mode $\Psi$ as \eqref{eq:ADHM_connection}.
 The connection $A = \Psi^\dag d \Psi$, which is anti-selfdual, transforms under the supergroup U($n_+|n_-$), and the Yang--Mills action is again evaluated using Osborn's formula~\cite{Osborn:1981yf}
\begin{align}
 - \frac{1}{16 \pi^2} \int d^4 x \, \str_N F_{\mu\nu} F^{\mu\nu}
 = \frac{1}{16 \pi^2} \int d^4 x \, \partial^2 \partial^2 \str_K \log \Delta^{-1}
 = k_+ - k_- =: \sk
 \, ,
 \label{eq:super_inst_num}
\end{align}
which shows that $(k_+|k_-)$-instanton configuration has a topological charge $\sk = k_+ - k_-$, implying the $k_\pm$-sectors describe $k_+$ positive and $k_-$ negative instantons, respectively.
Notice that the Killing form in the usual definition of the topological charge
has been replaced by the Killing form for the super Lie algebra given by the supertrace operator $\str$.

 We emphasize that one can consider an anti-selfdual configuration having a negative instanton number $\sk < 0$, which is possible only with the supergroup structure.
 A similar argument for the vortex system associated with the supergroup is found in~\cite{Okazaki:2017sbc}.

\section{Instanton counting with supergroups}\label{sec:counting}

In this Section, we perform the instanton counting to obtain the partition function for supergroup gauge theory.
We show several consistent approaches: The first is the direct integration over the ADHM quiver, which yields the LMNS-type contour integral formula.
The second is the equivariant localization with the sheaves on the ADHM moduli space.
The contribution to the partition function is localized on the fixed point under the equivariant action, and each fixed point corresponds to each pole in the contour integral formula.
The third focuses on the instanton partition function, and show the combinatorial formula in terms of the partition characterizing the fixed point.

\subsection{Integrating over ADHM moduli space}\label{sec:ADHM}

\subsubsection{$\mathrm{U}(n)$ theory}

Let $\CalM_{n,k}$ be the ADHM moduli space for $k$-instanton sector in U$(n)$ gauge theory, which is described by the ADHM variables $(B_1,B_2,I,J)$ satisfying the ADHM equations \eqref{eq:ADHMeq},
\begin{align}
 \CalM_{n,k} = \left\{ (B_1, B_2, I, J) \mid \mu_\BR = \mu_\BC = 0 \right\} /\,\mathrm{U}(k)
 \, .
\end{align}
The moduli space naively defined here has singularity, and thus a proper regularization is necessary to apply the equivariant localization:
With the stability condition, $K = \BC[B_1,B_2] I$, the regularized moduli space is given as follows,
\begin{align}
 \widetilde{\CalM}_{n,k} = \left\{ (B_1, B_2, I, J) \mid \mu_\BC = 0 \right\}/\!/\,\mathrm{GL}(k,\BC)
\end{align}
which is equivalent to the moduli space of the noncommutative instanton $\CalM_{n,k}^\zeta = \{ \mu_\BR = \zeta_{>0} \Id_K, \mu_\BC=0 \}/\mathrm{U}(k)$.
The equivariant actions on the ADHM variables are
\begin{align}
 (v, \nu) \cdot (B_1, B_2, I, J)
 = (v B_1 v^{-1}, v B_2 v^{-1}, v I \nu^{-1}, \nu J v^{-1})
\end{align}
for $v \in \mathrm{GL}(k,\BC)$ and $\nu \in \mathrm{GL}(n,\BC)$, and
\begin{align}
 (q_1, q_2) \cdot (B_1, B_2, I, J)
 = (q_1 B_1, q_2 B_2, I, q J)
\end{align}
where $(q_1, q_2) = (e^{\epsilon_{1}}, e^{\epsilon_2}) \in \BC^\times \times \BC^\times$ is the Cartan element of GL(2,$\BC$) group 
associated with the spacetime rotation, and we define $q := q_1 q_2$.
Thus the fixed point under the equivariant torus action is given by
\begin{align}
 q_1 B_1 = v B_1 v^{-1}
 \, , \qquad
 q_2 B_2 = v B_2 v^{-1}
 \, , \qquad
 I \nu = v I
 \, , \qquad
 q \nu^{-1} J = J v^{-1}
 \, .
 \label{eq:ADHM_fp}
\end{align}
Let $v = e^\phi$, $\nu = e^a$ where $\phi \in \operatorname{Lie}(\mathrm{GL}(k))$, $a = \operatorname{diag}(a_1,\ldots,a_n) \in \operatorname{Lie}(\mathrm{GL}(n))$ with $I = \bigoplus_{\alpha=1}^n I_\alpha$, $J = \bigoplus_{\alpha=1}^n J_\alpha$, then the infinitesimal analog is
\begin{align}
 \left[ \phi, B_{1} \right] = \epsilon_{1} B_{1}
 \, , \qquad
 \left[ \phi, B_2 \right] = \epsilon_2 B_2
 \, , \qquad 
 a_\alpha I_\alpha = \phi I_\alpha
 \, , \qquad
 (a_\alpha - \epsilon_+) J_\alpha = J_\alpha \phi
 \, ,
 \label{eq:ADHM_fp_inf} 
\end{align}
Namely, $I_\alpha$ $(J_\alpha)$ is the right (left) eigenvector of $\phi$.
One can show that
\begin{subequations}\label{eq:phi_ev}
\begin{align}
 \phi \left( B_1^{s_1-1} B_2^{s_2-1} I_\alpha \right)
 & =
 \left( a_\alpha + (s_1 - 1) \epsilon_1 + (s_2 - 1) \epsilon_2 \right) \left( B_1^{s_1-1} B_2^{s_2-1} I_\alpha \right)
 \, , \label{eq:phi_rev} \\
 \left( J_\alpha B_1^{s_1-1} B_2^{s_2-1} \right) \phi
 & = \left( a_\alpha - s_1 \epsilon_1 - s_2 \epsilon_2 \right) \left( J_\alpha B_1^{s_1-1} B_2^{s_2-1} \right)
 \label{eq:phi_lev}
\end{align}
\end{subequations}
for $\alpha = 1, \ldots, n$ and $s_1, s_2 =1, 2, \ldots, \infty$.
The order of multiplication of $B_{1,2}$ does not matter in the eigenvectors because $J_\alpha I_\alpha=0$ at the fixed point.
Since $\phi$ is $k$-dimensional, the eigenvector \eqref{eq:phi_rev} becomes trivial for sufficiently large $s_1, s_2$, which implies the stability condition
\begin{align}
 K = \bigoplus_{\alpha=1}^n \BC[B_1,B_2] I_\alpha = \BC[B_1,B_2] I
 \, .
\end{align}
Here we only focus on the right eigenvector of $\phi$, but the left eigenvector will also play a role for supergroup theory, as discussed later.

The instanton partition function for 4d $\CalN = 2$ pure gauge theory is obtained by the equivariant integral over the ADHM moduli space
\begin{align}
 Z^\text{4d} = \sum_k \fq^k \, Z_k^\text{4d}
 \qquad \text{with} \qquad
 Z_k^\text{4d} = \int_{\widetilde\CalM_{n,k}} 1
 \, .
\end{align}
Here the instanton fugacity $\fq = \exp \left( 2 \pi \iota \tau \right)$ is defined with the complexified coupling constant
\begin{align}
 \tau = \frac{\theta}{2\pi} + \iota \frac{4\pi}{g^2}
 \label{eq:cmplx_coupling}
\end{align}
where $\theta$ is the $\theta$-angle, and $g^2$ is the Yang--Mills coupling.
We also define the imaginary unit $\iota = \sqrt{-1}$.
We instead consider the K-theoretic 5d partition function, which is given by an integral over GL($k$) in terms of the Cartan elements $(v_a)_{a = 1 \ldots k}$ with the Vandermonde determinant as a Jacobian,
\begin{align}
 \text{Jacobian:} \quad
 \prod_{a \neq b}^k
 \left( 1 - \frac{v_a}{v_b} \right)
 \, .
\end{align}
The ADHM constraint leads to a similar factor in the integrand (See, for example,~\cite{Nekrasov:2004vw})
\begin{align}
 \text{ADHM:} \quad
 \prod_{a,b}^k
 \left( 1 - q \frac{v_a}{v_b} \right)
 = (1-q)^k \prod_{a \neq b}^k
 \left( 1 - q \frac{v_a}{v_b} \right)
 \, .
\end{align}
The contributions of the ADHM variables are correspondingly given by
\begin{subequations} 
\begin{align}
 B_{1,2}: \quad &
 \prod_{a,b}^k
 \left( 1 - q_{1,2} \frac{v_a}{v_b} \right)^{-1}
 = (1 - q_{1,2})^{-k} \prod_{a \neq b}^k
 \left( 1 - q_{1,2} \frac{v_a}{v_b} \right)^{-1}
 \\
 I: \quad &
 \prod_{a=1}^k \prod_{\alpha=1}^n
 \left( 1 - \frac{\nu_\alpha}{v_a} \right)^{-1} 
 \\
 J: \quad &
 \prod_{a=1}^k \prod_{\alpha=1}^n
 \left( 1 - q \frac{v_a}{\nu_\alpha} \right)^{-1}
\end{align}
\end{subequations}
where $(\nu_\alpha)_{a=1\ldots n}$ are the Cartan elements of GL($n$).
The K-theoretic 5d partition function is then given by a contour integral:
\begin{align}
 Z_k^\text{5d} & =
 \frac{1}{k!}
 \left( \frac{1-q}{(1-q_1)(1-q_2)} \right)^k
 \oint \prod_{a=1}^k
 \frac{dv_a}{2\pi \iota v_a}
 \prod_{a \neq b}^k
 \frac{(1 - v_a/v_b)(1 - q v_a/v_b)}
      {(1 - q_1 v_a/v_b)(1 - q_2 v_a/v_b)}
 \nonumber \\ & \hspace{17em}
 \times \prod_{a=1}^k \prod_{\alpha=1}^n
 \frac{1}{(1 - \nu_\alpha/v_a)(1 - q v_a/\nu_\alpha)}
 \nonumber \\
 & =: \oint \prod_{a=1}^k
 \frac{dv_a}{2\pi \iota v_a} \, z_{n,k}^{\text{5d}}(v,\nu)
 \label{eq:LMNS_K}
\end{align}
where the factor $k!$ is the size of GL$(k)$ Weyl group, and we put $\iota = \sqrt{-1}$.
We define
\begin{align}
 z_{n,k}^{\text{5d}}(v,\nu) & =
 \frac{1}{k!}
 \left( \frac{1-q}{(1-q_1)(1-q_2)} \right)^k
 \prod_{a \neq b}^k
 \msS\left( \frac{v_a}{v_b} \right)^{-1}
 \prod_{a=1}^k \prod_{\alpha=1}^n
 \frac{1}{(1 - \nu_\alpha/v_a)(1 - q v_a/\nu_\alpha)} 
\end{align}
with the $\msS$-factor
\begin{align}
 \msS(x) = \frac{(1 - q_1 x)(1 - q_2 x)}{(1 - x)(1 - q x)}
 \, ,
 \label{eq:S-factor}
\end{align}
which obeys 
\begin{align}
 \msS(x^{-1}) = \msS(q^{-1} x)
 \, .
 \label{eq:S-ref}
\end{align}
We remark that all the variables, $(q_{1,2}, v_a, \nu_\alpha) = (e^{\epsilon_{1,2}}, e^{\phi_a}, e^{a_\alpha})$, take a value in $\BC^\times$ in the K-theoretic partition function, while $(\epsilon_{1,2}, \phi_a, a_\alpha)$ are $\BC$-variables used in the 4d partition function in the following.

In the 4d limit, the contour integral \eqref{eq:LMNS_K} is reduced to the LMNS integral~\cite{Losev:1997tp,Moore:1997dj} 
\begin{align}
 Z_k^\text{4d}
 & =
 \frac{1}{k!} \left( \frac{\epsilon_+}{\epsilon_1 \epsilon_2} \right)^k
 \oint \prod_{a=1}^k \frac{d\phi_a}{2\pi \iota}
 \prod_{a < b}^k
 \frac{\phi_{ab}^2 (\phi_{ab}^2 - \epsilon_+^2)}{(\phi_{ab}^2 - \epsilon_1^2)(\phi_{ab}^2 - \epsilon_2^2)}
 \prod_{a=1}^k \prod_{\alpha=1}^n \frac{1}{(\phi_a - a_\alpha)(\phi_a - a_\alpha + \epsilon_+)}
 \nonumber \\
 & =: \oint \prod_{a=1}^k
 \frac{d\phi_a}{2\pi \iota} \, z_{n,k}^\text{4d}(\phi,a)
 \label{eq:cont_int4d}
\end{align}
with
\begin{align}
 z_{n,k}^\text{4d}(\phi,\nu)
 =
 \frac{1}{k!} \left( \frac{\epsilon_+}{\epsilon_1 \epsilon_2} \right)^k
 \prod_{a < b}^k
 \frac{\phi_{ab}^2 (\phi_{ab}^2 - \epsilon_+^2)}{(\phi_{ab}^2 - \epsilon_1^2)(\phi_{ab}^2 - \epsilon_2^2)}
 \prod_{a=1}^k \prod_{a=1}^n \frac{1}{(\phi_a - a_a)(\phi_a - a_a + \epsilon_+)}
\end{align}
where we define $\epsilon_+ = \epsilon_1 + \epsilon_2$ and $\phi_{ab} = \phi_a - \phi_b$.
The $\msS$-factor \eqref{eq:S-factor} is reduced in the 4d limit as
\begin{align}
 \msS(x) \ \longrightarrow \ \frac{(x + \epsilon_1)(x + \epsilon_2)}{x(x + \epsilon_+)}
 \, .
\end{align}
How to pick up proper poles and evaluate the residues in the contour integral \eqref{eq:cont_int4d} is explained as follows:
First, we rewrite the integral with the ordering,
\begin{align}
 \frac{1}{k!} \oint \prod_{a=1}^k \frac{d\phi_a}{2 \pi \iota}
 \ \longrightarrow \
 \oint \frac{d\phi_k}{2 \pi \iota} \cdots \oint \frac{d\phi_2}{2\pi \iota} \oint \frac{d\phi_1}{2\pi \iota}
\end{align}
Then the first variable $\phi_1$ picks up the pole at $\phi_1 = a_\alpha$, and the following ones are recursively determined as
\begin{align}
 \phi_{a} = a_{\beta (\neq \alpha)}
 \, , \quad
 \phi_{b (< a)} + \epsilon_1
 \, , \quad 
 \phi_{b (< a)} + \epsilon_2
 \, .
\end{align}
Namely, we take the poles at
\begin{align}
 \phi_a = a_\alpha
 \, , \quad
 \phi_a = \phi_b + \epsilon_1
 \, , \quad
 \phi_a = \phi_b + \epsilon_2
 \, .
 \label{eq:pole+}
\end{align}
In the end, the poles are parametrized using the $n$-tuple Young diagram $(\lambda_\alpha)_{\alpha = 1,\ldots, n}$, where each box $s = (s_1, s_2) \in \lambda_{\alpha}$ corresponds to the pole at
\begin{align}
 \phi_a = a_\alpha + (s_1 - 1) \epsilon_1 + (s_2 - 1) \epsilon_2 
 \, .
\end{align}
Remark that this coincides with the eigenvalue of $\phi$ shown in \eqref{eq:phi_rev}.
Evaluating the residues of such poles, we obtain the combinatorial expression of gauge theory partition function~\cite{Nekrasov:2002qd,Nekrasov:2003rj,nakajima-hilbert}.

\subsubsection{$\mathrm{U}(n_+|n_-)$ theory}

We generalize the previous argument to the case with supergroup symmetry.
In this case, the resolved ADHM moduli space is given as a supergroup quotient:
\begin{align}
 \widetilde{\CalM}_{n_+|n_-,k_+|k_-}
 =
 \left\{
 (B_1,B_2,I,J)
 \mid \mu_\BC = 0 
 \right\} /\!/\, \mathrm{GL}(k_+|k_-)
 \, .
\end{align}
The equivariant action to these ADHM variables are given by
\begin{align}
 (v, \nu) \cdot (B_1, B_2, I, J)
 = (v B_1 v^{-1}, v B_2 v^{-1}, v I \nu^{-1}, \nu J v^{-1})
\end{align}
for $v \in \mathrm{GL}(k_+|k_-)$ and $\nu \in \mathrm{GL}(n_+|n_-)$, and
\begin{align}
 (q_1, q_2) \cdot (B_1, B_2, I, J)
 = (q_1 B_1, q_2 B_2, I, q J)
\end{align}
as before.
Thus the fixed point equation is also the same as \eqref{eq:ADHM_fp}.
Recalling that, in this case, the vector space $K$ consists of the positive and negative parts \eqref{eq:supervect_sp}, we use the left and right eigenvectors \eqref{eq:phi_ev} to span them, yielding the stability condition,
\begin{align}
 K^+ = \bigoplus_{\alpha=1}^{n_+} \BC[B_1, B_2] \, I_\alpha^+
 = \BC[B_1, B_2] \, I^+
 \, , \quad
 K^- = \bigoplus_{\alpha=1}^{n_-} J_\alpha^- \, \BC[B_1, B_2]
 = J^- \, \BC[B_1, B_2]
 \, .
\end{align}
where we define
\begin{align}
 I = I^+ \oplus I^-
 \, , \qquad
 J = J^+ \oplus J^-
\end{align}
and
\begin{align}
 I^\sigma = \bigoplus_{\alpha=1}^{n_\sigma} I^\sigma_\alpha
 \, , \qquad
 J^\sigma = \bigoplus_{\alpha=1}^{n_\sigma} J^\sigma_\alpha
 \, ,
\end{align}
for $\sigma = \pm$.
We remark $I_\alpha^\sigma, J_\alpha^\sigma \in K, K^\vee$ with $J_\alpha^\sigma I_\alpha^\sigma = 0$ for $\sigma = \pm$, $\alpha = 1,\ldots,n_\sigma$ at the fixed point, so that $I^- = J^+ = 0$.

The partition function is obtained by replacing all the contributions with their super analogs.
Let $(v_a^\sigma)_{\sigma = \pm, a = 1\ldots k_\sigma}$ and $(\nu_a^\sigma)_{\sigma = \pm, a = 1\ldots n_\sigma}$ be the Cartan elements of GL($k_+|k_-$) and GL($n_+|n_-$).
 The Vandermonde determinant is replaced with the Cauchy determinant
 \begin{align}
 \text{Jacobian:} \quad  
 \prod_{a \neq b}^{k_+}
 \left( 1 - \frac{v_a^+}{v_b^+} \right)
 \prod_{a \neq b}^{k_-}
 \left( 1 - \frac{v_a^-}{v_b^-} \right)
 \prod_{a = 1}^{k_+} \prod_{b = 1}^{k_-}
 \left( 1 - \frac{v_a^+}{v_b^-} \right)^{-1}
 \left( 1 - \frac{v_b^-}{v_a^+} \right)^{-1}
 \, ,
 \end{align}
and the ADHM constraint gives rise to
\begin{align}
 \text{ADHM:} \quad
 (1 - q)^{k_++k_-}
 \prod_{a \neq b}^{k_+}
 \left( 1 - q \frac{v_a^+}{v_b^+} \right)
 \prod_{a \neq b}^{k_-}
 \left( 1 - q \frac{v_a^-}{v_b^-} \right)
 \prod_{a = 1}^{k_+} \prod_{b = 1}^{k_-}
 \left( 1 - q \frac{v_a^+}{v_b^-} \right)^{-1}
 \left( 1 - q \frac{v_b^-}{v_a^+} \right)^{-1}
 \, .
\end{align}
The ADHM variables contribute as follows:
\begin{subequations}
\begin{align}
 B_{1,2}: \quad &
 (1 - q_{1,2})^{-k_+-k_-}
 \prod_{\sigma = \pm}
 \prod_{a \neq b}^{k_\sigma}
 \left( 1 - q_{1,2} \frac{v^\sigma_a}{v^\sigma_b} \right)^{-1}
 \prod_{a = 1}^{k_+} \prod_{b = 1}^{k_-}
 \left( 1 - q_{1,2} \frac{v_a^+}{v_b^-} \right)
 \left( 1 - q_{1,2} \frac{v_b^-}{v_a^+} \right) 
 \\
 I: \quad &
 \prod_{\sigma = \pm}
 \prod_{a=1}^{k_\sigma} \prod_{\alpha=1}^{n_\sigma}
 \left( 1 - \frac{\nu_\alpha^\sigma}{v_a^\sigma} \right)^{-1}
 \prod_{a=1}^{k_+} \prod_{\alpha=1}^{n_-}
 \left( 1 - \frac{\nu_\alpha^-}{v_a^+}\right)
 \prod_{a=1}^{k_-} \prod_{\alpha=1}^{n_+}
 \left( 1 - \frac{\nu_\alpha^+}{v_a^-} \right)  
 \\
 J: \quad &
 \prod_{\sigma = \pm}
 \prod_{a=1}^{k_\sigma} \prod_{\alpha=1}^{n_\sigma}
 \left( 1 - q \frac{v_a^\sigma}{\nu_\alpha^\sigma} \right)^{-1}
 \prod_{a=1}^{k_+} \prod_{\alpha=1}^{n_-}
 \left( 1 - q \frac{v_a^+}{\nu_\alpha^-} \right)
 \prod_{a=1}^{k_-} \prod_{\alpha=1}^{n_+}
 \left( 1 - q \frac{v_a^-}{\nu_\alpha^+} \right) 
\end{align}
\end{subequations}
Thus the K-theoretic 5d instanton partition function is given as 
\begin{align}
 Z^\text{5d} = \sum_{k_+,k_-} \fq^{k_+-k_-} Z_{k_+|k_-}^\text{5d}
 \label{eq:5d_super_function}
\end{align}
with a super analog of \eqref{eq:LMNS_K}
\begin{align}
 Z_{k_+|k_-}^\text{5d}
 & =
 \oint \prod_{\sigma=\pm}
 \left[
 \prod_{a=1}^{k_\sigma}
 \frac{dv_a^\sigma}{2\pi \iota v_a^\sigma} \,
 z_{n_\sigma,k_\sigma}^\text{5d}(v^\sigma,\nu^\sigma)
 \right]
 \nonumber \\
 & \quad \times
 \prod_{a=1}^{k_+} \prod_{b=1}^{k_-}
 \msS \left( \frac{v_a^+}{v_b^-} \right)
 \msS \left( \frac{v_b^-}{v_a^+} \right)
 \prod_{a=1}^{k_+} \prod_{\alpha=1}^{n_-}
 \left( 1 - \frac{\nu_\alpha^-}{v_a^+} \right) 
 \left( 1 - q \frac{v_a^+}{\nu_\alpha^-} \right)
 \prod_{a=1}^{k_-} \prod_{\alpha=1}^{n_+}
 \left( 1 - \frac{\nu_\alpha^+}{v_a^-} \right)
 \left( 1 - q \frac{v_a^-}{\nu_\alpha^+} \right) 
 \, .
 \label{eq:cont-int-K}
\end{align}
This contour integral formula looks almost the same as the $\widehat{A}_1$ quiver gauge theory involving U($n_+$) and U($n_-$) vector multiplets with $k_{+,-}$ instantons and the bifundamental hypermultiplet~\cite{Shadchin:2005cc,Dijkgraaf:2016lym}.
However, the pole structure is actually different as explained in the following.

In the 4d limit, the contour integral \eqref{eq:cont-int-K} is reduced to a super analog of the LMNS formula,
\begin{align}
 Z_{k_+|k_-}^\text{4d} & =
 \oint \prod_{\sigma = \pm} \left[ \prod_{a=1}^{k_\sigma} \frac{d\phi_a^\sigma}{2\pi\iota} z_{n_\sigma,k_\sigma}^\text{4d}(\phi^\sigma,a^\sigma) \right]
 \prod_{a=1}^{k_+} \prod_{b=1}^{k_-}
 \frac{(\phi_{ab}^{+-2} - \epsilon_1^2)(\phi_{ab}^{+-2} - \epsilon_2^2)}
      {\phi_{ab}^{+-2}(\phi_{ab}^{+-2} - \epsilon_+^2)}
 \nonumber \\
 & \qquad \times
 \prod_{a=1}^{k_+} \prod_{\alpha=1}^{n_-}
 (\phi_a^+ - a_\alpha^-) (\phi_a^+ - a_\alpha^- + \epsilon_+) 
 \prod_{a=1}^{k_-} \prod_{\alpha=1}^{n_+}
 (\phi_a^- - a_\alpha^+) (\phi_a^- - a_\alpha^+ + \epsilon_+) 
\end{align}
where we define $\phi_{ab}^{\sigma\sigma'} = \phi_a^\sigma - \phi_{b}^{\sigma'}$.
In this case, there are two sets of the integral variables $(\phi_a^\sigma)_{a = 1,\ldots,k_\sigma}$ for $\sigma = \pm$, and thus the poles are labeled by $n_{+}$-tuple and $n_-$-tuple partitions $(\lambda_\alpha^\sigma)_{\alpha = 1,\ldots, n_\sigma}$ as follows:
\begin{subequations}\label{eq:v_pole}
\begin{align}
 \phi_a^+ & = a_\alpha^+ + (s_1 - 1) \epsilon_1 + (s_2 - 1) \epsilon_2
 \quad \text{for} \quad (s_1,s_2) \in \lambda_\alpha^+
 \label{eq:v+pole} \\
 \phi_{a'}^- & = a_{\alpha'}^-  - s_1 \epsilon_1 - s_2 \epsilon_2 
 \hspace{5.9em} \text{for} \quad (s_1,s_2) \in \lambda_{\alpha'}^-
 \label{eq:v-pole}
\end{align} 
\end{subequations}
which coincide with the eigenvalues of \eqref{eq:phi_ev}.
Namely, the $v^+$-variables pick up the poles at \eqref{eq:pole+}, while the poles for the $v^-$-variables take the other options:
\begin{align}
 \phi_a^- = a_\alpha^- - \epsilon_+
 \, , \quad
 \phi_a^- = \phi_b^- - \epsilon_1
 \, , \quad
 \phi_a^- = \phi_b^- - \epsilon_2
 \label{eq:pole-}
\end{align}
In other words, the direction of contours of the $v^-$-variables are opposite to that of the $v^+$-variables.
In addition, there are poles also at $\phi_a^+ = \phi_b^-$ and $\phi_a^+ = \phi_b^- \pm \epsilon_+$ in the integrand, however, due to the factors $\phi_a^+ - a_\alpha^-$, $\phi_a^+ - \phi_b^- \pm \epsilon_{1,2}$, etc, in the numerator, such a pole does not contribute to the integral.

\subsubsection{Non-unitarity and analytic continuation}\label{eq:rem}

The supergroup partition function \eqref{eq:5d_super_function} is a double infinite series over the positive and negative instanton numbers, $k_+$ and $k_-$.
Therefore it should be interpreted as a formal Laurent series because it contains both positive and negative powers of the coupling constant $\fq$.
This {\em unbounded} behavior of the partition function seems closely related to non-unitarity of the supergroup gauge theory:
The Yang--Mills action for the supergroup theory is defined with the supertrace,
\begin{align}
 S_\text{YM}
 = - \frac{1}{16 \pi^2} \int d^4 x \, \str F_{\mu\nu} F^{\mu\nu}
 = - \frac{1}{16 \pi^2} \int d^4 x \left( \tr F^{(+)}_{\mu\nu} F^{(+)\mu\nu} - \tr F^{(-)}_{\mu\nu} F^{(-)\mu\nu} \right) 
\end{align}
where $F^{(\pm)}$ is the field strength for the positive and negative nodes, respectively.
Therefore the action is not bounded in particular due to the negative node whose contribution is opposite to the positive one.

A similar situation is found in the $\U(n_+|n_-)$ supermatrix model~\cite{AlvarezGaume:1991zc,Yost:1991ht,Efetov:1996}, where the partition function is given by
\begin{align}
 Z = \int dX \, e^{-\str V(X)}
 \, .
\end{align}
Here $X$ is a size $(n_+|n_-)$ supermatrix whose positive and negative diagonal parts are $X^{(\pm)}$ with the potential function $V(x)$.
In this case, the exponential part $\str V(X) = \tr V(X^{(+)}) - \tr V(X^{(-)})$ is interpreted as the action, which is not bounded due to the same reason, so that the supermatrix model is a non-unitary model.

Such a non-unitary supermatrix model is obtained from a physical matrix model.
Consider a two-cut matrix model associated with the symmetry breaking $\U(n_+ + n_-) \to \U(n_+) \times \U(n_-)$.
Then the $\U(n_+|n_-)$ supermatrix model is obtained from this two-cut model through the analytic continuation $n_- \to - n_-$~\cite{Dijkgraaf:2002pp,Dijkgraaf:2003xk,Marino:2009jd}.
We can apply this analytic continuation argument to the supergroup gauge theory partition function.
Actually the pole structure of the contour integral~\eqref{eq:v_pole} suggests that the equivariant parameters are converted $\epsilon_{1,2} \to - \epsilon_{1,2}$ for the negative node.
Recalling the gauge theory partition function, which is a random partition model, can be regarded as a discrete version of the size $N$ matrix model under the identification $\epsilon_{1,2} \propto N^{-1}$~\cite{Eynard:2015aea}, the analytic continuation $N \to -N$ corresponding to converting the equivariant parameters $\epsilon_{1,2} \to - \epsilon_{1,2}$.
From this point of view, the supergroup gauge theory could make sense as the analytic continuation of the unitary gauge theory.
See also Sec.~\ref{sec:decoupling}.

\subsubsection{Bion-like configuration}\label{eq:bion}

Since the topological charge of $(k_+|k_-)$-instanton is given by $\sk = k_+ - k_-$, as shown in \eqref{eq:super_inst_num}, it is natural to rewrite the partition function as a summation over the topological charge
\begin{align}
 Z = \sum_{\sk \in \BZ} \fq^{\sk} \, Z_\sk
\end{align}
where
\begin{align}
 Z_\sk = \sum_{k_+ - k_- = \, \sk} Z_{k_+|k_-}
 \, .
\end{align}
For example, regarding the topologically trivial sector $\sk = 0$, we should incorporate infinitely many non-trivial contributions
\begin{align}
 Z_{\sk = 0} = \sum_{k = 0}^\infty Z_{k|k}
 \, .
\end{align}
From this point of view, $(k|k)$-instanton plays a role as the bion-like configuration, which contributes to the zero topological charge sector.

We remark that this zero charge sector is not as the ordinary bion contribution: the bion configuration, in a usual sense, consists of instantons and {\em anti-instantons}~\cite{Argyres:2012ka,Basar:2013eka}.
See also~\cite{Nekrasov:2018pqq}.
The latter contribution should be anti-holomorphic, so that $k$-bion configuration, namely $k$ instantons and $k$ anti-instantons, behaves as $\fq^k \bar{\fq}^k = \left| \fq \right|^{2k} = \left( e^{-\frac{8\pi^2}{g^2}} \right)^{2k}$
due to \eqref{eq:cmplx_coupling}.
Hence the bion gives rise to the non-perturbative contribution to the partition function even though it is topologically trivial.
The negative instanton, on the other hand, is counted as $\fq^{-k_-}$, instead of $\bar{\fq}^{k_-}$.
Therefore the $(k|k)$-instanton contributes to the topologically trivial $\fq^0$ sector in the partition function.

\subsection{Equivariant localization}\label{sec:full}

The partition function discussed in Sec.~\ref{sec:ADHM} was given as a contour integral, which ends up with a discrete sum over the residues.
Each factor is also derived from the equivariant localization, interpreted as a contribution of a fixed point in the ADHM moduli space under the equivariant action.
Here we apply the localization analysis to the ADHM moduli space with supergroup equivariant action.

We here consider generic quiver gauge theory.
Let $\Gamma$ be a quiver, consisting of nodes $i \in \Gamma_0 = \{\text{nodes}\}$ and edges $e \in \Gamma_1 = \{ \text{edges} \}$, so that $\Gamma = (\Gamma_0, \Gamma_1)$.
For each node, we assign super gauge group U$(n_{i,+}|n_{i,-})$.
We define instanton and framing bundles over the instanton moduli space evaluated at the fixed point for each node of quiver, $\bK = (\bK_i)$ and $\bN = (\bN_i)$ for $i \in \Gamma_0$.
Since these bundles are associated with the supergroups, the corresponding supercharacters are given by supertraces,
\begin{subequations} 
 \begin{align}
  \sch \bK_i & = \ch \bK_i^+ - \ch \bK_i^-
   \, , \\
   \sch \bN_i & = \ch \bN_i^+ - \ch \bN_i^-
  \, .
 \end{align} 
\end{subequations}
where each part is given by
\begin{align}
 \ch \bN_i^\sigma = \sum_{\alpha=1}^{n_{i,\sigma}} \nu_{i,\alpha}^\sigma
 \, ,
\end{align}
\begin{align}
 \ch \bK_i^+ = \sum_{\alpha=1}^{n_{i,+}} \sum_{(s_1,s_2) \in \lambda_{i,\alpha}^+} \nu_{i,\alpha}^+ q_1^{s_1-1} q_2^{s_2-1}
 \, , \quad
 \ch \bK_i^- = \sum_{\alpha=1}^{n_{i,-}} \sum_{(s_1,s_2) \in \lambda_{i,\alpha}^-} \nu_{i,\alpha}^- q_1^{-s_1} q_2^{-s_2}
 \, .
\end{align}
We remark that the Chern roots of $(\bK_i^\sigma)$ are determined by the fixed point condition under the equivariant action, corresponding to the possible poles shown in \eqref{eq:v_pole}.
Thus $(\bK_i^{\pm})$ plays a role as the dual of $(\bK_i^{\mp})$ as
\begin{align}
 \ch \bK_i^\pm[(\nu_{i,\alpha}^\pm)] = \ch \bK_i^{\mp\vee}[(\nu_{i,\alpha}^{\pm\vee} q)]
 \, .
\end{align}

We define the dynamical $x$-variables characterizing the instanton configuration%
\footnote{%
The pole structure \eqref{eq:v-pole} suggests to shift the Coulomb moduli for the negative node, $\nu_{i,\alpha}^- \to \nu_{i,\alpha}^- q^{-1}$, which leads to the shift of $x$-variables,
\begin{align}
 (x_{i,\alpha,k}^{-}, \mathring{x}_{i,\alpha,k}^{-})
 \ \longrightarrow \
 (q x_{i,\alpha,k}^{-}, q \mathring{x}_{i,\alpha,k}^{-})
 \, .
 \label{eq:x-shift}
\end{align}
We will use the shifted $x^-$-variables in the later Section.
}
\begin{subequations} 
\begin{align}
 \CalX_i^\sigma
 & = \left\{
 x_{i,\alpha,k}^\sigma = \nu_{i,\alpha}^\sigma q_2^{\sigma\lambda_{i,\alpha,k}^\sigma} q_1^{\sigma(k-1)} , \alpha = 1,\ldots,n_{i,\sigma}, k = 1, \ldots, \infty
 \right\}
 \, , \
 \CalX^\sigma = \bigsqcup_{i \in \Gamma_0} \CalX_i^\sigma
 \, . 
 \label{eq:Xset2} \\
 \mathring{\CalX}_i^\sigma
 & = \left\{
 x_{i,\alpha,k}^\sigma = \nu_{i,\alpha}^\sigma q_1^{\sigma(k-1)} , \alpha = 1,\ldots,n_{i,\sigma}, k = 1, \ldots, \infty
 \right\}
 \, , \
 \mathring{\CalX}^\sigma = \bigsqcup_{i \in \Gamma_0} \mathring{\CalX}_i^\sigma
 \, ,
 \label{eq:Xset0}
\end{align} 
\end{subequations}
and
\begin{align}
 \CalX = \CalX^+ \sqcup \CalX^-
 \, , \qquad
 \mathring{\CalX} = \mathring{\CalX}^+ \sqcup \mathring{\CalX}^-
 \, .
\end{align}
Namely the equivariant parameters for the positive node are $(q_1,q_2)$ and $(q_1^{-1},q_2^{-1})$ for the negative node.
Therefore the universal sheaf supercharacter is accordingly obtained as
\begin{align}
 \sch \bY_i
 & = \sch \bN_i - \ch \wedge \bQ \sch \bK_i
 \nonumber \\
 & =: \ch \bY_i^+ - \ch \bY_i^-
\end{align}
with
\begin{align}
 \ch \bY_i^\sigma = (1 - q_1^\sigma) X_{i,\sigma}
\end{align}
where we define
\begin{align}
 \ch \wedge \bQ = (1 - q_1)(1 - q_2)
\end{align}
and
\begin{align}
 X_{i,\sigma} = \sum_{x \in \CalX_i^\sigma} x
 \, , \qquad
 X_{i,\sigma}^\vee = \sum_{x \in \CalX_i^\sigma} x^{-1}
 \, .
\end{align}

\subsubsection{Vector multiplet}

The vector multiplet supercharacter is given using the universal sheaf
\begin{align}
 \sch \bV_i
 & = \frac{\sch \bY_i^\vee \sch \bY_i}{\ch \wedge \bQ}
 =: \sum_{\sigma,\sigma' = \pm} \sigma \sigma' \ch \bV_{i,\sigma\sigma'}
 \label{eq:chV}
\end{align}
with
\begin{subequations} 
\begin{align}
 \ch \bV_{i,++} =
 \frac{1-q_1^{-1}}{1-q_2}
 X_{i,+}^\vee X_{i,+}
 \, , \qquad
 \ch \bV_{i,+-} =
 q^{-1} \frac{1-q_1^{-1}}{1-q_2^{-1}}
 X_{i,+}^\vee X_{i,-}
 \, , \\
 \ch \bV_{i,-+} =
 \frac{1 - q_1}{1 - q_2}
 X_{i,-}^\vee X_{i,+} 
 \, , \qquad
 \ch \bV_{i,--} =
 \frac{1 - q_1^{-1}}{1 - q_2}
 X_{i,-}^\vee X_{i,-} 
 \, .
\end{align} 
\end{subequations}
We have to regularize, in particular, $\ch \bV_{i,++}$ and $\ch \bV_{i,--}$, by removing the diagonal contributions to avoid the zero mode:
\begin{align}
 \ch \bV_{i,\sigma\sigma}
 = \frac{1 - q_1^{-1}}{1 - q_2} \sum_{(x,x') \in \CalX_i^\sigma \times \CalX_i^\sigma} \frac{x}{x'}
 \ \stackrel{\text{reg}}{\longrightarrow} \
 \frac{1 - q_1^{-1}}{1 - q_2} \sum_{x \neq x'} \frac{x}{x'}
 \, .
\end{align}

The corresponding contribution to the full partition function is obtained by applying the index functor:
Given a character $\ch \bX = \sum_X x$, we apply the Dolbeault index,
\begin{align}
 \BI \left[ \bX \right] = \prod_{x \in X} \left( 1 - x^{-1} \right)
 \label{eq:ind5d}
\end{align}
which obeys the reflection
\begin{align}
 \BI \left[ \bX^\vee \right] =
 (-1)^{\rk \bX} \left( \det \bX \right) \BI \left[ \bX \right]
 \, .
 \label{eq:ind_ref}
\end{align}
Thus the vector multiplet contribution is given by
\begin{align}
 Z_i^\text{vec} =
 \BI \left[ \bV_i \right] =
 \prod_{\sigma,\sigma'=\pm} Z_{i,\sigma\sigma'}^\text{vec} 
\end{align}
where
\begin{subequations} 
\begin{align}
 Z_{i,++}^\text{vec} =
 \prod_{\substack{(x,x') \in \CalX_i^+ \times \CalX_i^+\\ x \neq x'}}
 \frac{(q x/x';q_2)_\infty}{(q_2 x/x';q_2)_\infty}
 \, , \qquad
 Z_{i,+-}^\text{vec} =
 \prod_{(x,x') \in \CalX_i^+ \times \CalX_i^-}
 \frac{(q q_1 x/x';q_2)_\infty}{(q x/x';q_2)_\infty}
 \, , \\ 
 Z_{i,-+}^\text{vec} =
 \prod_{(x,x') \in \CalX_i^- \times \CalX_i^+}
 \frac{(q_2 x/x';q_2)_\infty}{(q_1^{-1} q_2 x/x';q_2)_\infty}
 \, , \qquad
 Z_{i,--}^\text{vec} =
 \prod_{\substack{(x,x') \in \CalX_i^- \times \CalX_i^-\\ x \neq x'}}
 \frac{(q x/x';q_2)_\infty}{(q_2 x/x';q_2)_\infty}
 \, ,
\end{align} 
\end{subequations}
with the $q$-shifted factorial ($q$-Pochhammer) symbol
\begin{align}
 (z;q)_n = \prod_{k=0}^{n-1} \left( 1 - z q^k \right)
 \, .
 \label{eq:q-Pochhammer}
\end{align}
We also use the symbol $Z_i^\text{vec}[\CalX]$ in order to specify the instanton configuration.
Due to the sign factor, the off-diagonal factors $\CalX_i^+ \times \CalX_i^-$ and $\CalX_i^- \times \CalX_i^+$ contribute as bifundamental matters, which is consistent with the analysis in Sec.~\ref{sec:ADHM}.

Similarly we obtain the partition function for 6d gauge theory on $\BR^4 \times T^2$ by replacing the index \eqref{eq:ind5d} with the elliptic class~\cite{Kimura:2016dys}:
\begin{align}
 \BI_p \left[ \bX \right] = \prod_{x \in X} \theta(x^{-1};p)
\end{align}
obeying the same reflection formula \eqref{eq:ind_ref}, where the Jacobi theta function is defined
\begin{align}
 \theta(x;p) = (x;p)_\infty (px^{-1};p)_\infty
 \, .
\end{align}
The elliptic nome is defined as $p = \exp \left( 2 \pi \iota \tau \right)$ with the modulus parameter $\tau$ of the torus $T^2$, and we put $\iota = \sqrt{-1}$.
In this case, the full partition function is concisely expressed using the elliptic gamma function
\begin{align}
 \Gamma(z;p,q)
 = \prod_{n,m=0}^\infty
 \frac{1 - p^{n+1} q^{n+1} z^{-1}}{1 - p^n q^n z}
 \, .
 \label{eq:eGamma}
\end{align}
Hence we have
\begin{subequations} 
\begin{align}
 Z_{i,++}^\text{vec} =
 \prod_{\substack{(x,x') \in \CalX_i^+ \times \CalX_i^+\\ x \neq x'}}
 \frac{\Gamma(q_2 x/x';p,q_2)}{\Gamma(q x/x';p,q_2)}
 \, , \qquad
 Z_{i,+-}^\text{vec} =
 \prod_{(x,x') \in \CalX_i^+ \times \CalX_i^-}
 \frac{\Gamma(q x/x';p,q_2)}{\Gamma(q q_1 x/x';p,q_2)}
 \, , \\ 
 Z_{i,-+}^\text{vec} =
 \prod_{(x,x') \in \CalX_i^- \times \CalX_i^+}
 \frac{\Gamma(q_1^{-1} q_2 x/x';p,q_2)}{\Gamma(q_2 x/x';p,q_2)}
 \, , \qquad
 Z_{i,--}^\text{vec} =
 \prod_{\substack{(x,x') \in \CalX_i^- \times \CalX_i^-\\ x \neq x'}}
 \frac{\Gamma(q_2 x/x';p,q_2)}{\Gamma(q x/x';p,q_2)}
 \, .
\end{align} 
\end{subequations}

\subsubsection{Bifundamental hypermultiplet}

The bifundamental hypermultiplet contribution is similarly obtained
\begin{align}
 \sch \bH_{e:i \to j}
 & = - \ch \bM_{e} \frac{\sch \bY_i^\vee \sch \bY_j}{\ch \wedge \bQ}
 =: \sum_{\sigma,\sigma'=\pm} \sigma \sigma' \ch \bH_{e:i \to j,\sigma\sigma'}
 \label{eq:chH_bf}
\end{align}
where
\begin{subequations} 
\begin{align}
 \ch \bH_{e:i \to j,++} =
 - \mu_e \frac{1 - q_1^{-1}}{1 - q_2}
 X_{i,+}^\vee X_{j,+}
 \, , \quad
 \ch \bH_{e:i \to j,++} =
 - \mu_e q^{-1} \frac{1 - q_1^{-1}}{1 - q_2^{-1}}
 X_{i,+}^\vee X_{j,-} 
  \, , \\
 \ch \bH_{e:i \to j,-+} =
 - \mu_e \frac{1 - q_1}{1 - q_2}
 X_{i,-}^\vee X_{j,+} 
 \, , \qquad
 \ch \bH_{e:i \to j,--} =
 - \mu_e \frac{1 - q_1^{-1}}{1 - q_2}
 X_{i,-}^\vee X_{j,-} 
 \, .
\end{align} 
\end{subequations}
Thus the full partition function contribution is
\begin{align}
 Z_{e:i \to j}^\text{bf} =
 \BI \left[ \bH_{e:i \to j} \right] =
 \prod_{\sigma,\sigma'=\pm} Z_{e:i \to j,\sigma\sigma'}^\text{bf}
\end{align}
where for 5d theory
\begin{subequations} 
\begin{align}
 Z_{e:i \to j,++}^\text{bf} =
 \prod_{(x,x') \in \CalX_i^+ \times \CalX_j^+}
 \frac{(\mu_e^{-1}q_2 x/x';q_2)_\infty}{(\mu_e^{-1}q x/x';q_2)_\infty}
 \, , \quad
 Z_{e:i \to j,+-}^\text{bf} =
 \prod_{(x,x') \in \CalX_i^+ \times \CalX_j^-}
 \frac{(\mu_e^{-1}q x/x';q_2)_\infty}{(\mu_e^{-1}q q_1 x/x';q_2)_\infty}
 \, , \\ 
 Z_{e:i \to j,-+}^\text{bf} =
 \prod_{(x,x') \in \CalX_i^- \times \CalX_j^+}
 \frac{(\mu_e^{-1}q_1^{-1} q_2 x/x';q_2)_\infty}{(\mu_e^{-1}q_2 x/x';q_2)_\infty}
 \, , \quad
 Z_{e:i \to j,--}^\text{bf} =
 \prod_{(x,x') \in \CalX_i^- \times \CalX_j^-}
 \frac{(\mu_e^{-1}q_2 x/x';q_2)_\infty}{(\mu_e^{-1}q x/x';q_2)_\infty}
 \, ,
\end{align} 
\end{subequations}
and for 6d theory
\begin{subequations} 
\begin{align}
 Z_{e:i \to j,++}^\text{bf} =
 \prod_{(x,x') \in \CalX_i^+ \times \CalX_j^+}
 \frac{\Gamma(\mu_e^{-1}q x/x';p,q_2)}{\Gamma(\mu_e^{-1}q_2 x/x';p,q_2)}
 \, , \quad
 Z_{e:i \to j,+-}^\text{bf} =
 \prod_{(x,x') \in \CalX_i^+ \times \CalX_j^-}
 \frac{\Gamma(\mu_e^{-1}q q_1 x/x';p,q_2)}{\Gamma(\mu_e^{-1}q x/x';p,q_2)}
 \, , \\ 
 Z_{e:i \to j,-+}^\text{bf} =
 \prod_{(x,x') \in \CalX_i^- \times \CalX_j^+}
 \frac{\Gamma(\mu_e^{-1}q_2 x/x';p,q_2)}{\Gamma(\mu_e^{-1}q_1^{-1} q_2 x/x';p,q_2)}
 \, , \quad
 Z_{e:i \to j,--}^\text{bf} =
 \prod_{(x,x') \in \CalX_i^- \times \CalX_j^-}
 \frac{\Gamma(\mu_e^{-1}q x/x';p,q_2)}{\Gamma(\mu_e^{-1}q_2 x/x';p,q_2)}
 \, .
\end{align} 
\end{subequations}
In this case, the off-diagonal factors behave like the vector multiplet with the opposite sign to the ordinary bifundamental hypermultiplet contribution.

\subsubsection{Fundamental hypermultiplet}

The fundamental and antifundamental hypermultiplet contribution is obtained as follows:
\begin{subequations}\label{eq:chH_f}
\begin{align}
 \sch \bH_i^\text{f}
 & = - \frac{\sch \bY_i^\vee \sch \bM_i}{\ch \wedge \bQ}
 =: \sum_{\sigma,\sigma'=\pm} \sigma \sigma' \ch \bH_{i,\sigma\sigma'}^\text{f}
 \,, \\
 \sch \bH_i^\text{af}
 & = - \frac{\sch \widetilde{\bM}_i^\vee \sch \bY_i}{\ch \wedge \bQ}
 =: \sum_{\sigma,\sigma'=\pm} \sigma \sigma' \ch \bH_{i,\sigma\sigma'}^\text{af} 
\end{align}
\end{subequations}
where each contribution is given by
\begin{subequations}
\begin{align}
 \ch \bH_{i,+\sigma}^\text{f} =
 - \frac{q^{-1}}{1 - q_2^{-1}} X_{i,+}^\vee M_{i,\sigma}
 \, , \qquad
 \ch \bH_{i,-\sigma}^\text{f} =
 - \frac{1}{1 - q_2} X_{i,+}^\vee M_{i,\sigma}
 \, , \\
 \ch \bH_{i,\sigma+}^\text{af} =
 - \frac{q^{-1}}{1 - q_2} \widetilde{M}_{i,\sigma}^\vee X_{i,+}
 \, , \qquad
 \ch \bH_{i,\sigma-}^\text{af} =
 - \frac{q^{-1}}{1 - q_2^{-1}} \widetilde{M}_{i,\sigma}^\vee X_{i,-}
 \, .
\end{align} 
\end{subequations}
The characters associated with the flavor symmetry are
\begin{align}
 \sch \bM_{i} = M_{i,+} - M_{i,-}
 \, , \qquad
 \sch \widetilde{\bM}_{i} = \widetilde{M}_{i,+} - \widetilde{M}_{i,-} 
\end{align}
where
\begin{align}
 M_{i,\sigma} = \sum_{\mu \in \CalM_{i}^\sigma} \mu
 \, , \quad
 M_{i,\sigma}^\vee = \sum_{\mu \in \CalM_{i}^\sigma} \mu^{-1}
 \, , \quad
 \widetilde{M}_{i,\sigma} = \sum_{\mu \in \widetilde{\CalM}_{i}^\sigma} \mu
 \, , \quad
 \widetilde{M}_{i,\sigma}^\vee = \sum_{\mu \in \widetilde{\CalM}_{i}^\sigma} \mu^{-1}
\end{align}
with
\begin{align}
 \CalM_{i}^\sigma = \{ \mu_{i,f}^\sigma \,, f = 1, \ldots, n_{i,\sigma}^\text{f} \}
 \, , \qquad
 \widetilde{\CalM}_{i}^\sigma = \{ \tilde{\mu}_{i,f}^\sigma \,, f = 1, \ldots, n_{i,\sigma}^\text{af} \}
 \, .
\end{align}
The full partition function contribution is given by
\begin{align}
 Z_i^\text{(a)f} =
 \BI \left[ \bH_i^\text{(a)f} \right] =
 \prod_{\sigma,\sigma'=\pm}  Z_{i,\sigma\sigma'}^\text{(a)f}
\end{align}
where for 5d theory
\begin{subequations}
\begin{align}
 Z_{i,+\sigma}^\text{f} & =
 \prod_{(x,\mu) \in \CalX_i^+ \times \CalM_i^\sigma}
 \left( q \frac{x}{\mu};q_2 \right)_\infty^{-\sigma}
 \, , \qquad
 Z_{i,-\sigma}^\text{f} =
 \prod_{(x,\mu) \in \CalX_i^- \times \CalM_i^\sigma}
 \left( q_2 \frac{x}{\mu}; q_2 \right)_\infty^{-\sigma}
 \, , \\
 Z_{i,+\sigma}^\text{af} & =
 \prod_{(x,\mu) \in \CalX_i^+ \times \widetilde{\CalM}_i^\sigma}
 \left( q q_2 \frac{\mu}{x};q_2 \right)_\infty^{\sigma}
 \, , \qquad
 Z_{i,-\sigma}^\text{af} =
 \prod_{(x,\mu) \in \CalX_i^- \times \widetilde{\CalM}_i^\sigma}
 \left( q \frac{\mu}{x}; q_2 \right)_\infty^{\sigma}
 \, ,
\end{align} 
\end{subequations}
and for 6d theory
\begin{subequations} 
\begin{align}
 Z_{i,+\sigma}^\text{f} & =
 \prod_{(x,\mu) \in \CalX_i^+ \times \CalM_i^\sigma}
 \Gamma\left( q \frac{x}{\mu};p,q_2 \right)^{\sigma}
 \, , \qquad
 Z_{i,-\sigma}^\text{f} =
 \prod_{(x,\mu) \in \CalX_i^- \times \CalM_i^\sigma}
 \Gamma\left( q_2 \frac{x}{\mu}; p,q_2 \right)^{\sigma}
 \, , \\
 Z_{i,+\sigma}^\text{af} & =
 \prod_{(x,\mu) \in \CalX_i^+ \times \widetilde{\CalM}_i^\sigma}
 \Gamma\left( q q_2 \frac{\mu}{x};p,q_2 \right)^{-\sigma}
 \, , \qquad
 Z_{i,-\sigma}^\text{af} =
 \prod_{(x,\mu) \in \CalX_i^- \times \widetilde{\CalM}_i^\sigma}
 \Gamma\left( q \frac{\mu}{x}; p,q_2 \right)^{-\sigma}
 \, .
\end{align} 
\end{subequations}

\subsubsection{Topological term}

The topological term in the partition function counts the instanton number, which is given as the size of the partition:
\begin{align}
 Z_i^\text{top} = \fq_i^{|\lambda_i^+| - |\lambda_i^-|}
 \label{eq:pf_top1}
\end{align}
where we define
\begin{align}
 \fq_i = \exp \left( 2 \pi \iota \tau_i \right)
 \label{eq:cmplx_coupling_quiv}
\end{align}
for $i \in \Gamma_0$, similarly to \eqref{eq:cmplx_coupling}.
In terms of the $x$-variables, it is given by
\begin{align}
 |\lambda_i^\sigma| = \sum_{\alpha=1}^{n_{i,\sigma}} \sum_{k=1}^\infty \log_{q_2^\sigma} \frac{x_{i,\alpha,k}^\sigma}{\mathring{x}_{i,\alpha,k}^\sigma}
 \, ,
\end{align}
where the variable $(\mathring{x}_{i,\alpha,k}^\sigma)$ corresponds to the empty configuration $\lambda_i^\sigma = \emptyset$ defined in \eqref{eq:Xset0}.
Since \eqref{eq:cmplx_coupling_quiv}, the topological contribution \eqref{eq:pf_top1} is
\begin{align}
 Z_i^\text{top} =
 \exp
 \left(
  \frac{2 \pi \iota \tau_i}{\epsilon_2} \sum_{\sigma = \pm} \sum_{\alpha=1}^{n_{i,\sigma}} \sum_{k=1}^\infty \log \frac{x_{i,\alpha,k}^\sigma}{\mathring{x}_{i,\alpha,k}^\sigma}
 \right)
 \, .
 \label{eq:pf_top2} 
\end{align}


In addition, one can consider the Chern--Simons term in 5d gauge theory.
It is labeled by the integer, called the Chern--Simons level, assigned to each node, $(\kappa_i^\sigma)_{i \in \Gamma_0, \sigma =\pm}$.
The contribution to the partition function is given by
\begin{align}
 Z_i^\text{cs} = \prod_{\sigma = \pm} Z_{i,\sigma}^\text{cs}
\end{align}
with
\begin{align}
 Z_{i,\sigma}^\text{cs} & = \left( \det \bK_i^\sigma \right)^{\sigma \kappa_i^\sigma} =
 \begin{cases}
  \displaystyle
 \prod_{\alpha = 1}^{n_{i,+}}
 \prod_{s \in \lambda_{i,\alpha}^+}
  \left( \nu_{i,\alpha}^+ q_1^{s_1 - 1} q_2^{s_2 - 1} \right)^{\kappa_i^+}
  & (\sigma = +) \\[.5em]
  \displaystyle
 \prod_{\alpha = 1}^{n_{i,-}}
 \prod_{s \in \lambda_{i,\alpha}^-}
  \left( \nu_{i,\alpha}^- q_1^{-s_1} q_2^{-s_2} \right)^{-\kappa_i^-}
  & (\sigma = -  )
 \end{cases}
\end{align}

\subsection{Instanton partition function}

Instead of the full partition function, involving infinite products for the perturbative contribution, we consider the instanton partition function, which is a combinatorial finite product over the partition.

From the full vector multiplet character \eqref{eq:chV}, we obtain the instanton contribution
\begin{align}
 \sch \bV_i^\text{inst} & =
 - \sch \bN_i^\vee \sch \bK_i - q^{-1} \sch \bK_i^\vee \sch \bN_i
 + \ch \bQ^\vee 
 \sch \bK_i^\vee \sch \bK_i
 \nonumber \\
 & =:
 \sum_{\sigma,\sigma' = \pm} \sigma \sigma' \ch \bV_{i,\sigma\sigma'}^\text{inst}
 \label{eq:chV_inst}
\end{align}
where
\begin{align}
 \ch \bV_{i,\sigma\sigma'}^\text{inst} =
 - \ch \bN_i^{\sigma\vee} \ch \bK_i^{\sigma'} - q^{-1} \ch \bK_i^{\sigma\vee} \ch \bN_i^{\sigma'}
 + \ch \bQ^\vee 
 \ch \bK_i^{\sigma\vee} \ch \bK_i^{\sigma'}
 \, .
 \label{eq:chV_inst2} 
\end{align}
Similarly, the instanton part for the bifundamental and (anti)fundamental hypermultiplets is obtained from the full contribution \eqref{eq:chH_bf} and \eqref{eq:chH_f},
\begin{subequations} 
\begin{align}
 \sch \bH_{e:i \to j}^{\text{bf,inst}} & = \sum_{\sigma,\sigma'=\pm} \sigma \sigma' \ch \bH_{e:i \to j, \sigma\sigma'}^\text{bf,inst}
 \, , \\
 \sch \bH_{i}^{\text{(a)f,inst}} & = \sum_{\sigma,\sigma'=\pm} \sigma \sigma' \ch \bH_{i,\sigma\sigma'}^\text{(a)f,inst}
\end{align} 
\end{subequations}
where
\begin{align}
 \ch \bH_{e:i \to j, \sigma\sigma'}^\text{bf,inst} & =
 \mu_{e:i \to j}
 \left( \ch \bN_i^{\sigma\vee} \ch \bK_j^{\sigma'} + q^{-1} \ch \bK_i^{\sigma\vee} \ch \bN_j^{\sigma'} - \ch \bQ^\vee 
 \ch \bK_i^{\sigma\vee} \ch \bK_j^{\sigma'} \right)
 \, , 
\end{align}
and
\begin{subequations} 
\begin{align}
 \ch \bH_{i,\sigma\sigma'}^\text{f,inst} & = q^{-1} \ch \bK_i^{\sigma\vee} M_{i,\sigma'}
 \, , \\
 \ch \bH_{i,\sigma\sigma'}^\text{af,inst} & = \widetilde{M}_{i,\sigma}^\vee \ch \bK_i^{\sigma'}
 \, . 
\end{align} 
\end{subequations}

\subsubsection{Perturbative part}

The perturbative contribution for the vector multiplet is given as the remaining term in \eqref{eq:chV}, which is independent of the instanton configuration,
\begin{align}
 \sch \mathring{\bV}_i := \sch \bV_i - \sch \bV_i^\text{inst}
 = \frac{\sch \bN_i^\vee \sch \bN_i}{\ch \wedge \bQ}
 \, .
\end{align}
Thus the corresponding perturbative part of the 5d partition function is given by
\begin{align}
 \mathring{Z}_i^\text{vec} = \prod_{\sigma,\sigma' = \pm} \mathring{Z}_{i,\sigma\sigma'}^\text{vec}
\end{align}
where each contribution is given by 
\begin{align}
 \mathring{Z}_{i,\sigma\sigma'}^\text{vec}
 = \prod_{\alpha=1}^{n_{i,\sigma}} \prod_{\beta = 1}^{n_{i,\sigma'}} \prod_{k,k'=1}^\infty
 \left( 1 - \frac{\nu_{i,\alpha}^\sigma}{\nu_{i,\beta}^{\sigma'}} q_1^{k} q_2^{k'} \right)^{\sigma\sigma'}
 \label{eq:Zvec0}
\end{align}
for the region $|q_1|, |q_2| < 1$.
This double infinite product should be properly regularized:
We define a $q$-analog of the double gamma (Barnes) function,
\begin{align}
 \Gamma_2(z;q_1,q_2) =
 \begin{cases}
  \displaystyle
  \prod_{k,k'=0}^\infty (1 - z q_1^k q_2^{k'})^{-1} & (|z|, |q_1|, |q_2| < 1) \\
  \displaystyle
  \prod_{k,k'=0}^\infty (1 - z q_1^{-k-1} q_2^{k'}) & (|z| < 1, |q_1| >1, |q_2| < 1) \\
  \displaystyle
  \prod_{k,k'=0}^\infty (1 - z q_1^{-k-1} q_2^{-k'-1})^{-1} & (|z| < 1, |q_1|, |q_2| > 1)  
 \end{cases}
 \label{eq:qBarnes}
\end{align}
For the region $|z| > 1$, we use the reflection formula
\begin{align}
 \Gamma_2(z^{-1};q_1,q_2) = \Gamma(z^{-1};q_1,q_2) \Gamma_2(q z;q_1,q_2)
\end{align}
where $q = q_1 q_2$, and $\Gamma(z,q_1,q_2)$ is the elliptic gamma function~\eqref{eq:eGamma}.
Thus, in particular, if $|q_1| > 1$, $|q_2| < 1$, and $|\nu_{i,\alpha}^\sigma/\nu_{i,\beta}^{\sigma'}| < 1$, the diagonal contribution at $\sigma = \sigma'$ of the perturbative contribution \eqref{eq:Zvec0} becomes
\begin{align}
 \mathring{Z}_{i,\sigma\sigma}^\text{vec}
 = \prod_{\alpha<\beta}^{n_{i,\sigma}}
 \Gamma_2\left( q \frac{\nu_{i,\alpha}^\sigma}{\nu_{i,\beta}^\sigma};q_1, q_2 \right)^{-1}
 \Gamma_2\left( \frac{\nu_{i,\alpha}^\sigma}{\nu_{i,\beta}^\sigma};q_1, q_2 \right)^{-1} 
\end{align}
which is consistent with the convention of \cite{Alday:2009aq}.

The perturbative part for 6d theory is given by
\begin{align}
 \mathring{Z}_{i,\sigma\sigma}^\text{vec}
 & = \prod_{\alpha=1}^{n_{i,\sigma}} \prod_{\beta = 1}^{n_{i,\sigma'}} \prod_{k,k'=1}^\infty
 \theta\left( \frac{\nu_{i,\alpha}^\sigma}{\nu_{i,\beta}^{\sigma'}} q_1^{k} q_2^{k'};p \right)^{\sigma\sigma'}
 \, .
\end{align}
The diagonal part is written as follows:
\begin{align}
 \mathring{Z}_{i,\sigma\sigma}^\text{vec}
 = \prod_{\alpha<\beta}^{n_{i,\sigma}}
 \Gamma_2\left( q \frac{\nu_{i,\alpha}^\sigma}{\nu_{i,\beta}^\sigma};q_1, q_2, p \right)
 \Gamma_2\left( \frac{\nu_{i,\alpha}^\sigma}{\nu_{i,\beta}^\sigma};q_1, q_2, p \right)
\end{align}
where the elliptic double gamma function is defined as~\cite{Narukawa:2004AM}
\begin{align}
 \Gamma_2(z;q_1,q_2,q_3)
 & = \prod_{i,j,k=0}^\infty \left( 1 - z q_1^i q_2^j q_3^k \right) \left( 1 - z^{-1} q_1^{i+1} q_2^{j+1} q_3^{k+1} \right)
 \nonumber \\
 & =
 \exp \left( - \sum_{n \neq 0} \frac{z^n}{n (1 - q_1^n)(1 - q_2^n)(1 - q_3^n)} \right)
 \, .
 \label{eq:elliptic_double_gamma}
\end{align}

Similarly the perturbative contributions of the hypermultiplets are obtained from \eqref{eq:chH_bf} and \eqref{eq:chH_f} as
\begin{subequations}
\begin{align}
 \sch \mathring{\bH}_{e:i \to j}^\text{bf} & = - \ch \bM_e \frac{\sch \bN_i^\vee \sch \bN_j}{\ch \wedge \bQ} \\
 \sch \mathring{\bH}_i^\text{f} & = - \frac{\sch \bN_i^\vee \sch \bM_i}{\ch \wedge \bQ} \\
 \sch \mathring{\bH}_i^\text{af} & = - \frac{\sch \widetilde{\bM}_i^\vee \sch \bN_i}{\ch \wedge \bQ} 
\end{align} 
\end{subequations}
Hence the perturbative parts of the partition function are
\begin{align}
 \mathring{Z}_{e:i \to j}^\text{bf} = \prod_{\sigma,\sigma'=\pm} \mathring{Z}_{e:i \to j,\sigma\sigma'}^\text{bf}
 \, , \qquad
 \mathring{Z}_i^\text{f} = \prod_{\sigma,\sigma'=\pm} \mathring{Z}_{i,\sigma\sigma'}^\text{f}
 \, , \qquad
 \mathring{Z}_i^\text{af} = \prod_{\sigma,\sigma'=\pm} \mathring{Z}_{i,\sigma\sigma'}^\text{af} 
\end{align}
where
\begin{subequations} 
\begin{align}
 \mathring{Z}_{e:i \to j, \sigma\sigma'}^\text{bf}
 & = \prod_{\alpha=1}^{n_{i,\sigma}} \prod_{\beta=1}^{n_{j,\sigma'}} \prod_{k,k'=1}^\infty
 \left( 1 - \mu_e^{-1} \frac{\nu_{i,\alpha}^\sigma}{\nu_{j,\beta}^{\sigma'}} q_1^{k} q_2^{k'} \right)^{-\sigma\sigma'}
 = \prod_{\alpha=1}^{n_{i,\sigma}} \prod_{\beta=1}^{n_{j,\sigma'}}
 \Gamma_2  \left( \mu_e^{-1} \frac{\nu_{i,\alpha}^\sigma}{\nu_{j,\beta}^{\sigma'}}; q_1, q_2 \right)^{\sigma\sigma'}
 \\
 \mathring{Z}_{i,\sigma\sigma'}^\text{f}
 & = \prod_{\alpha=1}^{n_{i,\sigma}} \prod_{f=1}^{n_{i,\sigma'}^{\text{f}}} \prod_{k,k'=1}^\infty
 \left( 1 - \frac{\nu_{i,\alpha}^\sigma}{\mu_{i,f}^{\sigma'}} q_1^k q_2^{k'} \right)^{-\sigma\sigma'}
 = \prod_{\alpha=1}^{n_{i,\sigma}} \prod_{f=1}^{n_{i,\sigma'}^{\text{f}}}
 \Gamma_2 \left( \frac{\nu_{i,\alpha}^\sigma}{\mu_{i,f}^{\sigma'}}; q_1, q_2 \right)^{\sigma\sigma'}
 \\
 \mathring{Z}_{i,\sigma\sigma'}^\text{af}
 & = \prod_{\alpha=1}^{n_{i,\sigma}} \prod_{f=1}^{n_{i,\sigma'}^{\text{af}}} \prod_{k,k'=1}^\infty
 \left( 1 - \frac{\tilde{\mu}_{i,f}^{\sigma'}}{\nu_{i,\alpha}^\sigma} q_1^k q_2^{k'} \right)^{-\sigma\sigma'}
 = \prod_{\alpha=1}^{n_{i,\sigma}} \prod_{f=1}^{n_{i,\sigma'}^{\text{af}}} 
 \Gamma_2 \left( \frac{\tilde{\mu}_{i,f}^{\sigma'}}{\nu_{i,\alpha}^\sigma}; q_1, q_2 \right)^{\sigma\sigma'} 
\end{align}
\end{subequations}
The 6d perturbative partition function is similarly formulated using the elliptic double gamma function \eqref{eq:elliptic_double_gamma}.

\subsubsection{Instanton part}

The instanton part of the partition function is given by the index with the corresponding supercharacter~\eqref{eq:chV_inst},
\begin{align}
 Z_i^\text{vec,inst}
 = \BI[ \bV_i^\text{inst} ]
 = \prod_{\sigma, \sigma'=\pm} Z_{i,\sigma\sigma'}^\text{vec,inst}
 \, .
\end{align}
Applying the combinatorial formula shown in Appendix~\ref{sec:comb}, we obtain the diagonal parts of the vector multiplet contribution,
\begin{subequations} 
\begin{align}
 Z_{i,++}^{\text{vec,inst}} & =
 \prod_{\alpha,\beta}^{n_{i,+}}
 \CalZ_\text{diag}^\text{vec}(\nu_{i,\alpha}^+,\nu_{i,\beta}^+;\lambda_{i,\alpha}^+,\lambda_{i,\beta}^+)
 \, , \\
 Z_{i,--}^{\text{vec,inst}} & =
 \prod_{\alpha,\beta}^{n_{i,-}}
 \CalZ_\text{diag}^\text{vec}(\nu_{i,\alpha}^-,\nu_{i,\beta}^-;\lambda_{i,\beta}^-,\lambda_{i,\alpha}^-) 
\end{align} 
\end{subequations}
where the combinatorial factor is given by
\begin{align}
 \CalZ_\text{diag}^\text{vec}(\nu,\nu';\lambda_{\alpha},\lambda_\beta)
 & = \prod_{s \in \lambda_{\alpha}}
 \left(1 - \frac{\nu}{\nu'} q_1^{-\ell_{\beta}(s)} q_2^{a_{\alpha}(s)+1}\right)^{-1}
 \prod_{s \in \lambda_{\beta}}
 \left( 1 - \frac{\nu}{\nu'} q_1^{\ell_{\alpha}(s)+1} q_2^{-a_{\beta}(s)} \right)^{-1}
\end{align}
with the arm and leg lengths defined in \eqref{eq:arm_leg}.
The off-diagonal contributions are
\begin{align}
 Z_{i,\sigma\sigma'}^{\text{vec,inst}} & =
 \prod_{\alpha=1}^{n_{i,\sigma}} \prod_{\beta=1}^{n_{i,\sigma'}}
 \CalZ_{\sigma\sigma'}^\text{vec}(\nu_{i,\alpha}^\sigma,\nu_{i,\beta}^{\sigma'};\lambda_{i,\alpha}^\sigma,\lambda_{i,\beta}^{\sigma'})
\end{align}
for $\sigma \neq \sigma'$, where
\begin{subequations} 
\begin{align}
 \CalZ_{+-}^\text{vec}(\nu,\nu';\lambda_{\alpha},\lambda_\beta) & =
 \prod_{s_1=1}^{\lambda_{\alpha,1}^{\text{T}}} \prod_{s_2'=1}^{\lambda_{\beta,1}}
 \left( 1 - \frac{\nu}{\nu'} q_1^{\lambda_{\beta,s_2'}^{\text{T}}+s_1} q_2^{\lambda_{\alpha,s_1} + s_2'} \right)^{-1}
 \left( 1 - \frac{\nu}{\nu'} q_1^{s_1} q_2^{s_2'} \right)
 \nonumber \\
 & \quad \times
 \prod_{s \in \lambda_{\alpha}}
 \left(1 - \frac{\nu}{\nu'} q_1^{s_1} q_2^{\lambda_{\beta,1} + s_2} \right)
 \prod_{s' \in \lambda_{\beta}}
 \left( 1 - \frac{\nu}{\nu'} q_1^{\lambda_{\alpha,1}^{\text{T}} + s_1'} q_2^{s_2'} \right)
 \\
 \CalZ_{-+}^\text{vec}(\nu,\nu';\lambda_{\alpha},\lambda_\beta) & =
 \prod_{s_1=1}^{\lambda_{\alpha,1}^{\text{T}}} \prod_{s_2'=1}^{\lambda_{\beta,1}}
 \left( 1 - \frac{\nu}{\nu'} q_1^{-\lambda_{\beta,s_2'}^{\text{T}}-s_1+1} q_2^{-\lambda_{\alpha,s_1} - s_2' + 1} \right)^{-1}
 \left( 1 - \frac{\nu}{\nu'} q_1^{-s_1+1} q_2^{-s_2+1} \right)
 \nonumber \\
 & \quad \times
 \prod_{s \in \lambda_{\alpha}}
 \left(1 - \frac{\nu}{\nu'} q_1^{-s_1 + 1} q_2^{-\lambda_{\beta,1} - s_2 + 1} \right)
 \prod_{s' \in \lambda_{\beta}}
 \left( 1 - \frac{\nu}{\nu'} q_1^{-\lambda_{\alpha,1}^{\text{T}} - s_1' + 1} q_2^{- s_2' + 1} \right)
\end{align} 
\end{subequations}
The number of factors appearing in $\CalZ_{\sigma\sigma'}$ is as follows: $|\lambda_\alpha| + |\lambda_\beta|$ factors in the denominator of $\CalZ_{++}$ and $\CalZ_{--}$, $|\lambda_\alpha| + |\lambda_\beta| + \lambda_{\alpha,1}^\text{T} + \lambda_{\beta,1}$ factors in the numerator, $\lambda_{\alpha,1}^\text{T} + \lambda_{\beta,1}$ factors in the denominator of $\CalZ_{+-}$ and $\CalZ_{-+}$, so that the numbers of factors in the numerator and the denominator are balanced in total.

We remark that the diagonal contribution is given by the well-known combinatorial formula using the arm and leg lengths of the partition.
The off-diagonal contributions are still finite products written in terms of the partition, but do not have a compact formula similar to the diagonal ones.
This situation is similar to the BCD instanton partition function, involving $\phi_a + \phi_b$ in the contour integral~\cite{Marino:2004cn,Nekrasov:2004vw}.

The bifundamental hypermultiplet contribution is similarly given as follows:
\begin{align}
 Z_{e:i \to j}^{\text{bf,inst}}
 & = \BI[\bH_{e:i \to j}^\text{bf,inst}]
 = \prod_{\sigma,\sigma' = \pm} Z_{e:i \to j,\sigma\sigma'}^{\text{bf,inst}}
\end{align}
where
\begin{subequations}
\begin{align}
 Z_{e:i \to j,++}^{\text{bf,inst}}
 & = \prod_{\alpha=1}^{n_{i,\alpha}^+} \prod_{\beta=1}^{n_{j,\beta}^+}
 \CalZ_\text{diag}^\text{bf}(\nu_{i,\alpha}^+,\nu_{j,\beta}^+,\mu_{e:i \to j};\lambda_{i,\alpha}^+,\lambda_{j,\beta}^+)
 \\
 Z_{e:i \to j,--}^{\text{bf,inst}}
 & = \prod_{\alpha=1}^{n_{i,\alpha}^-} \prod_{\beta=1}^{n_{j,\beta}^-}
 \CalZ_\text{diag}^\text{bf}(\nu_{i,\alpha}^-,\nu_{j,\beta}^-,\mu_{e:i \to j};\lambda_{j,\beta}^-,\lambda_{i,\alpha}^-)
 \\
 Z_{e:i \to j,\sigma\sigma'}^{\text{bf,inst}}
 & = \prod_{\alpha=1}^{n_{i,\alpha}^\sigma} \prod_{\beta=1}^{n_{j,\beta}^{\sigma'}}
 \CalZ_{\sigma\sigma'}^\text{bf}(\nu_{i,\alpha}^\sigma,\nu_{j,\beta}^{\sigma'},\mu_{e:i \to j};\lambda_{i,\alpha}^\sigma,\lambda_{j,\beta}^{\sigma'})
 \qquad \text{for} \qquad \sigma \neq \sigma'
\end{align} 
\end{subequations}
with
\begin{subequations} 
\begin{align}
 \CalZ_\text{diag}^\text{bf}(\nu,\nu',\mu;\lambda_{\alpha},\lambda_\beta)
 & = \prod_{s \in \lambda_{\alpha}}
 \left(1 - \mu^{-1} \frac{\nu}{\nu'} q_1^{-\ell_{\beta}(s)} q_2^{a_{\alpha}(s)+1}\right)
 \prod_{s \in \lambda_{\beta}}
 \left( 1 - \mu^{-1} \frac{\nu}{\nu'} q_1^{\ell_{\alpha}(s)+1} q_2^{-a_{\beta}(s)} \right)
\end{align}
\begin{align}
 \CalZ_{+-}^\text{bf}(\nu,\nu',\mu;\lambda_{\alpha},\lambda_\beta) & =
 \prod_{s_1=1}^{\lambda_{\alpha,1}^{\text{T}}} \prod_{s_2'=1}^{\lambda_{\beta,1}}
 \left( 1 - \mu^{-1} \frac{\nu}{\nu'} q_1^{\lambda_{\beta,s_2'}^{\text{T}}+s_1} q_2^{\lambda_{\alpha,s_1} + s_2'} \right)
 \left( 1 - \mu^{-1} \frac{\nu}{\nu'} q_1^{s_1} q_2^{s_2'} \right)^{-1}
 \nonumber \\
 & \quad \times
 \prod_{s \in \lambda_{\alpha}}
 \left(1 - \mu^{-1} \frac{\nu}{\nu'} q_1^{s_1} q_2^{\lambda_{\beta,1} + s_2} \right)^{-1}
 \prod_{s' \in \lambda_{\beta}}
 \left( 1 - \mu^{-1} \frac{\nu}{\nu'} q_1^{\lambda_{\alpha,1}^{\text{T}} + s_1'} q_2^{s_2'} \right)^{-1}
 \\
 \CalZ_{-+}^\text{bf}(\nu,\nu',\mu;\lambda_{\alpha},\lambda_\beta) & =
 \prod_{s_1=1}^{\lambda_{\alpha,1}^{\text{T}}} \prod_{s_2'=1}^{\lambda_{\beta,1}}
 \left( 1 - \mu^{-1} \frac{\nu}{\nu'} q_1^{-\lambda_{\beta,s_2'}^{\text{T}}-s_1+1} q_2^{-\lambda_{\alpha,s_1} - s_2' + 1} \right)
 \left( 1 - \mu^{-1} \frac{\nu}{\nu'} q_1^{-s_1+1} q_2^{-s_2+1} \right)^{-1}
 \nonumber \\
 & \quad \times
 \prod_{s \in \lambda_{\alpha}}
 \left(1 - \mu^{-1} \frac{\nu}{\nu'} q_1^{-s_1 + 1} q_2^{-\lambda_{\beta,1} - s_2 + 1} \right)^{-1}
 \prod_{s' \in \lambda_{\beta}}
 \left( 1 - \mu^{-1} \frac{\nu}{\nu'} q_1^{-\lambda_{\alpha,1}^{\text{T}} - s_1' + 1} q_2^{- s_2' + 1} \right)^{-1}
\end{align} 
\end{subequations}
We remark that the total numbers of the factors appearing in the numerator and the denominator are balanced as well as the vector multiplet.

The (anti)fundamental hypermultiplet contribution to the instanton partition function is given by
\begin{align}
 Z_{i}^{\text{(a)f,inst}}
 & = \BI[\bH_{i}^\text{(a)f,inst}]
 = \prod_{\sigma,\sigma' = \pm} Z_{i,\sigma\sigma'}^{\text{(a)f,inst}}  
\end{align}
where
\begin{subequations}
\begin{align}
 Z_{i,+\sigma'}^{\text{f,inst}}
 & = \prod_{\alpha=1}^{n_{i,+}} \prod_{f=1}^{n^\text{f}_{i,\sigma'}} \prod_{s \in \lambda_{i,\alpha}^+}
 \left( 1 - \frac{\nu_{i,\alpha}^+}{\mu_{i,f}^{\sigma'}} q_1^{s_1} q_2^{s_2} \right)^{\sigma'}
 \\
 Z_{i,-\sigma'}^{\text{f,inst}}
 & = \prod_{\alpha=1}^{n_{i,-}} \prod_{f=1}^{n^\text{f}_{i,\sigma'}} \prod_{s \in \lambda_{i,\alpha}^-}
 \left( 1 - \frac{\nu_{i,\alpha}^-}{\mu_{i,f}^{\sigma'}} q_1^{-s_1+1} q_2^{-s_2+1} \right)^{-\sigma'} 
 \\
 Z_{i,+\sigma'}^{\text{af,inst}}
 & = \prod_{\alpha=1}^{n_{i,+}} \prod_{f=1}^{n^\text{af}_{i,\sigma'}} \prod_{s \in \lambda_{i,\alpha}^+}
 \left( 1 - \frac{\tilde{\mu}_{i,f}^{\sigma'}}{\nu_{i,\alpha}^+} q_1^{-s_1+1} q_2^{-s_2+1} \right)^{\sigma'}
 \\
 Z_{i,-\sigma'}^{\text{af,inst}}
 & = \prod_{\alpha=1}^{n_{i,-}} \prod_{f=1}^{n^\text{af}_{i,\sigma'}} \prod_{s \in \lambda_{i,\alpha}^-}
 \left( 1 - \frac{\tilde{\mu}_{i,f}^{\sigma'}}{\nu_{i,\alpha}^-} q_1^{s_1} q_2^{s_2} \right)^{-\sigma'}  
\end{align}
\end{subequations}

\section{Seiberg--Witten geometry and its quantization}\label{sec:qq-ch}

Seiberg--Witten theory is algebraic geometric description of the Coulomb branch of the vacua in 4d $\CalN = 2$ gauge theory~\cite{Seiberg:1994rs,Seiberg:1994aj}.
Such a geometry is actually reproduced from the instanton partition function in the classical limit $\epsilon_{1,2} \to 0$ $(q_{1,2} \to 1)$~\cite{Nekrasov:2002qd}.
Afterwards it has been pointed out that the equivariant parameter $\epsilon_{1,2}$ is not just for regularization, but plays a role as a quantum deformation parameter.
For generic $\epsilon_{1,2}$, the Seiberg--Witten geometry is described by the $qq$-character, which is a two-parameter deformation of the character associated with a quiver~\cite{Nekrasov:2015wsu}.
In this Section, we consider the instanton-adding/removing operation with the supergroup quiver gauge theory, and derive the $qq$-character thereof.
We show that the polynomiality of the $qq$-character is replaced with the rationality in the supergroup case.

\subsection{$\sY$-function}\label{sec:Y-func}

We define the $\sY$-function, which is a building block for the $qq$-character.
Given a Chern character $\ch \bX = \sum_X x$, the associated $\sY$-function is defined as
\begin{align}
 \sY_{i,X} & = \BI \left[ \bX^\vee \bY_i \right] =
 \prod_{x \in X} \prod_{\sigma = \pm} \prod_{x' \in \CalX_i^\sigma}
 \left( \frac{1 - x/x'}{1 - q_1^{-\sigma} x/x'} \right)^\sigma
 \, , \\
 \sY_{i,X}^\vee & = \BI \left[ \bY_i^\vee \bX \right] =
 \prod_{x \in X} \prod_{\sigma = \pm} \prod_{x' \in \CalX_i^\sigma}
 \left( \frac{1 - x'/x}{1 - q_1^\sigma x'/x} \right)^\sigma
 \, ,
\end{align}
which are interpreted as ratios of the positive and negative $\sY$-functions:
\begin{align}
 \sY_{i,X} = \frac{\sY_{i,X}^+}{\sY_{i,X}^-}
 \, , \qquad
 \sY_{i,X}^\vee = \frac{\sY_{i,X}^{+\vee}}{\sY_{i,X}^{-\vee}} 
 \label{eq:Y_ratio}
\end{align}
with
\begin{align}
 \sY_{i,X}^\sigma =
 \prod_{x \in X} \prod_{x' \in \CalX_i^\sigma}
 \frac{1 - x/x'}{1 - q_1^{-\sigma} x /x'}
 \, , \qquad
 \sY_{i,X}^{\sigma\vee} =
 \prod_{x \in X} \prod_{x' \in \CalX_i^\sigma}
 \frac{1 - x'/x}{1 - q_1^{\sigma} x' /x}
 \, .
 \label{eq:Y_pm}
\end{align}
We also use the symbol $\sY_{i,X}^\sigma[\CalX]$ to explicitly specify the instanton configuration.
The $\sY^\sigma$-function and its dual, $\sY_{i,X}^\sigma$ and $\sY_{i,X}^{\sigma\vee}$, have the same zeros at $x = x'$ and the same poles at $x = q_1^\sigma x'$ for $x' \in \CalX_i^\sigma$, so that they are identical up to a factor depending on the gauge group rank~\cite{Nekrasov:2013xda}.
We remark that there are no distinction between $\sY_{i,X}$ and $\sY_{i,X}^\vee$ in the 4d limit.

Although we have shown the infinite product formula for the $\sY$-function, it also has the finite product formula.
For simplicity, we consider a single variable $X = \{ x \}$.
Then it is given as follows:
\begin{align}
 \sY_{i,x}^+[\CalX] & =
 \prod_{\alpha=1}^{n_{i,+}}
 \left[
 \left( 1 - \frac{x}{\nu_{i,\alpha}^+} \right)
 \prod_{s \in \lambda_{i,\alpha}} \msS \left( \frac{x}{\nu_{i,\alpha}^+ q_1^{s_1} q_2^{s_2}} \right)
 \right]
 \\
 \sY_{i,x}^{+\vee}[\CalX] & =
 \prod_{\alpha=1}^{n_{i,+}}
 \left[
 \left( 1 - \frac{\nu_{i,\alpha}^+}{x} \right)
 \prod_{s \in \lambda_{i,\alpha}} \msS \left( \frac{\nu_{i,\alpha}^+ q_1^{s_1-1} q_2^{s_2-1}}{x} \right)
 \right]
 \\
 \sY_{i,x}^-[\CalX] & =
 \prod_{\alpha=1}^{n_{i,-}}
 \left[
 \left( 1 - \frac{x}{\nu_{i,\alpha}^-} \right)
 \prod_{s \in \lambda_{i,\alpha}} \msS \left( \frac{x}{\nu_{i,\alpha}^- q_1^{-s_1+1} q_2^{-s_2+1}} \right)
 \right]
 \\
 \sY_{i,x}^{-\vee}[\CalX] & =
 \prod_{\alpha=1}^{n_{i,-}}
 \left[
 \left( 1 - \frac{\nu_{i,\alpha}^-}{x} \right)
 \prod_{s \in \lambda_{i,\alpha}} \msS \left( \frac{\nu_{i,\alpha}^- q_1^{-s_1} q_2^{-s_2}}{x} \right)
 \right]  
\end{align}
with the $\msS$-factor \eqref{eq:S-factor}.
The asymptotic behavior of these functions is given by
\begin{align}
 \sY_{i,x}^\sigma
 & \ \longrightarrow \
 \begin{cases}
  1 & (x \to 0) \\
  (-x)^{n_{i,\sigma}} / \nu_i^\sigma & (x \to \infty)
 \end{cases}
 \\
 \sY_{i,x}^{\sigma\vee}
 & \ \longrightarrow \
 \begin{cases}
  \nu_i^\sigma / (-x)^{n_{i,\sigma}} & (x \to 0) \\
  1 & (x \to \infty)
 \end{cases} 
\end{align}
where we remark $x \in \BC^\times$ and define
\begin{align}
 \nu_{i}^\sigma = \prod_{\alpha=1}^{n_{i,\sigma}} \nu_{i,\alpha}^\sigma
 \, .
\end{align}

Due to the reflection formula \eqref{eq:S-ref}, the $\sY^\sigma$-function and its dual are converted to each other:
\begin{align}
 \sY_{i,x}^{\sigma\vee} =
 \frac{(-1)^{n_{i,\sigma}} \nu_{i}^\sigma}{x^{n_{i,\sigma}}} \,
 \sY_{i,x}^\sigma
 \, .
\end{align}
Since the $\sY$-function is given as a ratio of the $\sY^\sigma$-functions \eqref{eq:Y_ratio}, its dual is given by
\begin{align}
 \sY_{i,x}^\vee
 = (-1)^{n_{i,+} - n_{i,-}} \, \frac{\nu_i^+}{\nu_i^-} \, x^{-n_{i,+}+n_{i,-}} \, \sY_{i,x}
 \, .
\end{align}
Thus, if $n_{i,+} = n_{i,-}$, these two become equivalent to each other up to the constant,
\begin{align}
 \sY_{i,x}^\vee = \frac{\nu_i^+}{\nu_i^-} \, \sY_{i,x}
 \quad \text{for} \quad n_{i,+} = n_{i,-}
 \, .
\end{align}
We remark that, imposing the condition
\begin{align}
 \frac{\nu_i^+}{\nu_i^-} = 1
 \, ,
\end{align}
the supergroup U$(n_{i,+}|n_{i,-})$ becomes SU$(n_{i,+}|n_{i,-})$.
Therefore, in this case, we have
\begin{align}
 \sY_{i,x}^\vee = (-x)^{-n_{i,+}+n_{i,-}} \, \sY_{i,x}
 \ \stackrel{n_{i,+} = n_{i,-}}{\longrightarrow} \ \sY_{i,x}
 \, .
 \label{eq:Y_conv}
\end{align}

\subsection{iWeyl reflection}

We study behavior of the partition function by adding/removing an instanton.
The partition shift gives rise to the shift of the $x$-variables depending on $\sigma = \pm$ as follows:
\begin{subequations}
\begin{align}
 \lambda_{i,\alpha,k}^\sigma
 & \ \longrightarrow \
 \lambda_{i,\alpha,k}^\sigma + 1
 \quad \implies \quad
 x_{i,\alpha,k}^\sigma
 \ \longrightarrow \
 q_2^\sigma x_{i,\alpha,k}^\sigma =
 \begin{cases}
  q_2 x_{i,\alpha,k}^+ & (\sigma = +) \\[.5em]
  q_2^{-1} x_{i,\alpha,k}^- & (\sigma = -)
 \end{cases}
 \label{eq:shift+}
 \\
 \lambda_{i,\alpha,k}^\sigma
 & \ \longrightarrow \
 \lambda_{i,\alpha,k}^\sigma - 1
 \quad \implies \quad
 x_{i,\alpha,k}^\sigma
 \ \longrightarrow \
 q_2^{-\sigma} x_{i,\alpha,k}^\sigma =
 \begin{cases}
  q_2^{-1} x_{i,\alpha,k}^+ & (\sigma = +) \\[.5em]
  q_2 x_{i,\alpha,k}^- & (\sigma = -)
 \end{cases}
 \label{eq:shift-} 
\end{align} 
\end{subequations}
This behavior under the shift suggests that adding/removing an instanton for the positive node $\sigma = +$ is equivalent to removing/adding an instanton for the negative node $\sigma = -$.
Hence the plus ($\sigma = +$) sector and the minus ($\sigma = -$) sector describe the positive and negative instanton configuration.

Define the configuration obtained by adding/removing an instanton
\begin{align}
 \CalX_{\text{ad:}(i,\alpha,k,\sigma)} =
 \left( \CalX \backslash \{x_{i,\alpha,k}^\sigma\} \right) \sqcup \{ q_2^\sigma x_{i,\alpha,k}^\sigma \}
 \, , \quad 
 \CalX_{\text{rm:}(i,\alpha,k,\sigma)} =
 \left( \CalX \backslash \{x_{i,\alpha,k}^\sigma\} \right) \sqcup \{ q_2^{-\sigma} x_{i,\alpha,k}^\sigma \} 
 \, .
\end{align}
The vector multiplet partition function (together with the topological term) behaves as
\begin{subequations} 
\begin{align}
 \frac{Z^\text{vec}_{i}[\CalX_{\text{ad:}(i,\alpha,k,+)}]}{Z^\text{vec}_{i}[\CalX]} & =
 - \fq_i \,
 \frac{\sY_{i,qx}^-[\CalX] \sY_{i,x}^{-\vee}[\CalX]}{\sY_{i,qx}^+[\CalX_\text{ad}] \sY_{i,x}^{+\vee}[\CalX]}
 = - \fq_i \,
 \frac{1}{\sY_{i,qx}[\CalX_\text{ad}] \sY_{i,x}^{\vee}[\CalX]} 
 \\[.5em]
 \frac{Z^\text{vec}_{i}[\CalX_{\text{ad:}(i,\alpha,k,-)}]}{Z^\text{vec}_{i}[\CalX]} & =
 - \fq_i^{-1} \,
 \frac{\sY_{i,x}^+[\CalX] \sY_{i,q^{-1}x}^{+\vee}[\CalX]}{\sY_{i,x}^{-}[\CalX] \sY_{i,q^{-1}x}^{-\vee}[\CalX_\text{ad}]}
 =
 - \fq_i^{-1} \,
 \sY_{i,x}[\CalX] \sY_{i,q^{-1}x}^\vee[\CalX_\text{ad}]
  \\[.5em]
 \frac{Z^\text{vec}_{i}[\CalX_{\text{rm:}(i,\alpha,k,+)}]}{Z^\text{vec}_{i}[\CalX]} & =
 - \fq_i^{-1} \,
 \frac{\sY_{i,q_1 x}^+[\CalX] \sY_{i,q_2^{-1}x}^{+\vee}[\CalX_\text{rm}]}{\sY_{i,q_1 x}^{-}[\CalX] \sY_{i,q_2^{-1}x}^{-\vee}[\CalX]}
 =
 - \fq_i^{-1} \,
 \sY_{i,q_1 x}[\CalX] \sY_{i,q_2^{-1}x}^\vee[\CalX_\text{rm}] 
  \\[.5em]
 \frac{Z^\text{vec}_{i}[\CalX_{\text{rm:}(i,\alpha,k,-)}]}{Z^\text{vec}_{i}[\CalX]} & =
 - \fq_i \,
 \frac{\sY_{i,q_2 x}^-[\CalX_\text{rm}] \sY_{i,q_1^{-1}x}^{-\vee}[\CalX]}{\sY_{i,q_1 x}^+[\CalX] \sY_{i,q_1^{-1} x}^{+\vee}[\CalX]}
 = - \fq_i \,
 \frac{1}{\sY_{i,q_2 x}[\CalX_\text{rm}] \sY_{i,q_1^{-1}x}^{\vee}[\CalX]} 
\end{align} 
\end{subequations}
where $x = x_{i,\alpha,k}^\sigma$, and we omit the index $(i,\alpha,k,\sigma)$ specifying where to add/remove an instanton as far as no confusion.
We remark
\begin{align}
 \sY_{i,x}^\sigma[\CalX_{\text{ad/rm:}(i',\alpha,k,\sigma')}] = \sY_{i,x}^\sigma[\CalX]
 \qquad \text{for} \quad (i,\sigma) \neq (i',\sigma')
 \, .
\end{align}
Indeed adding/removing an instanton for the positive/negative node is equivalent to removing/adding an instanton for the negative/positive node.

Including the hypermultiplet and Chern--Simons contributions, the total partition function behaves under the shift as follows:
\begin{subequations}
\begin{align}
 \frac{Z[\CalX_{\text{ad:}(i,\alpha,k,+)}]}{Z[\CalX]} & =
 - \fq_i \, x^{\kappa_i^+} \,
 \frac{\sP_{i,qx} \widetilde{\sP}_{i,x}^\vee}{\sY_{i,qx}[\CalX_\text{ad}] \sY_{i,x}^{\vee}[\CalX]}
 \prod_{e:i \to j} \sY_{j,\mu_e^{-1} q x}[\CalX]
 \prod_{e:j \to i} \sY_{j,\mu_e x}^\vee[\CalX]
 \Bigg|_{x = x_{i,\alpha,k}^+}
 \label{eq:ref_full+}
 \\
 \frac{Z[\CalX_{\text{ad:}(i,\alpha,k,-)}]}{Z[\CalX]} & =
 - \fq_i^{-1} \, (q^{-1} x)^{-\kappa_i^-} \,
 \frac{\sY_{i,x}[\CalX] \sY_{i,q^{-1}x}^{\vee}[\CalX_\text{ad}]}{\sP_{i,x} \widetilde{\sP}_{i,q^{-1}x}^\vee}
 \prod_{e:i \to j} \sY_{j,\mu_e^{-1} x}[\CalX]^{-1}
 \prod_{e:j \to i} \sY_{j,\mu_e q^{-1} x}^{\vee}[\CalX]^{-1}
 \Bigg|_{x = x_{i,\alpha,k}^-}
 \label{eq:ref_full-} 
\end{align} 
\end{subequations}
where we define the matter function
\begin{align}
 \sP_{i,x} = \frac{\sP_{i,x}^+}{\sP_{i,x}^-}
 \, , \qquad
 \widetilde{\sP}_{i,x} = \frac{\widetilde{\sP}_{i,x}^+}{\widetilde{\sP}_{i,x}^-}
 \, , \qquad
 \sP_{i,x}^\vee = \frac{\sP_{i,x}^{+\vee}}{\sP_{i,x}^{-\vee}}
 \, , \qquad
 \widetilde{\sP}_{i,x}^\vee = \frac{\widetilde{\sP}_{i,x}^{+\vee}}{\widetilde{\sP}_{i,x}^{-\vee}}
 \, ,
 \label{eq:P_ratio}
\end{align}
with
\begin{subequations}
\begin{align}
 \sP_{i,x}^\sigma & = \prod_{\mu \in \CalM_i^\sigma} \left( 1 - \frac{x}{\mu} \right)
 \, , \qquad
 \widetilde{\sP}_{i,x}^\sigma = \prod_{\mu \in \widetilde{\CalM}_i^\sigma} \left( 1 - \frac{x}{\mu} \right)
 \, , \\
 \sP_{i,x}^{\sigma\vee} & = \prod_{\mu \in \CalM_i^\sigma} \left( 1 - \frac{\mu}{x} \right)
 \, , \qquad
 \widetilde{\sP}_{i,x}^{\sigma\vee} = \prod_{\mu \in \widetilde{\CalM}_i^\sigma} \left( 1 - \frac{\mu}{x} \right)
 \, .
\end{align}
\end{subequations}
As mentioned above, adding/removing for the positive/negative node is equivalent to removing/adding for the negative/positive node if the two Chern--Simons levels coincide $\kappa_i^+ = \kappa_i^-$.
Namely \eqref{eq:ref_full+} and \eqref{eq:ref_full-} are essentially inverse operations.

The analysis above shows that, using the full $\sY$-function and the matter function, consisting of both the positive and negative ones, we can apply the same argument to construct the $qq$-character with the supergroup gauge and flavor nodes as the ordinary (non-supergroup) gauge theory~\cite{Nekrasov:2015wsu}.
From the instanton-adding operation \eqref{eq:ref_full+}, one can show that the pole singularity of the $\sY^+$-function is cancelled as follows:
\begin{align}
 &
 \underset{x_{i,\alpha,k,+}}{\operatorname{Res}}
 \Bigg[
 \sY_{i,qx}[\CalX_{\text{ad:}(i,\alpha,k,+)}] Z[\CalX_{\text{ad:}(i,\alpha,k,+)}]
 \nonumber \\ & \hspace{10em} 
 + \fq_i \, x^{\kappa_i^+} \, 
 \frac{\sP_{i,qx} \widetilde{\sP}_{i,x}^\vee}{\sY_{i,x}^{\vee}[\CalX]}
 \prod_{e:i \to j} \sY_{j,\mu_e^{-1} q x}[\CalX]
 \prod_{e:j \to i} \sY_{j,\mu_e x}^\vee[\CalX] \,
 Z[\CalX]
 \Bigg] = 0
 \, .
 \label{eq:polefree_Y+}
\end{align}
We remark that the $\sY$-function is given as a ratio of the $\sY^\pm$-functions \eqref{eq:Y_ratio}.
Thus the pole singularity of the $\sY^+$-function is actually cancelled in \eqref{eq:polefree_Y+}, while there are still poles from the $\sY^-$-function.
Similarly from \eqref{eq:ref_full-}, we obtain the pole cancellation for the $\sY^-$-function:
\begin{align}
 &
 \underset{x_{i,\alpha,k,-}}{\operatorname{Res}}
 \Bigg[
 \frac{1}{\sY_{i,q^{-1}x}[\CalX_{\text{ad:}(i,\alpha,k,-)}]} \,
 Z[\CalX_{\text{ad:}(i,\alpha,k,-)}]
 \nonumber \\ & \hspace{3.5em}
 + \fq_i^{-1} \, (q^{-1} x)^{-\kappa_i^-} \,
 \frac{\sY_{i,x}[\CalX]}{\sP_{i,x} \widetilde{\sP}_{i,q^{-1}x}^\vee}
 \prod_{e:i \to j} \sY_{j,\mu_e^{-1} x}[\CalX]^{-1}
 \prod_{e:j \to i} \sY_{j,\mu_e q^{-1} x}^{\vee}[\CalX]^{-1} \,
 Z[\CalX]
 \Bigg] = 0
 \, .
\end{align}
As mentioned before, the latter is equivalent to the instanton-removing operation for the positive node when $\kappa_i^+ = \kappa_i^- =: \kappa_i$.

Therefore we can apply the same iWeyl reflection operation to construct the $qq$-character as discussed before~\cite{Nekrasov:2015wsu}:
\begin{align}
 \sT_{i,x} & = \sY_{i,x} 
 + \fq_i \, (q^{-1} x)^{\kappa_i} \, 
 \frac{\sP_{i,x} \widetilde{\sP}_{i,q^{-1}x}^\vee}{\sY_{i,q^{-1}x}^{\vee}[\CalX]}
 \prod_{e:i \to j} \sY_{j,\mu_e^{-1} x}[\CalX]
 \prod_{e:j \to i} \sY_{j,\mu_e q^{-1} x}^\vee[\CalX]
 + \cdots
\end{align}
For the non-supergroup theory, the gauge theory average of the $qq$-character turns out to be a regular function without any pole singularities, so that it is a polynomial, and its coefficients are interpreted as the Coulomb moduli, namely coordinates of the gauge theory moduli space at the Coulomb branch.
In contrast to this, for the supergroup gauge theory with a proper matter content, the $qq$-character average becomes a rational function, given as a ratio of polynomials
\begin{align}
 \left< \, \sT_{i,x} \, \right>
 = \frac{x^{n_{i,+}} + \cdots}{x^{n_{i,-}} + \cdots}
 =: T_{i,x}^{n_{i,+}|n_{i,-}}
\end{align}
where the operator average is taken with respect to the gauge theory partition function
\begin{align}
 \left< \, \CalO \, \right>
 = \frac{1}{Z} \sum_\CalX \CalO[\CalX] Z[\CalX]
 \, , \qquad
 Z = \sum_{\CalX} Z[\CalX]
 \, .
\end{align}
In this case, the coefficients in the numerator and the denominator become the Coulomb moduli of the positive and negative nodes of U$(n_{i,+}|n_{i,-})$, respectively.
Namely, the $qq$-character gives rise to the supercharacteristic polynomial of the adjoint scalar field in the $i$-th vector multiplet,
\begin{align}
 \left< \, \sT_{i,x} \, \right> =
 \begin{cases}
  \sdet \left( x \Id_N - \Phi_i \right) & (\text{4d}) \\[.5em] \displaystyle
  \sdet \left( \Id_N - e^{\Phi - \log x} \right) & (\text{5d})
 \end{cases}
\end{align}

In the classical limit $q_{1,2} \to 1$, the $qq$-character is reduced to the ordinary character associated with the quiver.
Thus the classical Seiberg--Witten geometry is similarly described with the fundamental characters of the quiver as the ordinary non-supergroup gauge theory~\cite{Nekrasov:2012xe}, but, in the supergroup theory, the Coulomb moduli are encoded as the coefficients of the rational function given as the gauge theory average of the fundamental characters.

\subsubsection{$A_1$ quiver}

Let us consider the simplest example $\Gamma = A_1$ which consists of a single gauge node U$(n_{1,+}|n_{1,-})$.
For simplicity, we consider the pure gauge theory with the Chern--Simons level $\kappa_1 = 0$.
In this case, the $qq$-character is given by
\begin{align}
 \sT_{1,x} = \sY_{1,x} + \fq_1 \, \frac{1}{\sY_{i,q^{-1}x}^\vee}
 \, .
 \label{eq:qq-ch_A1}
\end{align}
This expression itself is identical to the non-supergroup theory~\cite{Nekrasov:2015wsu}, but its gauge theory average becomes a rational function
\begin{align}
 \Big< \sT_{1,x} \Big> = T_{1,x}^{n_{1,+}|n_{1,-}}
 = \frac{x^{n_{1,+}} + \cdots}{x^{n_{1,-}} + \cdots}
 \, .
\end{align}
In the classical limit $q_{1,2} \to 1$, the $qq$-character is reduced to
\begin{align}
 y + \frac{\fq_1}{y}
 = T_{1,x}^{n_{1,+}|n_{1,-}}
\end{align}
where $y := \left< \, \sY_{1,x} \, \right>$.
This algebraic equation defines the classical Seiberg--Witten curve for pure SU($n_{1,+}|n_{1,-}$) SYM theory
\begin{subequations}\label{eq:SW-curve_cl}
 \begin{align}
 \Sigma = \left\{ (x,y) \mid H(x,y) = 0 \right\}
 \end{align}
 with
 \begin{align}
  H(x,y) = y + \frac{\fq_1}{y} - T_{1,x}^{n_{1,+}|n_{1,-}}
  \, .
 \end{align}
\end{subequations}
We also define a one-form, called the Seiberg--Witten differential on the curve $\Sigma$,
\begin{align}
 \lambda =
 \begin{cases}
  x \, d \log y & (\text{4d}) \\ \log x \, d \log y & (\text{5d})
 \end{cases}
 \label{eq:SW_1form}
\end{align}
and the associated symplectic two-form
\begin{align}
 \omega = d \lambda =
 \begin{cases}
  dx \wedge d \log y & (\text{4d}) \\
  d \log x \wedge d \log y & (\text{5d})
 \end{cases}
 \label{eq:symp_2form}
\end{align}
with
\begin{align}
 (x,y) \in
 \begin{cases}
  \BC \times \BC^\times & (\text{4d}) \\ \BC^\times \times \BC^\times & (\text{5d})
 \end{cases}
\end{align}
As pointed out in \cite{Dijkgraaf:2016lym}, the Seiberg--Witten curve \eqref{eq:SW-curve_cl} coincides with that for SU($n_{1,+}$) gauge theory with $2 n_{1,-}$ fundamental hypermultiplets by tuning the flavor fugacities to the Coulomb moduli for the negative node.
Actually, in this case, the iWeyl reflection with respect to $\sY^{+}$-function is given by
\begin{align}
 \sY_{1,x}^+ \ \longrightarrow \
 \fq_1 \, \frac{\sY_{1,x}^- \sY_{1,q^{-1} x}^{-\vee}}{\sY_{1,q^{-1} x}^{+\vee}}
 \, .
\end{align}
Applying the decoupling trick, discussed in Sec.~\ref{sec:decoupling}, the $\sY^-$-function is reduced to a degree $n_{1,-}$ polynomial, which just plays a role of the matter polynomial.
Hence this reflection is equivalent to that for SU($n_{1,+}$) gauge theory with $2 n_{1,-}$ fundamental matters.
The curve \eqref{eq:SW-curve_cl} describes the Riemann surface with genus $n_{1,+} - 1$ and $2 n_{1,-}$ punctures.
The cycle integral of the differential $\lambda$ gives rise to the positive Coulomb moduli, while the residue associated with the puncture provides the negative Coulomb moduli.

\subsubsection{$A_r$ quiver}\label{sec:Ar}

We then study the linear quiver theory $\Gamma = A_r$.
In this case, the iWeyl reflection is given by
\begin{align}
  \sY_{i,x}
 \ \longrightarrow \
 \fq_i \, \sP_{i,x} \widetilde{\sP}^\vee_{i,q^{-1} x} \frac{\sY_{i+1,\mu_{i \to i+1}^{-1} x} \sY^\vee_{i-1, \mu_{i-1 \to i}q^{-1} x}}{\sY_{i,q^{-1} x}^\vee}
 \qquad
 (i = 1,\ldots,r)
 \, ,
 \label{eq:Ar_ref}
\end{align}
where we define
\begin{align}
 \sY_{0,x} = \sY_{r+1,x} = 1
 \, .
\end{align}
Then one can construct the fundamental $qq$-characters $(\sT_{i,x})_{i = 1,\ldots,r}$ corresponding to the $i$-th antisymmetric representations of SU$(r+1)$~\cite{Nekrasov:2015wsu,Kimura:2015rgi}.

For simplicity, let us focus on the situation with $n_{i,\pm} = n_{\pm}$, $n_{i,\pm}^\text{(a)f}=0$, and putting $\mu_e = 0$ by the gauge transformation.
In this case, the iWeyl reflection \eqref{eq:Ar_ref} in terms of the $\sY^\pm$-functions is
\begin{subequations} 
 \begin{align}
  \sY_{1,x}^+
  & \ \longrightarrow \
  \fq_1
  \left(
  \frac{\sY_{1,x}^- \sY_{1,q^{-1} x}^{-\vee}}{\sY_{2, x}^-}
  \right)
  \frac{\sY_{2, x}^+}{\sY_{1,q^{-1} x}^{+\vee}}
  \, , \\
  \sY_{i,x}^+
  & \ \longrightarrow \
  \fq_i 
  \left(
  \frac{\sY_{i,x}^- \sY_{i,q^{-1} x}^{-\vee}}{\sY_{i+1, x}^- \sY^{-\vee}_{i-1, q^{-1} x}}
  \right)
  \frac{\sY_{i+1, x}^+ \sY^{+\vee}_{i-1, q^{-1} x}}{\sY_{i,q^{-1} x}^{+\vee}}
  \qquad (i = 2,\ldots,r-1)
  \, , \\
  \sY_{r,x}^+
  & \ \longrightarrow \
  \fq_r
  \left(
  \frac{\sY_{r,x}^- \sY_{r,q^{-1} x}^{-\vee}}{\sY^{-\vee}_{r-1, q^{-1} x}}
  \right)
  \frac{\sY^{+\vee}_{r-1, q^{-1} x}}{\sY_{r,q^{-1} x}^{+\vee}}  
 \end{align}
\end{subequations}
Then, applying the decoupling trick again, the $\sY^-$-functions are reduced to the matter polynomials.
Furthermore, tuning the negative Coulomb moduli as the common values, one can replace all the $\sY^-$-functions with a single polynomial, $(\sY_{i,x}^-, \sY_{i,x}^{-\vee}) \to (\sP_{x}^-,\sP_x^{-\vee})$, and obtain
\begin{subequations} 
 \begin{align}
  \sY_{1,x}^+
  & \ \longrightarrow \
  \fq_1 \, \sP_{q^{-1} x}^{-\vee} \,
  \frac{\sY_{2, x}^+}{\sY_{1,q^{-1} x}^{+\vee}}
  \, , \\
  \sY_{i,x}^+
  & \ \longrightarrow \
  \fq_i \,
  \frac{\sY_{i+1, x}^+ \sY^{+\vee}_{i-1, q^{-1} x}}{\sY_{i,q^{-1} x}^{+\vee}}
  \qquad (i = 2,\ldots,r-1)
  \, , \\
  \sY_{r,x}^+
  & \ \longrightarrow \
  \fq_r \, \sP_{x}^- \,
  \frac{\sY^{+\vee}_{r-1, q^{-1} x}}{\sY_{r,q^{-1} x}^{+\vee}}
  \, .
 \end{align}
\end{subequations}
These reflections are equivalent to those of $\mathrm{SU}(n_+) \times \cdots \times \mathrm{SU}(n_+)$ linear quiver gauge theory with fundamental and antinfundamental flavors attached to $r$-th and 1st nodes, which is consistent with the result based on the brane realization discussed in Sec.~\ref{sec:brane}.

\subsubsection{$\widehat{A}_0$ quiver}\label{sec:A0}

We consider the simplest affine quiver, $\Gamma = \widehat{A}_0$, corresponding to 4d $\CalN = 2^*$ (5d $\CalN=1^*$) theory, with supergroup gauge symmetry SU$(n_{1,+}|n_{1,-})$.
In this case, the iWeyl reflection is given by
\begin{align}
 \sY_{1,x} \ \longrightarrow \ \fq_1 \, \msS \left( \mu^{-1} \right) \frac{\sY_{1,\mu^{-1} x} \sY^\vee_{1,\mu q^{-1}}}{\sY^\vee_{1,q^{-1} x}}
 \, .
 \label{eq:A0_reflection}
\end{align}
Let us focus on SU$(n_1|n_1)$ theory for simplicity, where the $\sY$-function and its dual are equivalent.
Then the $qq$-character is given as an infinite sum over the partition~\cite{Nekrasov:2013xda,Nekrasov:2015wsu,Kimura:2015rgi},
\begin{align}
 \sT_{1,x}
 = \sum_{\lambda} \fq_1^{|\lambda|} \, Z_\lambda^{\widehat{A}_0}(q_3,q_4)
 :
 \prod_{s \in \partial_+ \lambda} \sY_{1,qx/\tilde{x}(s)}
 \prod_{s \in \partial_- \lambda} \sY_{1,x/\tilde{x}(s)} 
 :
\end{align}
where $\partial_+ \lambda$ and $\partial_- \lambda$ are the outer and innter boundaries of the partition $\lambda$, and we define
\begin{align}
 \tilde{x}(s) = q_3^{s_1 - 1} q_4^{s_2 - 1} q
\end{align}
with
\begin{align}
 q_3 = \mu^{-1}
 \, , \qquad
 q_4 = \mu q^{-1}
 \, .
\end{align}
These parameters play a role as the equivariant parameters for the transversal planes in the 8d setup, called the gauge origami~\cite{Nekrasov:2015wsu}.
Actually their product becomes unity,
\begin{align}
 q_1 q_2 q_3 q_4 = 1
 \, ,
\end{align}
which is interpreted as the Calabi--Yau condition in 8d. 
The factor $Z_\lambda^{\widehat{A}_0}(q_3, q_4)$ is the U(1) instanton partition function for $\widehat{A}_0$ quiver with the equivariant parameters $(q_3,q_4)$.

In terms of the $\sY^{\pm}$-functions, the iWeyl reflection \eqref{eq:A0_reflection} becomes
\begin{align}
 \sY_{1,x}^+ \ \longrightarrow \
 \fq_1 \, \msS \left( \mu^{-1} \right)
 \left( \frac{\sY_{1,x}^- \sY_{1,q^{-1}x}^{-\vee}}{\sY_{1,\mu^{-1} x}^- \sY_{1,\mu q^{-1}x}^{-\vee}} \right)
 \frac{\sY_{1,\mu^{-1} x}^+ \sY^{+\vee}_{1,\mu q^{-1}}}{\sY^{+\vee}_{1,q^{-1} x}}
 \, .
\end{align}
This implies that the Seiberg--Witten geometry of SU$(n_{1,+}|n_{1,-})$ $\widehat{A}_0$ quiver gauge theory is equivalent to that for the SU($n_{1,+}$) $\widehat{A}_0$ theory with $2 n_{1,-}$ {\em positive} fundamental hypermultiplets with the mass parameter $(\nu_{1,\alpha}^-)_{\alpha=1,\ldots,n_{1,-}}$, and $2 n_{1,-}$ {\em negative} fundamental matters with the mass parameters $(\mu \nu_{1,\alpha}^-)_{\alpha=1,\ldots,n_{1,-}}$.
These contributions from the positive and negative fundamental matters are not canceled with each other due to the finite adjoint mass parameter $\mu \neq 1$.

\subsubsection{Higher-order reflection and collision}

In general, we may consider the higher-order $qq$-character, corresponding to the higher-representation of the quiver, where the degree of the highest weight is larger than one.
In such a case, the $qq$-character starts with the product of the $\sY$-functions.
For example, the iWeyl reflection for pure U$(n+|n-)$ theory with the degree two weight is given as
\begin{align}
 \sY_{i,x} \sY_{i,x'}
 + \msS \left( \frac{x'}{x} \right) \frac{\sY_{i,x'}}{\sY^\vee_{i,q^{-1} x}}
 + \msS \left( \frac{x}{x'} \right) \frac{\sY_{i,x}}{\sY^\vee_{i,q^{-1} x'}}
 + \frac{1}{\sY^\vee_{i,q^{-1} x} \sY^\vee_{i,q^{-1} x'}}
\end{align}
where we have to insert the $\msS$-factor \eqref{eq:S-factor} to eliminate the pole singularity~\cite{Nekrasov:2015wsu,Kimura:2015rgi}.
In the colliding limit $x' \to x$, it involves a derivative term
\begin{align}
 \sY_{i,x}^2
 +
 \left( \mathfrak{c}(q_1,q_2) - \frac{(1 - q_1)(1 - q_2)}{1 - q} \partial_{\log x} \log \sY_{i,x} \sY_{i,q^{-1} x} \right) \frac{\sY_{i,x}}{\sY_{i,q^{-1} x}^\vee}
 + \frac{1}{\sY^{\vee2}_{i,q^{-1} x}}
\end{align}
where the factor $\mathfrak{c}(q_1,q_2)$ is given by
\begin{align}
 \mathfrak{c}(q_1,q_2)
 = \lim_{x \to 1} \left( \msS(x) + \msS(x^{-1}) \right)
 = \frac{1 - 6q + q^2 + (q_1 + q_2)(1 + q)}{(1-q)^2}
 \ \stackrel{q_{1,2} \to 1}{\longrightarrow} \ 2
 \, .
\end{align}
Appearance of such a derivative term is a specific feature of the $qq$-character, which appears for, e.g., the adjoint representation of $D_4$ quiver~\cite{Nekrasov:2015wsu,Kimura:2016ebq}.

\section{Bethe/Gauge correspondence}\label{sec:Bethe/Gauge}

It has been shown that the NS limit ($\epsilon_2 \to 0$) of gauge theory has a geometric correspondence to the quantum integrable system, a.k.a. the Bethe/Gauge correspondence~\cite{Nekrasov:2009rc}.
In such a limit, one can focus on the saddle point configuration with respect to the small parameter $\epsilon_2$, and the corresponding saddle point equation turns out to be Bethe equation of the associated quantum integrable system.
In this Section, we study the saddle point configuration of the supergroup gauge theory in the NS limit, and discuss its implication to quantum integrable system.

\subsection{Effective twisted superpotential from asymptotics}\label{sec:sup-pot}

We study the asymptotic behavior of the partition function in the NS limit $\epsilon_2 \to 0$.
In this limit, the partition function behaves as
\begin{align}
 Z
 \ \stackrel{\epsilon_2 \to 0}{\longrightarrow} \
 \exp \left( \frac{1}{\epsilon_2} \msW(\epsilon_1) + \cdots \right)
\end{align}
where the leading contribution is identified with the effective twisted superpotential.
This behavior suggests that one can apply the saddle point analysis with the small parameter $\epsilon_2$.

As shown in Sec.~\ref{sec:full}, the full partition function for 5d gauge theory is expressed using the $q$-shifted factorial \eqref{eq:q-Pochhammer} with $n \to \infty$, which asymptotically behaves
\begin{align}
 (z;q)_\infty
 = \exp \left[ \frac{1}{\epsilon} \operatorname{Li}_2(z) + O(\epsilon^0) \right]
 \qquad \text{for} \quad
 q = e^\epsilon
\end{align}
and we define the polylogarithm function
\begin{align}
 \operatorname{Li}_k(z) = \sum_{n=1}^\infty \frac{z^n}{n^k}
 \, .
\end{align}
Here the limit should be taken as $\epsilon \to 0^-$ since $|q| < 1$.

In the following we consider the asymptotic behavior of the partition function in the NS limit $\epsilon_2 \to 0$ with the function defined%
\footnote{%
For 6d theory, we apply the same analysis just by replacing the $L$-function with its elliptic analog:
\begin{align}
 L(z;q;p) = \operatorname{Li}_2(qz;p) - \operatorname{Li}_2(z;p)
\end{align}
and the elliptic analog of the polylogarithm
\begin{align}
 \operatorname{Li}_k(z;p) = \sum_{m \neq 0} \frac{z^m}{m^k (1 - p^m)}
 \, ,
\end{align}
since the elliptic gamma function asymptotically behaves as
\begin{align}
 \Gamma(z;p,q) = \exp \left( - \frac{1}{\epsilon} \operatorname{Li}_2(z;p) + O(\epsilon^0) \right)
\end{align}
with $q = e^\epsilon$.
}
\begin{align}
 L(z;q)
 = \operatorname{Li}_2(qz) - \operatorname{Li}_2(z)
\end{align}
with the reflection
\begin{align}
 L(z;q^{-1}) = - L(q^{-1}z;q)
 \, .
\end{align}
The leading contribution in the limit $\epsilon_2 \to 0$ would be identified with the effective twisted superpotential as summarized in the following:

\subsubsection*{Vector multiplet}
\begin{subequations} 
\begin{align}
 Z_{i,++}^\text{vec} & \ \longrightarrow \
 \exp \left[
 \frac{1}{\epsilon_2} \sum_{\substack{(x,x') \in \CalX_i^+ \times \CalX_i^+\\ x \neq x'}} L\left( \frac{x}{x'};q_1 \right) 
 \right]
 \\[.5em]
 Z_{i,+-}^\text{vec} & \ \longrightarrow \
 \exp \left[
 \frac{1}{\epsilon_2} \sum_{(x,x') \in \CalX_i^+ \times \CalX_i^-}
 L \left( q_1 \frac{x}{x'};q_1 \right)
 \right]
\end{align}
\begin{align}
 Z_{i,-+}^\text{vec} & \ \longrightarrow \
 \exp \left[
 \frac{1}{\epsilon_2} \sum_{(x,x') \in \CalX_i^- \times \CalX_i^+}
 L \left( q_1^{-1} \frac{x}{x'};q_1 \right)
 \right]
 \\[.5em]
 Z_{i,--}^\text{vec} & \ \longrightarrow \
 \exp \left[
 \frac{1}{\epsilon_2} \sum_{\substack{(x,x') \in \CalX_i^- \times \CalX_i^-\\ x \neq x'}} L\left( \frac{x}{x'};q_1 \right)
 \right] 
\end{align}
\end{subequations}

\subsubsection*{Bifundamental hypermultiplet}

\begin{subequations}
\begin{align}
 Z_{e:i \to j,++}^\text{bf} & \ \longrightarrow \
 \exp \left[
 - \frac{1}{\epsilon_2} \sum_{(x,x') \in \CalX_i^+ \times \CalX_j^+} L \left( \mu_e^{-1} \frac{x}{x'} ;q_1 \right)
 \right]
 \\[.5em]
 Z_{e:i \to j,+-}^\text{bf} & \ \longrightarrow \
 \exp \left[
 - \frac{1}{\epsilon_2} \sum_{(x,x') \in \CalX_i^+ \times \CalX_j^-} L \left( \mu_e^{-1} q_1 \frac{x}{x'} ;q_1 \right)
 \right]
\end{align}
\begin{align}
 Z_{e:i \to j,-+}^\text{bf} & \ \longrightarrow \
 \exp \left[
 - \frac{1}{\epsilon_2} \sum_{(x,x') \in \CalX_i^- \times \CalX_j^+} L \left( \mu_e^{-1} q_1^{-1} \frac{x}{x'} ;q_1 \right)
 \right]
 \\[.5em]
 Z_{e:i \to j,--}^\text{bf} & \ \longrightarrow \
 \exp \left[
 - \frac{1}{\epsilon_2} \sum_{(x,x') \in \CalX_i^- \times \CalX_j^-} L \left( \mu_e^{-1} \frac{x}{x'} ;q_1 \right)
 \right] 
\end{align}
\end{subequations}

\subsubsection*{Fundamental and antifundamental hypermultiplet}
\begin{subequations}
 \begin{align}
 Z_{i,+\sigma}^\text{f} & \ \longrightarrow \
 \exp \left[
  - \frac{\sigma}{\epsilon_2} \sum_{(x,\mu) \in \CalX_i^+ \times \CalM_i^\sigma} \operatorname{Li}_2 \left( q_1 \frac{x}{\mu} \right)
 \right]
 \\[.5em]
 Z_{i,-\sigma}^\text{f} & \ \longrightarrow \
 \exp \left[
  - \frac{\sigma}{\epsilon_2} \sum_{(x,\mu) \in \CalX_i^- \times \CalM_i^\sigma} \operatorname{Li}_2 \left( \frac{x}{\mu} \right)
 \right]
\end{align}
\begin{align}
 Z_{i,+\sigma}^\text{af} & \ \longrightarrow \
 \exp \left[
  \frac{\sigma}{\epsilon_2} \sum_{(x,\mu) \in \CalX_i^+ \times \widetilde{\CalM}_i^\sigma} \operatorname{Li}_2 \left( q_1 \frac{\mu}{x} \right)
 \right]
 \\[.5em]
 Z_{i,-\sigma}^\text{af} & \ \longrightarrow \
 \exp \left[
  \frac{\sigma}{\epsilon_2} \sum_{(x,\mu) \in \CalX_i^- \times \widetilde{\CalM}_i^\sigma} \operatorname{Li}_2 \left( q_1 \frac{\mu}{x} \right)
 \right]
\end{align}
\end{subequations}

\subsection{Bethe equation}

\subsubsection{Preliminary}

We first quickly review the Bethe equation for $G$-symmetric spin chain.
The Bethe equation for spin chain with symmetry $G$ is given as follows:
\begin{align}
 \prod_{a=1}^{L_i}
 \frac{[x_{i,k} - \xi_{i,a} + s_{i,a} \hbar]}
      {[x_{i,k} - \xi_{i,a} - s_{i,a} \hbar]}
 = - \fq_i
 \prod_{(j,k') (\neq (i,k))}
 \frac{[x_{i,k} - x_{j,k'} - c_{ij} \hbar/2]}
      {[x_{i,k} - x_{j,k'} + c_{ij} \hbar/2]}
 \label{eq:BE1}
\end{align}
with the Cartan matrix $(c_{ij})_{i,j = 1,\ldots, \rk G}$,%
\footnote{%
For the non-simply-laced case, the Cartan matrix in the Bethe equation \eqref{eq:BE1} is replaced by its symmetrization $(b_{ij})_{i,j=1,\ldots,\rk G}$~\cite{Reshetikhin:1987bz}.
See \cite{Kimura:2017hez} for its gauge theory realization.
}
the inhomogeneous parameters $(\xi_{i,a})_{i = 1,\ldots, \rk G, a = 1,\ldots, L_i}$, the spins $(s_{i,a})_{i = 1,\ldots, \rk G, a = 1,\ldots, L_i}$, the twist parameters $(\fq_i)_{i = 1,\ldots,\rk G}$, and the Bethe roots~$(x_{i,k})_{i=1,\ldots,\rk G, k = 1,\ldots, N_i}$.
In general, one can assign different lengths $(L_i)_{i = 1,\ldots, \rk G}$ for each node.
The number of Bethe roots $(N_i)_{i = 1,\ldots, \rk G}$ corresponds to that of magnons in the spin chain.
We define the odd function
\begin{align}
 [x] =
 \begin{cases}
  x & (\text{rational}) \\
  2 \sinh \frac{x}{2} & (\text{trigonometric}) \\
  \theta_1(x) & (\text{elliptic})
 \end{cases}
 \label{eq:[x]}
\end{align}
obeying
\begin{align}
 [-x] = -[x]
 \, .
\end{align}
The three possibilities of $[x]$-function (rational/trigonometric/elliptic) correspond to the anisotropy of spin chain (XXX/XXZ/XYZ).

The Bethe equation \eqref{eq:BE1} is rephrased in a compact form using the $Q$-function
\begin{align}
 \frac{a_i(x_{i,k})}{d_i(x_{i,k})}
 = \fq_i \, \prod_{j=1}^{\rk G}
 \frac{Q_j(x_{i,k} - c_{ij}\hbar/2)}{Q_j(x_{i,k} + c_{ij}\hbar/2)}
 \label{eq:BE2}
\end{align}
where we define
\begin{align}
 Q_i(x) = \prod_{k=1}^{N_i} [x - x_{i,k}]
\end{align}
and
\begin{align}
 a_i(x) = \prod_{a=1}^{L_i} [x - \xi_{i,a} + s_{i,a} \hbar]
 \, , \qquad
 d_i(x) = \prod_{a=1}^{L_i} [x - \xi_{i,a} - s_{i,a} \hbar]
 \, .
\end{align}
We remark that the functions $a_i(x)$ and $d_i(x)$ can be also expressed in terms of the Drinfeld polynomial.

\subsubsection{Gauge theory derivation}

To obtain the on-shell value of the twisted superpotential $\msW(\epsilon_1)$ from the asymptotic behavior of the partition function discussed in Sec.~\ref{sec:sup-pot}, we have to evaluate it with the critical configuration $\CalX_*$ defined with the saddle point equation
\begin{align}
 \exp \left( \epsilon_2 \frac{\partial}{\partial \log x} \log Z[\CalX_*] \right) = 1
 \, ,
 \label{eq:saddle1}
\end{align}
which is essentially the twisted F-term condition with respect to the twisted superpotential, and thus the critical configuration gives rise to the SUSY vacuum.
Actually, in the limit $\epsilon_2 \to 0$, the critical configuration dominates in the partition function:
\begin{align}
 Z = \sum_{\CalX} Z[\CalX]
 \quad \stackrel{\epsilon_2 \to 0}{\longrightarrow} \quad
 Z[\CalX_*]
 = \exp \left( \frac{1}{\epsilon_2} \msW(\epsilon_1) + \cdots \right)
 \, .
\end{align}

Since the dynamical $x$-variable is defined as \eqref{eq:Xset2}, the saddle point condition \eqref{eq:saddle1} is equivalent to the invariance under the partition shift.
Namely, from \eqref{eq:ref_full+} and \eqref{eq:ref_full-}, we obtain the saddle point equations in the limit $q_2 \to 1$ with $q_1$ fixed,
\begin{align}
 1 & =
 - \fq_i \, x^{\kappa_i} \,
 \frac{\sP_{i,q_1 x} \widetilde{\sP}_{i,x}^\vee}{\sY_{i,q_1x} \sY_{i,x}^{\vee}}
 \prod_{e:i \to j} \sY_{j,\mu_e^{-1} q_1 x}
 \prod_{e:j \to i} \sY_{j,\mu_e x}^\vee
 \Bigg|_{\CalX_*,x = x_{i,\alpha,k}^+}
 \label{eq:saddle_full+}
 \\
 1 & =
 - \fq_i^{-1} \, (q_1^{-1} x)^{-\kappa_i} \,
 \frac{\sY_{i,x}\sY_{i,q_1^{-1}x}^{\vee}}{\sP_{i,x} \widetilde{\sP}_{i,q_1^{-1}x}^\vee}
 \prod_{e:i \to j} \sY_{j,\mu_e^{-1} x}^{-1}
 \prod_{e:j \to i} \sY_{j,\mu_e q_1^{-1} x}^{\vee-1}
 \Bigg|_{\CalX_*,x = x_{i,\alpha,k}^-}
 \label{eq:saddle_full-}
\end{align}
where we assume $\kappa_i^+ = \kappa_i^- =: \kappa_i$.
Instead of these expressions, we use the following ones obtained by rescaling the $\sY$-functions with the $\Gamma$-factors without modifying the singularity of them,
\begin{align}
 1 & =
 - \fq_i \, x^{\kappa_i} \,
 \frac{\widetilde{\sP}_{i,x}^\vee}{\sP_{i,x}}
 \frac{1}{\sY_{i,q_1 x} \sY_{i,x}^{\vee}}
 \prod_{e:i \to j} \sY_{j,\mu_e^{-1} q_1 x}
 \prod_{e:j \to i} \sY_{j,\mu_e x}^\vee
 \Bigg|_{\CalX_*,x = x_{i,\alpha,k}^+}
 \label{eq:saddle_full2+}
 \\
 1 & =
 - \fq_i^{-1} \, (q_1^{-1} x)^{-\kappa_i} \,
 \frac{\sP_{i,q_1^{-1} x}}{\widetilde{\sP}_{i,q_1^{-1} x}^\vee}
 \sY_{i,x}\sY_{i,q_1^{-1}x}^{\vee}
 \prod_{e:i \to j} \sY_{j,\mu_e^{-1} x}^{-1}
 \prod_{e:j \to i} \sY_{j,\mu_e q_1^{-1} x}^{\vee-1}
 \Bigg|_{\CalX_*,x = x_{i,\alpha,k}^-}
 \label{eq:saddle_full2-}
\end{align}

Let us then discuss how the gauge theory vacuum determined by the saddle point equation is related to the Bethe equation.
It has been shown that the saddle point equation of the $\Gamma$-quiver gauge theory turns out to be the Bethe equation of $\Gamma$-symmetric spin chain~\cite{Nekrasov:2013xda}.
In the present case with supergroup gauge symmetry, we obtain a generalization of the Bethe equation \eqref{eq:BE2}.

The $\sY$-function with generic $(q_1,q_2)$ defined in Sec.~\ref{sec:Y-func} is rewritten in terms of $\sQ^\pm$-functions having zeros at $x \in \CalX_i^\sigma$,
\begin{align}
 \sQ_{i,x}^\sigma = \prod_{x' \in \CalX_i^\sigma}
 \left( 1 - \frac{x}{x'} \right)
 \, , \qquad
 \sQ_{i,x}^{\sigma\vee} = \prod_{x' \in \CalX_i^\sigma}
 \left( 1 - \frac{x'}{x} \right)
 \, ,
\end{align}
as follows:
\begin{align}
 \sY_{i,x} = \frac{\sQ^+_{i,x} \, \sQ^-_{i,q_2^{-1} x}}{\sQ^+_{i,q_1^{-1}x} \sQ^-_{i,q^{-1} x}}
 \, , \qquad
 \sY_{i,x}^\vee = \frac{\sQ^{+\vee}_{i,x} \, \sQ^{-\vee}_{i,q_2^{-1} x}}{\sQ^{+\vee}_{i,q_1^{-1}x} \sQ^{-\vee}_{i,q^{-1} x}}
 \, ,
\end{align}
where we shift the $x^-$-variables as $x \to q x$ for $x \in \CalX_i^-$ for convenience.
See also the argument around \eqref{eq:x-shift}.
In the NS limit $q_2 \to 1$, the $\sY$-function is reduced to
\begin{align}
 \sY_{i,x} = \frac{\sQ_{i,x}}{\sQ_{i,q_1^{-1}x}}
 \, , \qquad
 \sY_{i,x}^\vee = \frac{\sQ^\vee_{i,x}}{\sQ^\vee_{i,q_1^{-1}x}}
 \, ,
\end{align}
where the total $\sQ$-function is defined as the product of $\sQ^{\pm}$-functions,
\begin{align}
 \sQ_{i,x} = \sQ_{i,x}^+ \sQ_{i,x}^-
 \, , \qquad
 \sQ_{i,x}^\vee = \sQ_{i,x}^{+\vee} \sQ_{i,x}^{-\vee}
 \, .
\end{align}
Then from the saddle point equation \eqref{eq:saddle_full2+}, we obtain
\begin{align}
 \frac{\sP_{i,x}}{\widetilde{\sP}_{i,x}^\vee}
 = \fq_i \, x^{\kappa_i} \,
 \frac{\sQ_{i,q_1^{-1}x}^\vee}{\sQ_{i,q_1 x}}
 \prod_{e:i \to j}
 \frac{\sQ_{i,\mu_e^{-1} q_1^{1/2}x}}{\sQ_{i,\mu_e^{-1} q_1^{-1/2}x}}
 \prod_{e:j \to i}
 \frac{\sQ_{i,\mu_e q_1^{1/2}x}^\vee}{\sQ_{i,\mu_e q_1^{-1/2}x}^\vee}
 \qquad \text{for} \quad x \in \CalX_i^+
\end{align}
where we shift the bifundamental mass parameters $\mu_e \to \mu_e q_1^{1/2}$ to obtain a more symmetric expression.
For the conformal case, we can rewrite the $\sQ$-function and the $\sP$-function in terms of the $[x]$-function \eqref{eq:[x]}, namely the Dirac index convention, so that it is formally reduced to the Bethe equation~\eqref{eq:BE2} by identifying $\epsilon_1 = \hbar$ ($q_1 = e^\hbar$).
In this correspondence, the $a_i$ and $d_i$ functions are given by the matter functions, $\sP_{i,x}$ and $\widetilde{\sP}_{i,x}$, the twist parameters are given by the gauge coupling constants $(\fq_i)$, and the Bethe roots correspond to the instanton configuration $x \in \CalX$.
We remark that the naive saddle point equation of gauge theory involves infinitely many $x$-variables, but it can be truncated at the root of Higgs branch by tuning the Coulomb moduli with the fundamental mass parameters~\cite{Dorey:2011pa,Chen:2011sj}.

\subsubsection{Generalization: a coupled system}

A crucial difference between the ordinary case and the supergroup case is that, in the latter case, 
the $\sP$-function \eqref{eq:P_ratio} is not a polynomial, but a ratio of polynomials, namely a rational function.
The gauge theory analysis suggests the following Bethe equation:
\begin{subequations}\label{eq:BEs}
\begin{align}
 \frac{a_i^+(x)}{a_i^-(x)}
 \frac{d_i^-(x)}{d_i^+(x)} 
 & = \fq_i \, \prod_{j=1}^{\rk G}
 \frac{Q_j^+(x - c_{ij}\hbar/2)}{Q_j^+(x + c_{ij}\hbar/2)}
 \frac{Q_j^-(x - c_{ij}\hbar/2)}{Q_j^-(x + c_{ij}\hbar/2)}
 \qquad \text{for} \quad x \in \CalX_i^+ \sqcup \CalX_i^-
\end{align}
\end{subequations}
where we define
\begin{align}
 Q_i^\sigma(x) = \prod_{k=1}^{N_{i,\sigma}} [x - x_{i,k}^\sigma]
\end{align}
and
\begin{subequations} 
 \begin{align}
 a_i^+(x) = \prod_{a=1}^{L_{i,+}} [x - \xi_{i,a}^+ + s_{i,a}^+ \hbar]
 \, , \qquad
 d_i^+(x) = \prod_{a=1}^{L_{i,+}} [x - \xi_{i,a}^+ - s_{i,a}^+ \hbar]
 \, ,\\
 a_i^-(x) = \prod_{a=1}^{L_{i,-}} [x - \xi_{i,a}^- - s_{i,a}^- \hbar]
 \, , \qquad
 d_i^-(x) = \prod_{a=1}^{L_{i,-}} [x - \xi_{i,a}^- + s_{i,a}^- \hbar]
 \, .  
 \end{align}
\end{subequations} 
Introducing the total $Q$-function
\begin{align}
 Q_i(x) = Q_i^+(x) Q_i^-(x)
\end{align}
and rational functions,
\begin{align}
 a_i(x) = \frac{a_i^+(x)}{a_i^-(x)}
 \, , \qquad
 d_i(x) = \frac{d_i^+(x)}{d_i^-(x)}
 \, ,
 \label{eq:BE_meros}
\end{align}
the Bethe equation \eqref{eq:BEs} formally agrees with the ordinary expression \eqref{eq:BE2}.
In this case, however, there are inhomogeneous parameters both in the numerator and the denominator of $a_i(x)$ and $d_i(x)$.
This could be interpreted as analytic continuation $L_{i,-} \to - L_{i,-}$ of the spin chain with the length $L_{i,+} + L_{i,-}$:
The {\em positive} inhomogeneous parameters are assigned to $L_{i,+}$ sites, while the {\em negative} ones are for $L_{i,-}$ sites, which would be thought of as the spectral dual of the situation studied in~\cite{Orlando:2010uu,Nekrasov:2018gne,Zenkevich:2018fzl}.
See also the argument in Sec.~\ref{sec:QIS}.

Another interpretation is as follows:
Define the polynomial functions,
\begin{align}
 \tilde{a}_i(x) = a_i^+(x) d_i^-(x)
 \, , \qquad
 \tilde{d}_i(x) = d_i^+(x) a_i^- (x)
 \, .
\end{align}
Then the Bethe equation \eqref{eq:BEs} itself coincides with the ordinary one \eqref{eq:BE2} where the magnon number is $N_{i,+} + N_{i,-}$, and the spin chain length is $L_i + L_i$.
However, it would be different from the ordinary quantum spin system since the underlying $T$-function is a rational function as mentioned in Sec.~\ref{sec:TQ-rel}.

\subsection{TQ-relation}\label{sec:TQ-rel}

In Sec.~\ref{sec:qq-ch}, we have discussed the doubly quantum Seiberg--Witten geometry, where the $qq$-character plays a key role.
Such a two-parameter deformation of the character is then reduced to the so-called $q$-character~\cite{Knight:1995JA,Frenkel:1998}, which is a one-parameter deformation, in the NS limit $\epsilon_2 \to 0$.
The $q$-character gives rise to the transfer matrix of the associated quantum integrable system, and obeys functional relations, e.g., T-system, TQ-relation, etc.
In particular, the TQ-relation is naturally interpreted as a quantization of the algebraic curve characterizing the Seiberg--Witten geometry.
We study the role of these with the supergroup gauge theory.

\subsubsection{U$(n|n)$ pure Yang--Mills theory}\label{sec:TQ-rel_YM}

Let us focus on $A_1$ quiver with U$(n|n)$ gauge node, namely U$(n|n)$ pure Yang--Mills theory, for simplicity.
In this case, there is no distinction between the $\sY$-function and its dual $\sY^\vee$ as shown in \eqref{eq:Y_conv}.
Thus the $qq$-character \eqref{eq:qq-ch_A1} is reduced to the fundamental $q$-character of $A_1$ quiver:
\begin{align}
 \Big< \sY_{1,x} \Big> + \Big< \frac{\fq_1}{\sY_{1,q^{-1}x}} \Big>
 = T_{1,x}^{n|n}
 \, .
\end{align}
Since, in the NS limit, the average is replaced by that evaluated with the critical configuration, we have
\begin{align}
 \sY_{1,x}[\CalX_*] + \frac{\fq_1}{\sY_{1,q_1^{-1}x}[\CalX_*]}
 = T_{1,x}^{n|n}
 \, .
\end{align}
In terms of the $\sQ$-function, it is equivalent to
\begin{align}
 \sQ_{1,q_1 x} + \fq_1 \sQ_{1,q_1^{-1},x} = T_{1,x}^{n|n} \, \sQ_{1,x}
 \label{eq:TQ-rel1}
\end{align}
where the $\sQ$-function is accordingly evaluated with the critical configuration $\CalX_*$.
This is the TQ-relation for the relativistic Toda chain, but again $T$-function is not a polynomial, but a rational function.
We remark that, since the $\sQ$-function is an entire function, the LHS of \eqref{eq:TQ-rel1} is an entire function.
Hence the RHS should be also an entire function although $T_{1,x}^{n|n}$ is a rational function.

Furthermore, the TQ-relation \eqref{eq:TQ-rel1} is rephrased in the operator formalism
\begin{align}
 H(\hat{x},\hat{y}) \, \sQ_{1,x} = 0
\end{align}
where $H(x,y)$ is the two variable algebraic function used to define the Seiberg--Witten curve \eqref{eq:SW-curve_cl}.
$(\hat{x},\hat{y})$ is the operator pair
\begin{align}
 \hat{x} = x
 \, , \qquad
 \hat{y} = \exp \left( \epsilon_1 \frac{\partial}{\partial \log x} \right)
 \, .
\end{align}
which is obtained through the canonical quantization with respect to the symplectic two-form \eqref{eq:symp_2form}
\begin{align}
 \left[ \log \hat{x}, \, \log \hat{y} \right] = - \epsilon_1
 \, .
\end{align}
In this sense, the TQ-relation is interpreted as quantization of the Seiberg--Witten curve, defined as the kernel of the operator $H(\hat{x},\hat{y})$ in the Hilbert space, while the classical curve is defined as the zero locus of the algebraic function $H(x,y)$, which is actually the Lagrangian submanifold of $T^* \Sigma$.

\subsubsection{U$(n|n)$ SQCD}

In the presence of the fundamental hypermultiplets, the TQ-relation is promoted to
\begin{align}
 a_1(x) \sQ_{1,q_1 x} + \fq_1 d_1(x) \sQ_{1,q_1^{-1},x} = T_{1,x}^{n|n} \, \sQ_{1,x}
\end{align}
which reproduces the TQ-relation for SU(2)-spin chain, where the rational functions, $a_1(x)$ and $d_1(x)$, are given by the matter functions $\sP_{1,x}$ and $\widetilde{\sP}_{1,x}$.
This TQ-relation is rephrased as
\begin{align}
 a_1^+(x) d_1^-(x) \sQ_{1,q_1 x} + \fq_1 a_1^-(x) d_1^+(x) \sQ_{1,q_1^{-1},x} = a_1^-(x) d_1^-(x) T_{1,x}^{n|n} \, \sQ_{1,x}
 \, .
\end{align}
Then, as in the previous case \eqref{eq:TQ-rel1}, this is a functional relation on entire functions although the $T$-function is a rational function.

\section{Brane construction}\label{sec:brane}

Following the approach by Dijkgraaf et al.~\cite{Dijkgraaf:2016lym}, we consider the brane construction of the supergroup gauge theory, in particular, of the linear quiver type.
First of all, 4d $\CalN=2$ SU$(n_+|n_-)$ Yang--Mills theory is realized as world-volume theory of positive and negative branes, D4$^+$ and D4$^-$, suspended between two separated NS5 brane:
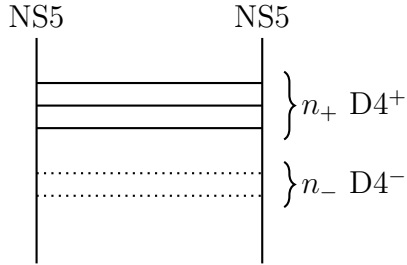
\begin{center}
  \begin{tikzpicture}[thick,scale=1.5]

   \draw (-1,-1) -- ++(0,2) node [above] {NS5};
   \draw (1,-1) -- ++(0,2) node [above] {NS5};

   \draw (-1,.2) -- ++(2,0);
   \draw (-1,.4) -- ++(2,0);
   \draw (-1,.6) -- ++(2,0);

   \draw [dotted] (-1,-.2) -- ++(2,0);
   \draw [dotted] (-1,-.4) -- ++(2,0);

   \draw [decorate,decoration={brace,amplitude=4pt,mirror,raise=4pt},yshift=0pt] (1.1,.1) -- ++(0,.6) node [black,midway,xshift=2.6em] {$n_+$ D4$^+$};

   \draw [decorate,decoration={brace,amplitude=4pt,mirror,raise=4pt},yshift=0pt] (1.1,-.5) -- ++(0,.4) node [black,midway,xshift=2.6em] {$n_-$ D4$^-$};      
   
  \end{tikzpicture}  
\end{center}
where D4$^+$ and D4$^-$ branes are depicted as horizontal solid and dotted lines.
It has been pointed out in~\cite{Dijkgraaf:2016lym} that the negative branes are removed through gauging process:
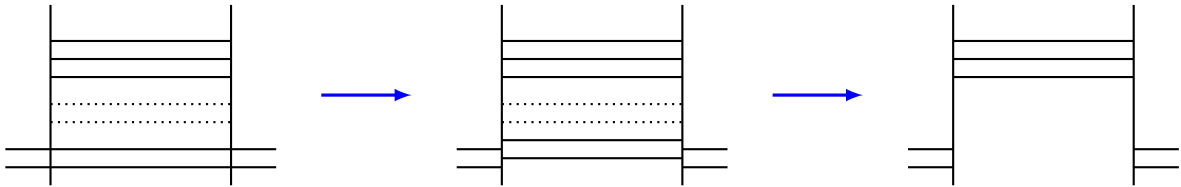
\begin{center}
  \begin{tikzpicture}[thick,scale=1.2]

   \draw (-1,-1) -- ++(0,2) node [above] {};
   \draw (1,-1) -- ++(0,2) node [above] {};

   \draw (-1,.2) -- ++(2,0);
   \draw (-1,.4) -- ++(2,0);
   \draw (-1,.6) -- ++(2,0);

   \draw [dotted] (-1,-.1) -- ++(2,0);
   \draw [dotted] (-1,-.3) -- ++(2,0);

   \draw (-1.5,-.6) -- ++(3,0);
   \draw (-1.5,-.8) -- ++(3,0);

   \draw[-latex,blue,very thick] (2,0) -- ++(.5,0) node [above] {} -- ++(.5,0);

   \begin{scope}[shift={(5,0)}]

   \draw (-1,-1) -- ++(0,2) node [above] {};
   \draw (1,-1) -- ++(0,2) node [above] {};

   \draw (-1,.2) -- ++(2,0);
   \draw (-1,.4) -- ++(2,0);
   \draw (-1,.6) -- ++(2,0);

   \draw [dotted] (-1,-.1) -- ++(2,0);
   \draw [dotted] (-1,-.3) -- ++(2,0);

    \draw (-1.5,-.6) -- ++(.5,0);
    \draw (-1.5,-.8) -- ++(.5,0);
    \draw (1,-.6) -- ++(.5,0);
    \draw (1,-.8) -- ++(.5,0);       

    \draw (-1,-.5) -- ++(2,0);
    \draw (-1,-.7) -- ++(2,0);    
    
    \draw[-latex,blue,very thick] (2,0) -- ++(.5,0) node [above] {} -- ++(.5,0);
    
   \end{scope}

   \begin{scope}[shift={(10,0)}]

   \draw (-1,-1) -- ++(0,2) node [above] {};
   \draw (1,-1) -- ++(0,2) node [above] {};

   \draw (-1,.2) -- ++(2,0);
   \draw (-1,.4) -- ++(2,0);
   \draw (-1,.6) -- ++(2,0);

    
    \draw (-1.5,-.6) -- ++(.5,0);
    \draw (-1.5,-.8) -- ++(.5,0);
    \draw (1,-.6) -- ++(.5,0);
    \draw (1,-.8) -- ++(.5,0);       
    
   \end{scope}   
   
  \end{tikzpicture}  
\end{center}
and the resulting configuration is equivalent to 4d SU$(n_+)$ theory with $n_\text{F} = 2 n_-$ flavors.
More precisely, there are $n_-$ fundamental and $n_-$ anti-fundamental hypermultiplets having the same masses because we imposed {\em horizontal} D4$^+$ branes before gauging.

\subsection{$A_r$ quiver}

This argument is easily generalized to linear quiver theory.
The brane configuration of $A_r$ quiver with gauge group SU$(n_+|n_-)$ is shown as follows:
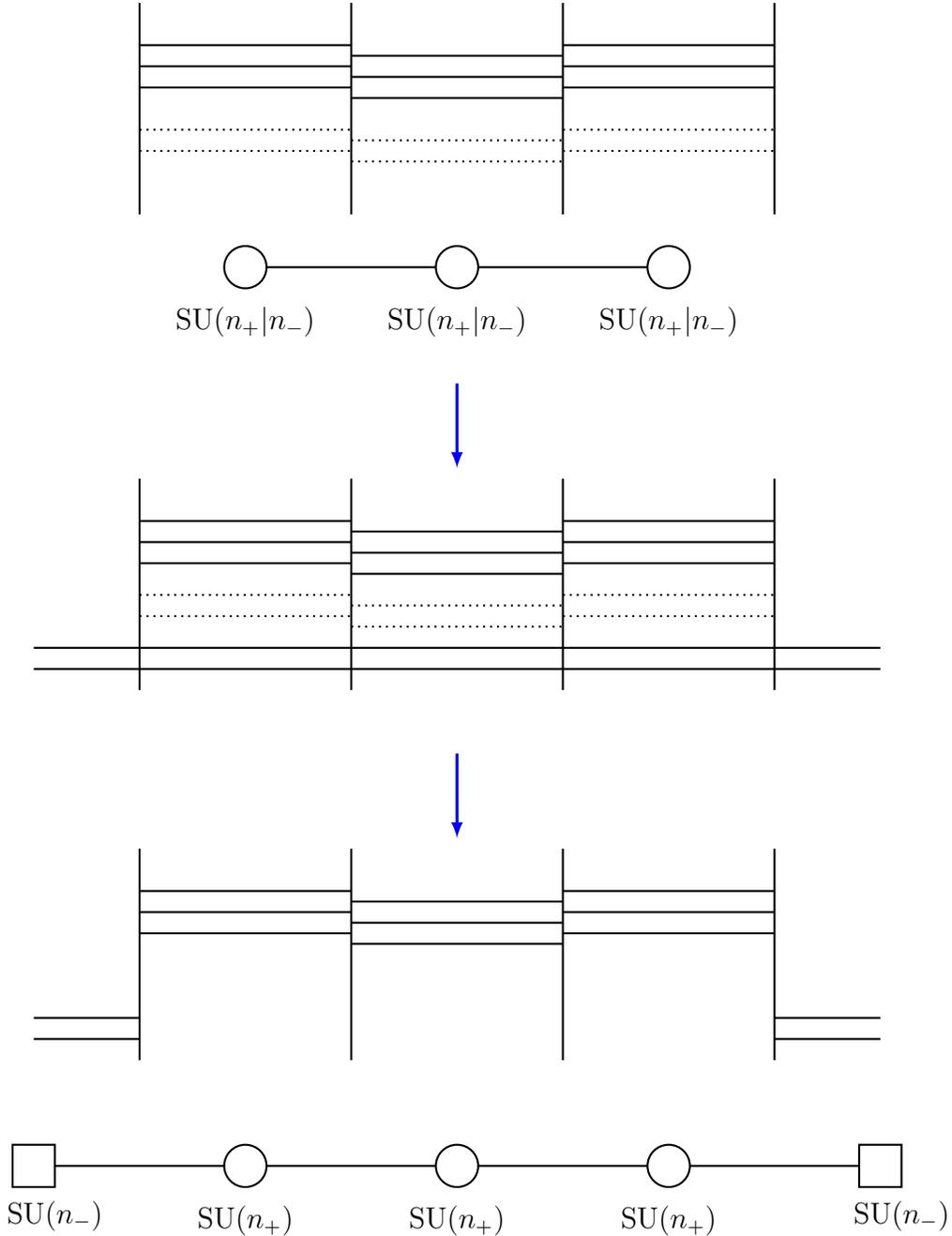
\begin{center}
  \begin{tikzpicture}[thick,scale=1.5]

   \draw (-1,-1) -- ++(0,2);
   \draw (1,-1) -- ++(0,2);
   \draw (3,-1) -- ++(0,2);
   \draw (5,-1) -- ++(0,2);

   \draw (-1,.2) -- ++(2,0);
   \draw (-1,.4) -- ++(2,0);
   \draw (-1,.6) -- ++(2,0);

   \draw (1,.1) -- ++(2,0);
   \draw (1,.3) -- ++(2,0);
   \draw (1,.5) -- ++(2,0);

   \draw (3,.2) -- ++(2,0);
   \draw (3,.4) -- ++(2,0);
   \draw (3,.6) -- ++(2,0);
   
   \draw [dotted] (-1,-.2) -- ++(2,0);
   \draw [dotted] (-1,-.4) -- ++(2,0);

   \draw [dotted] (1,-.3) -- ++(2,0);
   \draw [dotted] (1,-.5) -- ++(2,0);

   \draw [dotted] (3,-.2) -- ++(2,0);
   \draw [dotted] (3,-.4) -- ++(2,0);

    \begin{scope}[shift={(0,-1.5)}]

    \draw (0,0) -- (4,0);

    \filldraw[fill=white,draw=black] (0,0) circle (.2) node [below=1em] {SU$(n_+|n_-)$};
    \filldraw[fill=white,draw=black] (2,0) circle (.2) node [below=1em] {SU$(n_+|n_-)$};;
    \filldraw[fill=white,draw=black] (4,0) circle (.2) node [below=1em] {SU$(n_+|n_-)$};;

   \draw [-latex,blue,very thick] (2,-1.1) -- ++(0,-.8);
     
    \end{scope}
   
   \begin{scope}[shift={(0,-4.5)}]

   \draw (-1,-1) -- ++(0,2);
   \draw (1,-1) -- ++(0,2);
   \draw (3,-1) -- ++(0,2);
   \draw (5,-1) -- ++(0,2);

   \draw (-1,.2) -- ++(2,0);
   \draw (-1,.4) -- ++(2,0);
   \draw (-1,.6) -- ++(2,0);

   \draw (1,.1) -- ++(2,0);
   \draw (1,.3) -- ++(2,0);
   \draw (1,.5) -- ++(2,0);

   \draw (3,.2) -- ++(2,0);
   \draw (3,.4) -- ++(2,0);
   \draw (3,.6) -- ++(2,0);
   
   \draw [dotted] (-1,-.1) -- ++(2,0);
   \draw [dotted] (-1,-.3) -- ++(2,0);

   \draw [dotted] (1,-.2) -- ++(2,0);
   \draw [dotted] (1,-.4) -- ++(2,0);

   \draw [dotted] (3,-.1) -- ++(2,0);
    \draw [dotted] (3,-.3) -- ++(2,0);

   \draw (-2,-.6) -- ++(8,0);
    \draw (-2,-.8) -- ++(8,0);

   \draw [-latex,blue,very thick] (2,-1.6) -- ++(0,-.8);    
    
   \end{scope}

   \begin{scope}[shift={(0,-8)}]

   \draw (-1,-1) -- ++(0,2);
   \draw (1,-1) -- ++(0,2);
   \draw (3,-1) -- ++(0,2);
   \draw (5,-1) -- ++(0,2);

   \draw (-1,.2) -- ++(2,0);
   \draw (-1,.4) -- ++(2,0);
   \draw (-1,.6) -- ++(2,0);

   \draw (1,.1) -- ++(2,0);
   \draw (1,.3) -- ++(2,0);
   \draw (1,.5) -- ++(2,0);

   \draw (3,.2) -- ++(2,0);
   \draw (3,.4) -- ++(2,0);
   \draw (3,.6) -- ++(2,0);
   
%
%

   \draw (-2,-.6) -- ++(1,0);
    \draw (-2,-.8) -- ++(1,0);

   \draw (5,-.6) -- ++(1,0);
   \draw (5,-.8) -- ++(1,0);    
    
   \end{scope}   

   \begin{scope}[shift={(0,-10)}]

    \draw (-2,0) -- (6,0);

    \filldraw[fill=white,draw=black] (0,0) circle (.2) node [below=1em] {SU$(n_+)$};
    \filldraw[fill=white,draw=black] (2,0) circle (.2) node [below=1em] {SU$(n_+)$};;
    \filldraw[fill=white,draw=black] (4,0) circle (.2) node [below=1em] {SU$(n_+)$};;

    \filldraw[fill=white,draw=black] (-2.2,-.2) rectangle ++(.4,.4) node [below=1.55em] {SU$(n_-)$};;
    \filldraw[fill=white,draw=black] (5.8,-.2) rectangle ++(.4,.4) node [below=1.55em] {SU$(n_-)$};;

   \end{scope}
   
  \end{tikzpicture}
\end{center}
For the moment, we assign the same super gauge group to all the gauge node for simplicity.
The negative branes are annihilated by gauging, and the resulting theory is the linear quiver theory with gauge groups SU$(n_+)$ and flavor nodes SU$(n_-)$ attached to the left and right most nodes.
We can consider the situation with different super gauge groups assigned to each node.
However, in such a case, it's not possible to annihilate all the negative branes at the same time in general.
See also Sec.~\ref{sec:decoupling}.

\subsection{$D_r$ quiver}\label{sec:brane_D}

The procedure discussed above has a natural generalization to $D_r$ and $\widehat{D}_r$ quivers, by introducing the ON$^0$ plane to the configuration~\cite{Kapustin:1998fa,Hanany:1999sj,Hayashi:2015vhy}.
For $D_r$ quiver, the brane configuration is given as follows:
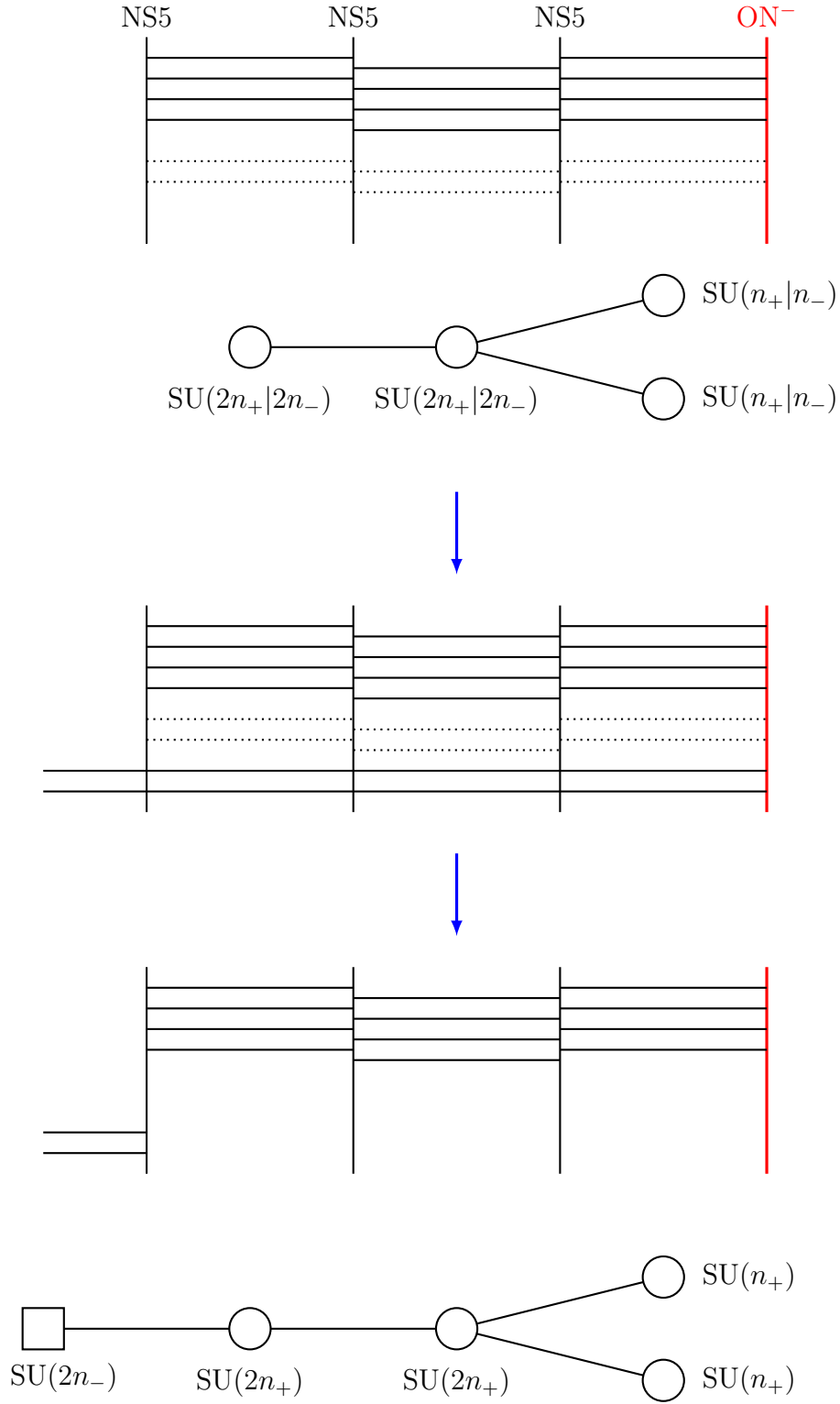
\begin{center}
  \begin{tikzpicture}[thick,scale=1.5]

   \draw (-1,-1) -- ++(0,2) node [above] {NS5};
   \draw (1,-1) -- ++(0,2) node [above] {NS5};
   \draw (3,-1) -- ++(0,2) node [above] {NS5};
   \draw[red, very thick] (5,-1) -- ++(0,2) node [above] {ON$^-$};

   \draw (-1,.2) -- ++(2,0);
   \draw (-1,.4) -- ++(2,0);
   \draw (-1,.6) -- ++(2,0);
   \draw (-1,.8) -- ++(2,0);   

   \draw (1,.1) -- ++(2,0);
   \draw (1,.3) -- ++(2,0);
   \draw (1,.5) -- ++(2,0);
   \draw (1,.7) -- ++(2,0);   

   \draw (3,.2) -- ++(2,0);
   \draw (3,.4) -- ++(2,0);
   \draw (3,.6) -- ++(2,0);
   \draw (3,.8) -- ++(2,0);   
   
   \draw [dotted] (-1,-.2) -- ++(2,0);
   \draw [dotted] (-1,-.4) -- ++(2,0);

   \draw [dotted] (1,-.3) -- ++(2,0);
   \draw [dotted] (1,-.5) -- ++(2,0);

   \draw [dotted] (3,-.2) -- ++(2,0);
   \draw [dotted] (3,-.4) -- ++(2,0);

    \begin{scope}[shift={(0,-2)}]

     \draw (0,0) -- (2,0);
     \draw (4,.5) -- (2,0) -- (4,-.5);

    \filldraw[fill=white,draw=black] (0,0) circle (.2) node [below=1em] {SU$(2n_+|2n_-)$};
    \filldraw[fill=white,draw=black] (2,0) circle (.2) node [below=1em] {SU$(2n_+|2n_-)$};;
     \filldraw[fill=white,draw=black] (4,.5) circle (.2) node [right=1em] {SU$(n_+|n_-)$};;
    \filldraw[fill=white,draw=black] (4,-.5) circle (.2) node [right=1em] {SU$(n_+|n_-)$};;

   \draw [-latex,blue,very thick] (2,-1.4) -- ++(0,-.8);
     
    \end{scope}
   
   \begin{scope}[shift={(0,-5.5)}]

   \draw (-1,-1) -- ++(0,2);
   \draw (1,-1) -- ++(0,2);
   \draw (3,-1) -- ++(0,2);
   \draw[red,very thick] (5,-1) -- ++(0,2);

   \draw (-1,.2) -- ++(2,0);
   \draw (-1,.4) -- ++(2,0);
   \draw (-1,.6) -- ++(2,0);
   \draw (-1,.8) -- ++(2,0);    

   \draw (1,.1) -- ++(2,0);
   \draw (1,.3) -- ++(2,0);
   \draw (1,.5) -- ++(2,0);
   \draw (1,.7) -- ++(2,0);    

   \draw (3,.2) -- ++(2,0);
   \draw (3,.4) -- ++(2,0);
   \draw (3,.6) -- ++(2,0);
   \draw (3,.8) -- ++(2,0);
    
   \draw [dotted] (-1,-.1) -- ++(2,0);
   \draw [dotted] (-1,-.3) -- ++(2,0);

   \draw [dotted] (1,-.2) -- ++(2,0);
   \draw [dotted] (1,-.4) -- ++(2,0);

   \draw [dotted] (3,-.1) -- ++(2,0);
    \draw [dotted] (3,-.3) -- ++(2,0);

   \draw (-2,-.6) -- ++(7,0);
    \draw (-2,-.8) -- ++(7,0);

   \draw [-latex,blue,very thick] (2,-1.4) -- ++(0,-.8);    
    
   \end{scope}

   \begin{scope}[shift={(0,-9)}]

   \draw (-1,-1) -- ++(0,2);
   \draw (1,-1) -- ++(0,2);
   \draw (3,-1) -- ++(0,2);
   \draw[red, very thick] (5,-1) -- ++(0,2);

   \draw (-1,.2) -- ++(2,0);
   \draw (-1,.4) -- ++(2,0);
   \draw (-1,.6) -- ++(2,0);
   \draw (-1,.8) -- ++(2,0);
    
   \draw (1,.1) -- ++(2,0);
   \draw (1,.3) -- ++(2,0);
   \draw (1,.5) -- ++(2,0);
   \draw (1,.7) -- ++(2,0);
    
   \draw (3,.2) -- ++(2,0);
   \draw (3,.4) -- ++(2,0);
   \draw (3,.6) -- ++(2,0);
   \draw (3,.8) -- ++(2,0);
    
%
%

   \draw (-2,-.6) -- ++(1,0);
    \draw (-2,-.8) -- ++(1,0);

    
   \end{scope}   

   \begin{scope}[shift={(0,-11.5)}]

     \draw (-2,0) -- (2,0);
     \draw (4,.5) -- (2,0) -- (4,-.5);

    \filldraw[fill=white,draw=black] (0,0) circle (.2) node [below=1em] {SU$(2n_+)$};
    \filldraw[fill=white,draw=black] (2,0) circle (.2) node [below=1em] {SU$(2n_+)$};
    \filldraw[fill=white,draw=black] (4,.5) circle (.2) node [right=1em] {SU$(n_+)$};
    \filldraw[fill=white,draw=black] (4,-.5) circle (.2) node [right=1em] {SU$(n_+)$};    

    \filldraw[fill=white,draw=black] (-2.2,-.2) rectangle ++(.4,.4) node [below=1.55em] {SU$(2n_-)$};;

   \end{scope}
   
  \end{tikzpicture}
\end{center}
Thus the configuration is reduced to that for $D_r$ quiver with non-supergroup gauge nodes.
Such a reduction is consistent with another approach discussed in Sec.~\ref{sec:decoupling_D}.
We can similarly deal with $\widehat{D}_r$ quiver by imposing ON$^0$ planes on the both ends of the configuration.

\section{Decoupling trick}\label{sec:decoupling}

\subsection{Vector multiplet}

The SU$(n_+|n_-)$ vector multiplet consists of SU$(n_+)$ and SU$(n_-)$ vectors and two bifundamentals of $\mathrm{SU}(n_+) \times \mathrm{SU}(n_-)$ and $\mathrm{SU}(n_-) \times \mathrm{SU}(n_+)$, where two gauge couplings, $\fq_+$ and $\fq_-$, should obey~\cite{Dijkgraaf:2016lym}:
\begin{align}
 \fQ := \fq_+ \fq_- = 1
 \, .
 \label{eq:super_cond}
\end{align}
For example, $A_1$ quiver with super gauge group is equivalent to $\widehat{A}_1$ quiver, and the Seiberg--Witten geometry exhibits the modular property with the gauge coupling product $\fQ$~\cite{Nekrasov:2012xe}.
In this sense, the super group condition $\fQ \to 1$ seems singular because it's a boundary of the convergence radius.
Instead of such a singular limit, we alternatively consider the condition $\fQ \to 0$ using the modular transformation, which implies decoupling the SU$(n_-)$ vector multiplet: $\fq_- = 0$.%
\footnote{%
Precisely speaking, there are two possibilities: $\fq_+ = 0$ or $\fq_- = 0$.
We assume $\fq_+ \neq 0$ since $\fq_+$ would be interpreted as the physical gauge coupling.}
In this limit, $\widehat{A}_1$ quiver is reduced to $A_1$ quiver with two SU$(n_-)$ flavor nodes:
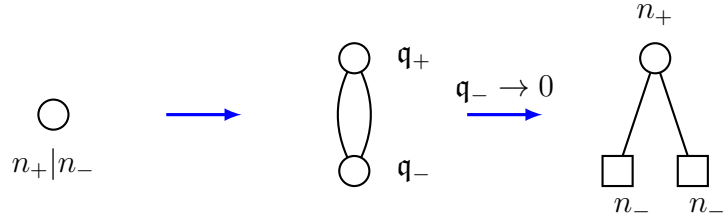
\begin{center}
  \begin{tikzpicture}[thick]

   \filldraw[fill=white,draw=black] (0,0) circle (.2) node [below=.7em] {$n_+|n_-$};

   \draw[-latex,very thick,blue] (1.5,0) -- ++(1,0);

     \begin{scope}[shift={(4,0)}]

      \draw (0,.75) to [bend right] (0,-.75);
      \draw (0,.75) to [bend left] (0,-.75);      

     \filldraw[fill=white,draw=black] (0,.75) circle (.2) node [right=1em] {$\fq_+$}; 
     \filldraw[fill=white,draw=black] (0,-.75) circle (.2) node [right=1em] {$\fq_-$}; 

      \draw[-latex,very thick,blue] (1.5,0) -- ++(.5,0) node [above] {\textcolor{black}{$\fq_- \to 0$}} -- ++(.5,0);
    
     \end{scope}

   \begin{scope}[shift={(8,0)}]

    \draw (0,.75) -- ++(-.5,-1.5) node (l) {};
    \draw (0,.75) -- ++(.5,-1.5) node (r) {};    
        
    \filldraw[fill=white,draw=black] (0,.75) circle (.2) node [above=.7em] {$n_+$};
    \filldraw[fill=white,draw=black] (l)++(-.2,-.2) rectangle ++(.4,.4) node [below=1em] {$n_-$};
    \filldraw[fill=white,draw=black] (r)++(-.2,-.2) rectangle ++(.4,.4) node [below=1em] {$n_-$};

   \end{scope}   
  \end{tikzpicture}  
\end{center}
This is consistent with the previous argument in Sec.~\ref{sec:brane}.

\subsection{Bifundamental hypermultiplet}

In order to apply this argument to generic quiver, we consider the hypermultiplet in the bifundamental representation of $\mathrm{SU}(n_{i,+}|n_{i,-}) \times \mathrm{SU}(n_{j,+}|n_{j,-})$.
We consider $A_2$ quiver as follows:
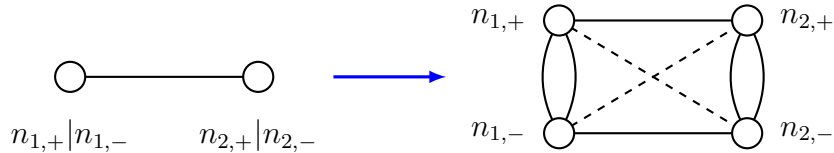
\begin{center}
 \begin{tikzpicture}[thick]

  \draw (0,0) -- (2.5,0);
  \filldraw[fill=white,draw=black] (0,0) circle (.2) node [below=1em] {$n_{1,+}|n_{1,-}$};
  \filldraw[fill=white,draw=black] (2.5,0) circle (.2) node [below=1em] {$n_{2,+}|n_{2,-}$};  

  \draw [blue,-latex, very thick] (3.5,0) -- ++(1.5,0);
  
  \begin{scope}[shift={(6.5,0)}]

   \draw (0,.75) node (i1) {} to [bend right] (0,-.75) node (i2) {};
   \draw (0,.75) to [bend left] (0,-.75);
   
   \draw (2.5,.75) node (j1) {} to [bend right] (2.5,-.75) node (j2) {};
   \draw (2.5,.75) to [bend left] (2.5,-.75);

   \draw (i1) -- ++(2.5,0);
   \draw (0,-.75) -- ++(2.5,0);

   \draw [dashed] (i1) -- (j2);
   \draw [dashed] (i2) -- (j1);   
   
     \filldraw[fill=white,draw=black] (0,.75) circle (.2) node [left=.7em] {$n_{1,+}$};
     \filldraw[fill=white,draw=black] (0,-.75) circle (.2) node [left=.7em] {$n_{1,-}$};

   \filldraw[fill=white,draw=black] (2.5,.75) circle (.2) node [right=.7em] {$n_{2,+}$};
   \filldraw[fill=white,draw=black] (2.5,-.75) circle (.2) node [right=.7em] {$n_{2,-}$};

  \end{scope}

 \end{tikzpicture}
\end{center}
where the solid lines are the {\em positive} bifundamental hypermultiplets, while the dashed lines are the {\em negative} (vector-like) bifundamental multiplets.
Then, turning off the gauge couplings, $\fq_{1,-}$ and $\fq_{2,-}$, with the assumption $n_{1,-} = n_{2,-} =: n_-$, and all the mass parameters (Coulomb moduli for the negative nodes) coincide with each other, it becomes
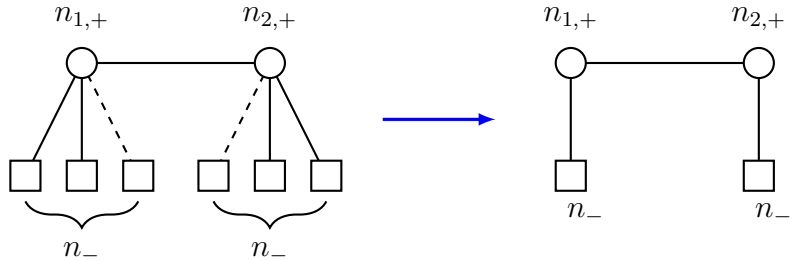
\begin{center}
 \begin{tikzpicture}[thick]

  \draw (0,0) node (11) {} --++(2.5,0) node (21) {};

  \draw (11)--++(-.75,-1.5) node (12a) {};
  \draw (11)--++(0,-1.5) node (12b) {};
  \draw[dashed] (11)--++(.75,-1.5) node (12c) {};

  \draw[dashed] (21)--++(-.75,-1.5) node (22a) {};
  \draw (21)--++(0,-1.5) node (22b) {};
  \draw (21)--++(.75,-1.5) node (22c) {};

  \filldraw[fill=white,draw=black] (11) circle (.2) node [above=.7em] {$n_{1,+}$};
  \filldraw[fill=white,draw=black] (21) circle (.2) node [above=.7em] {$n_{2,+}$};    

  \filldraw[fill=white,draw=black] (12a)++(-.2,-.2) rectangle ++(.4,.4);
  \filldraw[fill=white,draw=black] (12b)++(-.2,-.2) rectangle ++(.4,.4);
  \filldraw[fill=white,draw=black] (12c)++(-.2,-.2) rectangle ++(.4,.4);

  \filldraw[fill=white,draw=black] (22a)++(-.2,-.2) rectangle ++(.4,.4);
  \filldraw[fill=white,draw=black] (22b)++(-.2,-.2) rectangle ++(.4,.4);
  \filldraw[fill=white,draw=black] (22c)++(-.2,-.2) rectangle ++(.4,.4); 

  \draw [decorate,decoration={brace,amplitude=10pt,mirror,raise=4pt},yshift=0pt]
  (-.75,-1.7) -- ++(1.5,0) node [black,midway,xshift=0cm,yshift=-2em] {$n_{-}$};

  \draw [decorate,decoration={brace,amplitude=10pt,mirror,raise=4pt},yshift=0pt]
  (1.75,-1.7) -- ++(1.5,0) node [black,midway,xshift=0cm,yshift=-2em] {$n_{-}$};  
  
  \draw[very thick,blue,-latex] (4,-.75) --++(1.5,0);

   \begin{scope}[shift={(6.5,0)}]

  \draw (0,0) node (11) {} --++(2.5,0) node (21) {};

  \draw (11)--++(0,-1.5) node (12b) {};
  \draw (21)--++(0,-1.5) node (22b) {};

  \filldraw[fill=white,draw=black] (11) circle (.2) node [above=.7em] {$n_{1,+}$};
  \filldraw[fill=white,draw=black] (21) circle (.2) node [above=.7em] {$n_{2,+}$};    

  \filldraw[fill=white,draw=black] (12b)++(-.2,-.2) rectangle ++(.4,.4) node [below=1em] {$n_-$}; 
  \filldraw[fill=white,draw=black] (22b)++(-.2,-.2) rectangle ++(.4,.4) node [below=1em] {$n_-$};
   
   \end{scope}
  
 \end{tikzpicture}
\end{center}
which is consistent with the brane construction discussed in Sec.~\ref{sec:brane}.
If $n_{1,-} \neq n_{2,-}$, such a cancellation does not occur, and the flavor nodes become different from each other.
This procedure is naturally generalized to $A_r$ quiver theory.

\subsection{$D_r$ quiver}\label{sec:decoupling_D}

Let us apply the decoupling trick to $D_r$ quiver with $r = 4$ as an example.
Splitting the SU$(n_{i,+}|n_{i,-})$ gauge nodes into positive SU$(n_{i,+})$ and negative SU$(n_{i,-})$ nodes, we obtain the following:
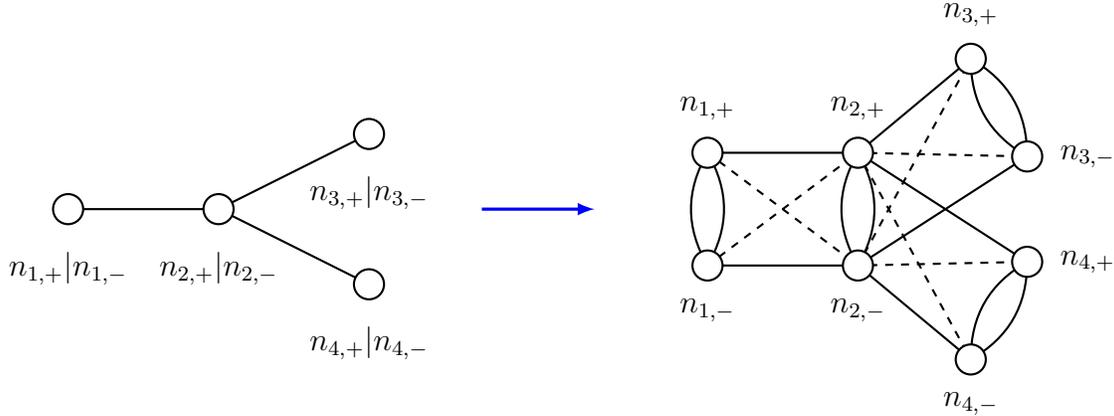
\begin{center}
  \begin{tikzpicture}[thick]
   
   \draw (0,0) -- (2.,0);
   \draw (4,1) -- (2,0) -- (4,-1);
   
   \filldraw[fill=white,draw=black] (0,0) circle (.2) node [below=1em] {$n_{1,+}|n_{1,-}$};
   \filldraw[fill=white,draw=black] (2,0) circle (.2) node [below=1em] {$n_{2,+}|n_{2,-}$};
   \filldraw[fill=white,draw=black] (4,1) circle (.2) node [below=1em] {$n_{3,+}|n_{3,-}$};
   \filldraw[fill=white,draw=black] (4,-1) circle (.2) node [below=1em] {$n_{4,+}|n_{4,-}$};     

   \draw [blue,-latex, very thick] (5.5,0) -- ++(1.5,0);

  \begin{scope}[shift={(8.5,0)}]

   \draw (0,.75) node (i1) {} to [bend right] (0,-.75) node (i2) {};
   \draw (0,.75) to [bend left] (0,-.75);
   
   \draw (2.,.75) node (j1) {} to [bend right] (2.,-.75) node (j2) {};
   \draw (2.,.75) to [bend left] (2.,-.75);

   \draw (3.5,2) node (k1) {} to [bend right] ++(-60:1.5) node (k2) {};
   \draw (3.5,2) to [bend left] ++(-60:1.5);

   \draw (3.5,-2) node (l1) {} to [bend right] ++(60:1.5) node (l2) {};
   \draw (3.5,-2) to [bend left] ++(60:1.5);      

   \draw (i1) -- ++(2.,0);
   \draw (i2) -- ++(2.,0);
   \draw (j1) -- (k1);
   \draw (j1) -- (l2);
   \draw (j2) -- (k2);
   \draw (j2) -- (l1);      

   \draw [dashed] (i1) -- (j2);
   \draw [dashed] (i2) -- (j1);
   \draw [dashed] (j1) -- (k2);
   \draw [dashed] (j1) -- (l1);
   \draw [dashed] (j2) -- (k1);
   \draw [dashed] (j2) -- (l2);         
   
   \filldraw[fill=white,draw=black] (i1) circle (.2) node [above=.7em] {$n_{1,+}$};
   \filldraw[fill=white,draw=black] (i2) circle (.2) node [below=.7em] {$n_{1,-}$};     
   \filldraw[fill=white,draw=black] (j1) circle (.2) node [above=.7em] {$n_{2,+}$};
   \filldraw[fill=white,draw=black] (j2) circle (.2) node [below=.7em] {$n_{2,-}$};     
   \filldraw[fill=white,draw=black] (k1) circle (.2) node [above=.7em] {$n_{3,+}$};
   \filldraw[fill=white,draw=black] (k2) circle (.2) node [right=.7em] {$n_{3,-}$};

   \filldraw[fill=white,draw=black] (l2) circle (.2) node [right=.7em] {$n_{4,+}$};
   \filldraw[fill=white,draw=black] (l1) circle (.2) node [below=.7em] {$n_{4,-}$};  
   
  \end{scope}   
   
  \end{tikzpicture}
\end{center}
We then turn off the gauge couplings for the negative nodes, $\fq_{i,-} \to 0$.
Applying the condition $n_{1,-} = n_{2,-} = 2 n_{3,-} = 2 n_{4,-} =: 2 n_-$, and tuning the Coulomb moduli for the negative nodes, we obtain $D_4$ quiver configuration with a single SU$(2n_-)$ flavor node:
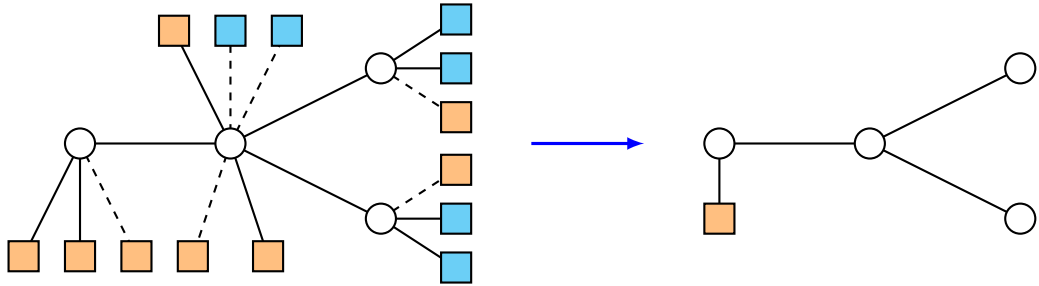
\begin{center}
 \begin{tikzpicture}[thick]

  \draw (0,0) node (11) {} --++(2.,0) node (21) {};
  \draw (4,1) node (31) {} -- (21) -- (4,-1) node (41) {};

  \draw (11)--++(-.75,-1.5) node (12a) {};
  \draw (11)--++(0,-1.5) node (12b) {};
  \draw[dashed] (11)--++(.75,-1.5) node (12c) {};

  \draw[dashed] (21)--++(-.5,-1.5) node (22a) {};
  \draw (21)--++(.5,-1.5) node (22b) {};
  \draw (21)--++(-.75,1.5) node (22d) {};
  \draw[dashed] (21)--++(0,1.5) node (22e) {};
  \draw[dashed] (21)--++(.75,1.5) node (22f) {};  

  \draw (31) --++(1,.65) node (32a) {};
  \draw (31) --++(1,0) node (32b) {};
  \draw[dashed] (31) --++(1,-.65) node (32c) {};

  \draw[dashed] (41) --++(1,.65) node (42a) {};
  \draw (41) --++(1,0) node (42b) {};
  \draw (41) --++(1,-.65) node (42c) {};    

  \filldraw[fill=white,draw=black] (11) circle (.2); 
  \filldraw[fill=white,draw=black] (21) circle (.2); 
  \filldraw[fill=white,draw=black] (31) circle (.2); 
  \filldraw[fill=white,draw=black] (41) circle (.2); 

  \filldraw[fill=orange!50,draw=black] (12a)++(-.2,-.2) rectangle ++(.4,.4);
  \filldraw[fill=orange!50,draw=black] (12b)++(-.2,-.2) rectangle ++(.4,.4);
  \filldraw[fill=orange!50,draw=black] (12c)++(-.2,-.2) rectangle ++(.4,.4);

  \filldraw[fill=orange!50,draw=black] (22a)++(-.2,-.2) rectangle ++(.4,.4);
  \filldraw[fill=orange!50,draw=black] (22b)++(-.2,-.2) rectangle ++(.4,.4);
  \filldraw[fill=orange!50,draw=black] (22d)++(-.2,-.2) rectangle ++(.4,.4);
  \filldraw[fill=cyan!50,draw=black] (22e)++(-.2,-.2) rectangle ++(.4,.4);
  \filldraw[fill=cyan!50,draw=black] (22f)++(-.2,-.2) rectangle ++(.4,.4);  
  
  \filldraw[fill=cyan!50,draw=black] (32a)++(-.2,-.2) rectangle ++(.4,.4);
  \filldraw[fill=cyan!50,draw=black] (32b)++(-.2,-.2) rectangle ++(.4,.4);
  \filldraw[fill=orange!50,draw=black] (32c)++(-.2,-.2) rectangle ++(.4,.4);

  \filldraw[fill=orange!50,draw=black] (42a)++(-.2,-.2) rectangle ++(.4,.4);
  \filldraw[fill=cyan!50,draw=black] (42b)++(-.2,-.2) rectangle ++(.4,.4);
  \filldraw[fill=cyan!50,draw=black] (42c)++(-.2,-.2) rectangle ++(.4,.4);    
  
  \draw[very thick,blue,-latex] (6,0) --++(1.5,0);

   \begin{scope}[shift={(8.5,0)}]

    \draw (0,0) node (11) {} --++(2.,0) node (21) {};
    \draw (4,1) node (31) {} -- (21) -- (4,-1) node (41) {};

  \draw (11)--++(0,-1.) node (12b) {};

    \filldraw[fill=white,draw=black] (11) circle (.2); 
    \filldraw[fill=white,draw=black] (21) circle (.2);
    \filldraw[fill=white,draw=black] (31) circle (.2);
    \filldraw[fill=white,draw=black] (41) circle (.2);     

  \filldraw[fill=orange!50,draw=black] (12b)++(-.2,-.2) rectangle ++(.4,.4);
   
   \end{scope}
  
 \end{tikzpicture}
\end{center}
where \ \tikz[baseline=(00.base),thick] \filldraw[fill=orange!50,draw=black] (0,0)
 node (00) [above = .5em,right=1em] {= SU$(2n_-)$} rectangle ++(.4,.4); and \ \tikz[baseline=(00.base),thick] \filldraw[fill=cyan!50,draw=black] (0,0)
 node (00) [above = .5em,right=1em] {= SU$(n_-)$} rectangle ++(.4,.4); flavor nodes.
This is consistent with the brane configuration discussed in Sec.~\ref{sec:brane_D}.
 
\subsection{$\widehat{A}_0$ quiver}

Let us then consider the affine quiver $\Gamma = \widehat{A}_0$.
Here we have a parameter $\mu \in \BC^\times$ assigned to the loop edge, which is the multiplicative adjoint mass parameter.
Applying the decoupling trick, it becomes:
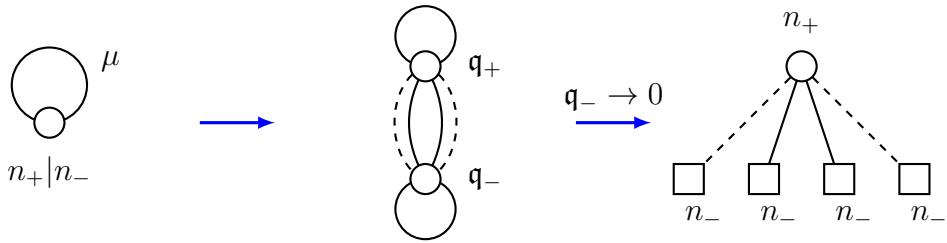
\begin{center}
  \begin{tikzpicture}[thick]

   \draw (0,0) arc (-90:270:.5) node [above right=1.3em] {$\mu$};
   
   \filldraw[fill=white,draw=black] (0,0) circle (.2) node [below=.7em] {$n_+|n_-$};

   \draw[-latex,very thick,blue] (2,0) -- ++(1,0);

     \begin{scope}[shift={(5,0)}]

      \draw (0,.75) arc (-90:270:.4);
      \draw (0,-.75) arc (90:450:.4);      
      
      \draw (0,.75) to [bend right] (0,-.75);
      \draw (0,.75) to [bend left] (0,-.75);

      \draw[dashed] (0,.75) to [bend right=70] (0,-.75);
      \draw[dashed] (0,.75) to [bend left=70] (0,-.75);            

     \filldraw[fill=white,draw=black] (0,.75) circle (.2) node [right=1em] {$\fq_+$}; 
     \filldraw[fill=white,draw=black] (0,-.75) circle (.2) node [right=1em] {$\fq_-$}; 

      \draw[-latex,very thick,blue] (2,0) -- ++(.5,0) node [above] {\textcolor{black}{$\fq_- \to 0$}} -- ++(.5,0);
    
     \end{scope}

   \begin{scope}[shift={(10,0)}]

    \draw (0,.75) -- ++(-.5,-1.5) node (l) {};
    \draw (0,.75) -- ++(.5,-1.5) node (r) {};

    \draw[dashed] (0,.75) -- ++(-1.5,-1.5) node (l-) {};
    \draw[dashed] (0,.75) -- ++(1.5,-1.5) node (r-) {};        
        
    \filldraw[fill=white,draw=black] (0,.75) circle (.2) node [above=.7em] {$n_+$};
    \filldraw[fill=white,draw=black] (l)++(-.2,-.2) rectangle ++(.4,.4) node [below=1em] {$n_-$};
    \filldraw[fill=white,draw=black] (r)++(-.2,-.2) rectangle ++(.4,.4) node [below=1em] {$n_-$};

    \filldraw[fill=white,draw=black] (l-)++(-.2,-.2) rectangle ++(.4,.4) node [below=1em] {$n_-$};
    \filldraw[fill=white,draw=black] (r-)++(-.2,-.2) rectangle ++(.4,.4) node [below=1em] {$n_-$};

   \end{scope}   
  \end{tikzpicture}  
\end{center}
Then we have the SU$(n_+)$ gauge node, and with $2 n_-$ {\em positive} and $2 n_-$ {\em negative} fundamental matters.
In this case, due to the adjoint mass parameter, the positive and negative fundamentals cannot be canceled with each other.
This is consistent with the argument discussed in Sec.~\ref{sec:A0}.

\section{Outlook}\label{sec:outlook}

Let us conclude with outlooks for the study of supergroup gauge theory.

\subsection{Classical supergroup gauge theory}

We have shown the super analog of LMNS integral formula for the instanton partition function in Sec.~\ref{sec:ADHM}.
It is natural to generalize this result to other (classical) supergroups, i.e., OSp($n|m$), in a similar way to \cite{Marino:2004cn,Nekrasov:2004vw,Hwang:2014uwa}.
The resultant super-instanton partition function would be given by a multi-variable contour integral over Cartan elements of the corresponding supergroup. Also the case of the super Lie algebra
$D(2|1, \alpha)$ could be a particular interesting, since the bosonic part
of this algebra contains three copies of $\mathfrak{sl}_2$ and the fermionic
part transforms as a tri-fundamental representation. 

\subsection{Localization on a curved compact manifold}

In this paper, we have considered the supergroup gauge theory partition function on the non-compact manifolds, $\BR^4 \times \CalE$, where the elliptic curve is $\CalE = \text{pt} \times \text{pt}$ (4d), $S^1 \times \text{pt}$ (5d), and $S^1 \times S^1 = T^2$ (6d).
It would be interesting to study the partition function on a curved compact manifold with the supersymmetric localization~\cite{Pestun:2007rz,Pestun:2016zxk}.
Our result implies that the equivariant localization is similarly applicable to supergroup theory.
Hence the resultant partition function would be reduced to the integral over Cartan part of the corresponding supergroup, namely a supermatrix integral.
We shall see the holomorphic factorization of the four-sphere partition function, $Z_{S^4} = \int da \, |Z_{\BR^4}|^2$, as well for the supergroup gauge theory, and such a factorization is also expected for lower dimensional theories.
See \cite{Gu:2018xzx} for a related result in two dimensions.
The Harish-Chandra--Itzykson--Zuber integral for supergroup U($n_+|n_-$)~\cite{Guhr:1991JMP,Alfaro:1994ca,Guhr:1996CMP} would be interpreted as a zero-dimensional example.

\subsection{Quantum inverse scattering method}\label{sec:QIS}

In Sec.~\ref{sec:Bethe/Gauge}, we have discussed quantum integrability arising in the NS limit of supergroup gauge theory.
In this case, the polynomials, $a(x)$ and $d(x)$, used to diagonalize the transfer matrix, are replaced with rational functions, ratio of polynomials, $a^\pm(x)$ and $d^\pm(x)$.
This implies that the eigenvalues of $A$ and $D$ operators in the monodromy matrix $\CalT = \begin{pmatrix} A & B \\ C & D \end{pmatrix}$ would be rational functions, namely $a(x) = a^+(x)/a^-(x)$ and $d(x) = d^+(x)/d^-(x)$, as in \eqref{eq:BE_meros}, and the eigenvalue of the transfer matrix $\sT = \Tr \CalT$ becomes also a rational function $T(x) = a(x) + d(x)$.
We can turn on the twist parameter $\fq$ in general, but do not have it for simplicity.
Since we now have two sets of inhomogeneous parameters, the corresponding monodromy matrix of $(L_+|L_-)$-chain may consist of two $\CalL$ matrices, $\CalL^+$ and $\CalL^-$, as follows:
\begin{align}
 \CalT(x) & = \CalL_{L_-}^-(x - \xi_{L_-}^-) \cdots \CalL_1^-(x - \xi_1^-) \CalL_{L_+}^+(x - \xi_{L_+}^+) \cdots \CalL_1^+(x - \xi_1^+)
 \nonumber \\
 & = \CalT_-(x) \CalT_+(x)
 \, .
\end{align}
This configuration looks similar to the $(L_++L_-)$-chain system, but it should be interpreted as its analytic continuation $L_- \to - L_-$ since $a(x)$ and $d(x)$ are ratios of polynomials of degree $L_+$ and $L_-$.
Similarly, we have two sets of Bethe roots in this case.
Thus the $(N_+|N_-)$-magnon Bethe state is generated by the $B^\pm$ operator, which is a off-diagonal part of $\CalT^\pm$,
\begin{align}
 \ket{\Psi(\{x_k^+, x_{k'}^{-}\})} = B^-(x_{N_-}^-) \cdots B^-(x_{1}^-) B^+(x_{N_+}^+) \cdots B^+(x_{1}^+) \ket{0}
 \, .
\end{align}

Let us mention a possible connection with the Bethe/Gauge correspondence for the spin chain with supergroup symmetry~\cite{Orlando:2010uu,Nekrasov:2018gne,Zenkevich:2018fzl}.
Through the Bethe/Gauge correspondence, the $G$-symmetric spin chain model corresponds to $\Gamma$-quiver gauge theory where the Dynkin diagram of $G$ coincides with quiver $\Gamma$.
The construction of~\cite{Orlando:2010uu,Nekrasov:2018gne,Zenkevich:2018fzl} is based on the gauge theory whose quiver structure is given by the Dynkin diagram of the Lie superalgebra, but 
with ordinary, non supergroup gauge symmetry of the supersymmetric gauge theory. 
An important point in this construction is that the negative equivariant parameters (adjoint masses) should be assigned for the gauge nodes corresponding to the additional nodes in the super Dynkin diagram.
This situation is actually dual to the setup studied in this paper through the so-called spectral duality~\cite{Mironov:2012uh,Mironov:2012ba,Mironov:2013xva,Mironov:2016cyq}, which exchanges the gauge and quiver structures.
It'd be interesting to study details of such a super analog of the spectral duality.

\subsection{AGT correspondence}

 The AGT correspondence claims that the partition function of 4d $\CalN=2$ theory with $G$ gauge symmetry on $\BR^4_{\epsilon_{1,2}}$ is identical to the conformal block of W($G$)-algebra~\cite{Alday:2009aq,Wyllard:2009hg}.
 Similarly its K-theory analog is the correspondence between the 5d partition function compactified on a circle $\BR^4_{\epsilon_{1,2}} \times S^1$ and the conformal block of $q$-deformed W($G$)-algebra~\cite{Awata:2009ur}.
 This correspondence is expected to be generalized to the situation studied in this paper:
 The partition function for 4d and 5d theory with supergroup gauge symmetry would be identified with the conformal block of the corresponding W-algebra associated with supergroup.
 A possible realization of such an algebra is the Drinfeld--Sokolov reduction of affine Lie superalgebra.
 Such a W-algebraic structure should be also interpreted as a further quantization of quantum algebra associated with quantum integrability discussed in Sec.~\ref{sec:Bethe/Gauge}.
 See~\cite{Bershtein:2018SM,Litvinov:2016mgi,Litvinov:2019rlv} for the studies along this direction.

 \subsection{Quiver variety}

 The construction of super-instantons shown in Sec.~\ref{sec:counting} suggests a generalization of quiver variety equipped with supergroup structure~\cite{Nakajima1994,Nakajima:1998DM}.
 The ordinary quiver variety denoted by $\CalM^\Gamma_{\vec{n},\vec{k}}$ is defined for a quiver $\Gamma$ with two integral vectors $\vec{n}, \vec{k} \in \BZ_{\ge 0}^{|\Gamma_0|}$.
 A super analog of the quiver variety is defined for a quiver $\Gamma$ with two graded vectors, $\vec{n}^\sigma, \vec{k}^\sigma \in \BZ_{\ge 0}^{|\Gamma_0|}$ for $\sigma = \pm$.
 Let $K = (K_i)_{i \in \Gamma_0}$ and $N = (N_i)_{i \in \Gamma_0}$ with $K_i = \BC^{k_{i,+}|k_{i,-}}$ and $N_i = \BC^{n_{i,+}|n_{i,-}}$ be the graded vector spaces for $i \in \Gamma_0$, and define maps
 \begin{align}
  I_i: \ N_i \to K_i
  \, , \qquad
  J_i: \ K_i \to N_i
  \, , \qquad
  B_{e:i \to j}: \ K_i \to K_j
  \, , \qquad
  \overline{B}_{e:i \to j}: \ K_j \to K_i
  \, ,
 \end{align}
with the supergroup action $\mathrm{U}(K) = \prod_{i \in \Gamma_0} \mathrm{U}(k_{i,+}|k_{i,-})$
\begin{align}
  \mathrm{U}(K): \
 \left( \left( B_e, \overline{B}_e \right)_{e: i \to j}, \left( I_i, J_i \right)_{i \in \Gamma_0} \right)
 \longrightarrow
 \left( \left( v_j B_e v_i^{-1}, v_i \overline{B}_e v_j^{-1} \right)_{e: i \to j}, \left( v_i I_i, J_i v_i^{-1} \right)_{i \in \Gamma_0} \right)  
\end{align}
where $v_i \in \mathrm{U}(n_{i,+}|n_{i,-})$.
As mentioned in Sec.~\ref{sec:ADHM0}, these maps are realized as supermatrices.
A super analog of the quiver variety $\CalM_{\vec{n}^\sigma,\vec{k}^\sigma}^\Gamma$, which is a hyper-K\"ahler supermanifold, is defined as a supergroup quotient
 \begin{align}
  \CalM^\Gamma_{\vec{n}^\sigma,\vec{k}^\sigma} = \vec{\mu}^{-1}(\vec{\zeta}) / \mathrm{U}(K)
 \end{align}
 where the moment maps $\vec{\mu} = (\mu_{i,\BR}, \mu_{i,\BC})_{i \in \Gamma_0}$ are defined
 \begin{subequations} 
 \begin{align}
  \mu_{i,\BR} & =
  I_i I_i^\dag - J_i^\dag J_i
  - \sum_{e:i \to j} \left( B_e^\dag B_e - \overline{B}_e \overline{B}_e^\dag \right)
  + \sum_{e:j \to i} \left( B_e B_e^\dag - \overline{B}_e^\dag \overline{B}_e \right)
  \\
  \mu_{i,\BC} & =
  I_i J_i
  - \sum_{e:i \to j} \overline{B}_e B_e
  + \sum_{e:j \to i} B_e \overline{B}_e
 \end{align}  
 \end{subequations}
 and we typically consider $\vec{\zeta} = (\zeta_i \Id_{K_i}, 0)_{i \in \Gamma_0}$.
 For $\Gamma = \widehat{A}_0$, it is reduced to the super ADHM moduli space discussed in Sec.~\ref{sec:ADHM}.
 Such a supergroup quiver variety would be realized as the instanton moduli space for supergroup gauge theory on the ALE space $\BC^2/\Gamma'$ if $\Gamma = \widehat{\Gamma}'$.
 It is also interesting to study the realization of $\CalM_{\vec{n}^\sigma,\vec{k}^\sigma}^\Gamma$ as the Higgs branch of the corresponding 3d quiver gauge theory.

\appendix

\section{Combinatorial calculus}\label{sec:comb}

We summarize the combinatorics calculus of the partition for the instanton partition function.
We denote the transposed partition of $\lambda$ by $\lambda^\text{T}$.
Summation over the partition is expressed in the following two ways,
\begin{align}
 \sum_{s \in \lambda}
 = \sum_{s_1=1}^{\lambda_1^\text{T}} \sum_{s_2 = 1}^{\lambda_{s_1}}
 = \sum_{s_2=1}^{\lambda_1} \sum_{s_1 = 1}^{\lambda_{s_2}^\text{T}} 
\end{align}

\subsection{U($n$) theory}\label{sec:comb_U(n)}

We consider the instanton contribution to the Chern character of the bifundamental hypermultiplet
\begin{align}
 \ch \bH_{e:i \to j}^{\text{bf,\,inst}} & =
 - \mu_e (1 - q_1^{-1})(1 - q_2^{-1}) \ch \bK_i^\vee \ch \bK_j
 + \mu_e \ch \bN_i^\vee \ch \bK_j + \mu_e q^{-1} \ch \bK_i^\vee \ch \bN_j
 \nonumber \\
 & =: \sum_{\alpha=1}^{n_i} \sum_{\beta=1}^{n_j} \mu_e \frac{\nu_{j,\beta}}{\nu_{i,\alpha}} \, \Xi[\lambda_{i,\alpha},\lambda_{j,\beta}] 
\end{align}
where we define
\begin{align}
 \Xi[\lambda_{\alpha},\lambda_{\beta}]
 & =
 - (1 - q_1^{-1})(1 - q_2^{-1}) \sum_{s \in \lambda_{\alpha}} \sum_{s' \in \lambda_{\beta}} q_1^{-s_1+s_1'} q_2^{-s_2+s_2'}
 + \sum_{s \in \lambda_{\alpha}} q_1^{-s_1} s_2^{-s_2}
 + \sum_{s' \in \lambda_{\beta}} q_1^{s_1'-1} s_2^{s_2'-1}
 \label{eq:comb1}
\end{align}
From this expression, we obtain a combinatorial formula
(See, for example, \cite{nakajima-hilbert})
\begin{align}
 \Xi[\lambda_{\alpha},\lambda_{\beta}]
 & =
 \sum_{s \in \lambda_{\alpha}} q_1^{\ell_{\beta}(s)} q_2^{-a_{\alpha}(s)-1}
 + \sum_{s \in \lambda_{\beta}} q_1^{-\ell_{\alpha}(s)-1} q_2^{a_{\beta}(s)}
 \, .
 \label{eq:comb2}
\end{align}
where the arm and leg lengths for each box $s = (s_1,s_2)$ in the partition $\lambda_{\alpha}$ are defined as
\begin{align}
 a_{\alpha}(s) = \lambda_{\alpha,s_1} - s_2
 \, , \qquad
 \ell_{\alpha}(s) = \lambda_{\alpha,s_2}^\text{T} - s_1
 \, .
 \label{eq:arm_leg}
\end{align}
We remark
\begin{align}
 q \, \Xi[\lambda_\alpha,\lambda_\beta]\Big|_{q_1,q_2}
 = \Xi[\lambda_\beta,\lambda_\alpha]\Big|_{q_1^{-1},q_2^{-1}}
 \, .
 \label{eq:comb_inv}
\end{align}
The vector multiplet contribution has a similar expression
\begin{align}
 \ch \bV_i^\text{inst} & =
 (1 - q_1^{-1})(1 - q_2^{-1}) \ch \bK_i^\vee \ch \bK_j
 - \ch \bN_i^\vee \ch \bK_j - q^{-1} \ch \bK_i^\vee \ch \bN_j
 \nonumber \\
 & =
 - \sum_{\alpha,\beta=1}^{n_i}
 \frac{\nu_{i,\beta}}{\nu_{i,\alpha}} \, \Xi[\lambda_{i,\alpha},\lambda_{i,\beta}]
 \, .
\end{align}

\subsubsection{Proof of the formula \eqref{eq:comb2}}\label{sec:comb_sub}

We prove the combinatorial formula \eqref{eq:comb2}.
We partially perform the summation for the first term in \eqref{eq:comb1},
\begin{align}
 &
 - (1 - q_1^{-1})(1 - q_2^{-1}) \sum_{s \in \lambda_{\alpha}} \sum_{s' \in \lambda_{\beta}} q_1^{-s_1+s_1'} q_2^{-s_2+s_2'}
 =
 \sum_{s_1=1}^{\lambda_{\alpha,1}^\text{T}} \sum_{s_2'=1}^{\lambda_{\beta,1}}
 (1 - q_1^{\lambda_{\beta,s_2'}^\text{T}}) q_1^{-s_1}
 (1 - q_2^{-\lambda_{\alpha,s_1}}) q_2^{s_2'-1}
 \nonumber \\
 & =
 \sum_{s_1=1}^{\lambda_{\alpha,1}^\text{T}} \sum_{s_2'=1}^{\lambda_{\beta,1}}
 \left[
 q_1^{\lambda_{\beta,s_2'}^\text{T} - s_1} q_2^{-\lambda_{\alpha,s_1}+s_2'-1}
 - q_1^{- s_1} q_2^{s_2'-1}
 + (1 - q_1^{\lambda_{\beta,s_2'}^\text{T}}) q_1^{-s_1} q_2^{s_2'-1}
 + q_1^{-s_1} (1 - q_2^{-\lambda_{\alpha,s_1}}) q_2^{s_2'-1}
 \right]
 \, .
 \label{eq:comb3}
\end{align}
The third and fourth terms in \eqref{eq:comb3} are then given by
\begin{subequations}
 \begin{align}
  \sum_{s_1=1}^{\lambda_{\alpha,1}^\text{T}} \sum_{s_2'=1}^{\lambda_{\beta,1}}
  (1 - q_1^{\lambda_{\beta,s_2'}^\text{T}}) q_1^{-s_1} q_2^{s_2'-1}
  & =
  \sum_{s_1=1}^{\lambda_{\alpha,1}^\text{T}} \sum_{s_2'=1}^{\lambda_{\beta,1}}
  \sum_{s_1'=1}^{\lambda_{\beta,s_2'}^\text{T}}
  (1 - q_1) q_1^{-s_1+s_1'-1} q_2^{s_2'-1}
  \nonumber \\
  & =
  - \sum_{s' \in \lambda_\beta} (1 - q_1^{-\lambda_{\alpha,1}^\text{T}}) q_1^{s_1'-1} q_2^{s_2'-1}
  \, ,
 \end{align}
 \begin{align}
  \sum_{s_1=1}^{\lambda_{\alpha,1}^\text{T}} \sum_{s_2'=1}^{\lambda_{\beta,1}}
  q_1^{-s_1} (1 - q_2^{-\lambda_{\alpha,s_1}}) q_2^{s_2'-1}
  & =
  \sum_{s_1=1}^{\lambda_{\alpha,1}^\text{T}} \sum_{s_2'=1}^{\lambda_{\beta,1}}
  \sum_{s_2=1}^{\lambda_{\alpha,s_1}}
  q_1^{-s_1} (1 - q_2^{-1}) q_2^{-s_2+s_2'}
  \nonumber \\
  & =
  - \sum_{s \in \lambda_\alpha}
  q_1^{-s_1} (1 - q_2^{\lambda_{\beta,1}}) q_2^{-s_2}
  \, .
 \end{align}
\end{subequations}
Combining them together, \eqref{eq:comb1} becomes
\begin{align}
 \Xi[\lambda_\alpha,\lambda_\beta]
 & =
 \sum_{s_1=1}^{\lambda_{\alpha,1}^\text{T}} \sum_{s_2'=1}^{\lambda_{\beta,1}}
 \left[
 q_1^{\lambda_{\beta,s_2'}^\text{T} - s_1} q_2^{-\lambda_{\alpha,s_1}+s_2'-1}
 - q_1^{- s_1} q_2^{s_2'-1}
 \right]
 + \sum_{s \in \lambda_\alpha} q_1^{-s_1} q_2^{\lambda_{\beta,1}-s_2}
 + \sum_{s' \in \lambda_\beta} q_1^{-\lambda_{\alpha,1}^\text{T} + s_1' - 1} q_2^{s_2'-1}
 \, .
 \label{eq:comb4} 
\end{align}

We divide it into the negative and positive parts
\begin{align}
 \Xi[\lambda_\alpha,\lambda_\beta]
 = \Xi_{q_2^{<0}}[\lambda_\alpha,\lambda_\beta] + \Xi_{q_2^{\ge 0}}[\lambda_\alpha,\lambda_\beta]
\end{align}
where $\Xi_{q_2^{<0}}$ consists of monomials with negative powers of $q_2$, while $\Xi_{q_2^{\ge 0}}$ consists of positive ones.
Let us focus on $\Xi_{q_2^{<0}}$ with \eqref{eq:comb4}.
\begin{itemize}
 \item For $\lambda_{\beta,1} > \lambda_{\alpha,s_1}$, the first term in \eqref{eq:comb4} can contribute to the negative part $\Xi_{q_2^{<0}}$.
 \item For $\lambda_{\beta,1} \le \lambda_{\alpha,s_1}$, the first and third terms can contribute to $\Xi_{q_2^{<0}}$.
\end{itemize}
In both cases, the negative part $\Xi_{q_2^{<0}}$ is given by
\begin{align}
 \Xi_{q_2^{<0}}[\lambda_\alpha,\lambda_\beta] = \sum_{s \in \lambda_\alpha}
 q_1^{\lambda_{\beta,s_2}^\text{T} - s_1} q_2^{-\lambda_{\alpha,s_1} + s_2 - 1}
 = \sum_{s \in \lambda_\alpha} q_1^{\ell_\beta(s)} q_2^{-a_\alpha(s) - 1}
 \, .
\end{align}
We can similarly obtain the positive part $\Xi_{q_2^{\ge 0}}$ by utilizing the formula \eqref{eq:comb_inv},
\begin{align}
 \Xi_{q_2^{\ge 0}}[\lambda_\alpha,\lambda_\beta] = 
 \sum_{s \in \lambda_{\beta}} q_1^{-\ell_{\alpha}(s)-1} q_2^{a_{\beta}(s)}
 \, .
\end{align}
This proves the formula \eqref{eq:comb2}.

\subsection{U($n_+|n_-$) theory}

For the supergroup theory, we consider the following contribution to the Chern character
\begin{align}
 \ch \bH_{e:i \to j,\sigma\sigma'}^{\text{bf,\,inst}} & =
 - \mu_e (1 - q_1^{-1})(1 - q_2^{-1}) \ch \bK_i^{\sigma\vee} \ch \bK_j^{\sigma'}
 + \mu_e \ch \bN_i^{\sigma\vee} \ch \bK_j^{\sigma'} + \mu_e q^{-1} \ch \bK_i^{\sigma\vee} \ch \bN_j^{\sigma'}
 \nonumber \\
 & =: \sum_{\alpha=1}^{n_i} \sum_{\beta=1}^{n_j} \mu_e \frac{\nu_{j,\beta}}{\nu_{i,\alpha}} \, \Xi_{\sigma\sigma'}[\lambda_{i,\alpha}^\sigma,\lambda_{j,\beta}^{\sigma'}]
\end{align}
where the diagonal factors are written using \eqref{eq:comb1} as
\begin{align}
 \Xi_{++}[\lambda_\alpha,\lambda_\beta] = \Xi[\lambda_\alpha,\lambda_\beta]
 \, , \qquad
 \Xi_{--}[\lambda_\alpha,\lambda_\beta] = \Xi[\lambda_\beta,\lambda_\alpha]
 \, .
\end{align}
The vector multiplet contribution \eqref{eq:chV_inst2} is given by
\begin{align}
 \ch \bV_{i,\sigma\sigma'}^\text{inst} & =
 - \sum_{\alpha=1}^{n_i} \sum_{\beta=1}^{n_i} \frac{\nu_{j,\beta}}{\nu_{i,\alpha}} \, \Xi_{\sigma\sigma'}[\lambda_{i,\alpha}^\sigma,\lambda_{i,\beta}^{\sigma'}]
 \, .
\end{align}
The off-diagonal factors are 
\begin{subequations}
\begin{align}
 \Xi_{+-}[\lambda_\alpha,\lambda_\beta] & =
 - (1 - q_1^{-1})(1 - q_2^{-1}) \sum_{s \in \lambda_\alpha} \sum_{s' \in \lambda_\beta} q_1^{-s_1-s_1'+1} q_2^{-s_2-s_2'+1}
 + \sum_{s \in \lambda_\alpha} q_1^{-s_1} s_2^{-s_2}
 + \sum_{s' \in \lambda_\beta} q_1^{-s_1'} s_2^{-s_2'}
 \label{eq:comb+-1} \\
 \Xi_{-+}[\lambda_\alpha,\lambda_\beta] & =
 - (1 - q_1^{-1})(1 - q_2^{-1}) \sum_{s \in \lambda_\alpha} \sum_{s' \in \lambda_\beta} q_1^{s_1+s_1'-1} q_2^{s_2+s_2'-1}
 + \sum_{s \in \lambda_\alpha} q_1^{s_1-1} s_2^{s_2-1}
 + \sum_{s' \in \lambda_\beta} q_1^{s_1'-1} s_2^{s_2'-1} 
\end{align}
\end{subequations}
We remark that these off-diagonal factors are symmetric under $\lambda_\alpha \leftrightarrow \lambda_\beta$,
\begin{align}
 \Xi_{+-}[\lambda_\alpha,\lambda_\beta] = \Xi_{+-}[\lambda_\beta,\lambda_\alpha]
 \, , \qquad
 \Xi_{-+}[\lambda_\alpha,\lambda_\beta] = \Xi_{-+}[\lambda_\beta,\lambda_\alpha]
 \, ,
\end{align}
and
\begin{align}
 q \, \Xi_{+-}[\lambda_\alpha,\lambda_\beta]\Big|_{q_1,q_2}
 = \Xi_{-+}[\lambda_\alpha,\lambda_\beta]\Big|_{q_1^{-1},q_2^{-1}}
\end{align}

We apply a similar computation to \eqref{eq:comb+-1} as discussed in Appendix~\ref{sec:comb_U(n)}.
The first term in \eqref{eq:comb+-1} yields
\begin{align}
 &
 - (1 - q_1^{-1})(1 - q_2^{-1}) \sum_{s \in \lambda_\alpha} \sum_{s' \in \lambda_\beta} q_1^{-s_1-s_1'+1} q_2^{-s_2-s_2'+1}
 =
 - \sum_{s_1=1}^{\lambda_{\alpha,1}^\text{T}} \sum_{s_2'=1}^{\lambda_{\beta,1}}
 (1 - q_1^{-\lambda_{\beta,s_2'}^\text{T}}) q_1^{-s_1}
 (1 - q_2^{-\lambda_{\alpha,s_1}}) q_2^{-s_2'}
 \nonumber \\
 & =
 - \sum_{s_1=1}^{\lambda_{\alpha,1}^\text{T}} \sum_{s_2'=1}^{\lambda_{\beta,1}}
 \left[
 q_1^{-\lambda_{\beta,s_2'}^\text{T}-s_1} q_2^{-\lambda_{\alpha,s_1}-s_2'}
 - q_1^{-s_1} q_2^{-s_2'}
 + (1 - q_1^{-\lambda_{\beta,s_2'}^\text{T}}) q_1^{-s_1} q_2^{-s_2'}
 + q_1^{-s_1} (1 - q_2^{-\lambda_{\alpha,s_1}}) q_2^{-s_2'}
 \right]
 \, .
 \label{eq:comb+-2}
\end{align}
The third and fourth terms in \eqref{eq:comb+-2} are given by
\begin{subequations} 
 \begin{align}
  - \sum_{s_1=1}^{\lambda_{\alpha,1}^\text{T}} \sum_{s_2'=1}^{\lambda_{\beta,1}}
  (1 - q_1^{-\lambda_{\beta,s_2'}^\text{T}}) q_1^{-s_1} q_2^{-s_2'}
  & = - \sum_{s' \in \lambda_\beta}
  (1 - q_1^{-\lambda_{\alpha,1}^\text{T}}) q_1^{-s_1'} q_2^{-s_2'}
  \, , \\
  - \sum_{s_1=1}^{\lambda_{\alpha,1}^\text{T}} \sum_{s_2'=1}^{\lambda_{\beta,1}}
  q_1^{-s_1} (1 - q_2^{-\lambda_{\alpha,s_1}}) q_2^{-s_2'}
  & = - \sum_{s \in \lambda_\alpha} q_1^{-s_1} (1 - q_2^{-\lambda_{\beta,1}}) q_2^{-s_2}
  \, .
 \end{align}
\end{subequations}
Hence we obtain
\begin{align}
 \Xi_{+-}[\lambda_\alpha,\lambda_\beta]
 =
 - \sum_{s_1=1}^{\lambda_{\alpha,1}^\text{T}} \sum_{s_2'=1}^{\lambda_{\beta,1}}
 \left[
 q_1^{-\lambda_{\beta,s_2'}^\text{T}-s_1} q_2^{-\lambda_{\alpha,s_1}-s_2'}
 - q_1^{-s_1} q_2^{-s_2'}
 \right]
 + \sum_{s \in \lambda_\alpha} q_1^{-s_1} q_2^{-\lambda_{\beta,1}-s_2}
 + \sum_{s' \in \lambda_\beta} q_1^{-\lambda_{\alpha,1}^\text{T}-s_1'} q_2^{-s_2'}
 \, ,
\end{align}
and similarly
\begin{align}
 \Xi_{-+}[\lambda_\alpha,\lambda_\beta]
 =  - \sum_{s_1=1}^{\lambda_{\alpha,1}^\text{T}} \sum_{s_2'=1}^{\lambda_{\beta,1}}
 \left[
 q_1^{\lambda_{\beta,s_2'}^\text{T}+s_1-1} q_2^{\lambda_{\alpha,s_1}+s_2'-1}
 - q_1^{s_1-1} q_2^{s_2'-1}
 \right]
 + \sum_{s \in \lambda_\alpha} q_1^{s_1-1} q_2^{\lambda_{\beta,1}+s_2-1}
 + \sum_{s' \in \lambda_\beta} q_1^{\lambda_{\alpha,1}^\text{T}+s_1'-1} q_2^{s_2'-1}
\end{align}
In contrast to the diagonal part $\Xi_{++(--)}$, further simplification does not occur for these off-diagonal ones.
The situation seems similar to the BCD instanton partition function, involving $\phi_a + \phi_b$ in the contour integral, as mentioned in~\cite{Nekrasov:2004vw}. See also~\cite{Marino:2004cn}.
The Jeffrey--Kirwan residue operation is still applicable in such a case~\cite{Hwang:2014uwa,Nakamura:2015zsa}.


\bibliographystyle{utphysurl} \bibliography{wquiver} 
\end{document}